\documentclass[prd,superscriptaddress,preprintnumbers,tightenlines,nofootinbib, eqsecnum]{revtex4-2}

\usepackage{amsmath}
\usepackage{amsfonts}
\usepackage{amssymb}
\usepackage{physics}
\usepackage{bm}
\usepackage{mathrsfs}
\usepackage{graphicx}
\usepackage[colorlinks]{hyperref}
\usepackage[usenames]{color}
\usepackage{stackengine}
\usepackage{tikz}
\usepackage{diagbox}
\usepackage{mathtools}

\definecolor{darkgreen}{rgb}{0,0.5,0}
\hypersetup{urlcolor=darkgreen}
\usepackage{empheq}
\usepackage{ulem}
\usepackage{orcidlink}
\usepackage{multirow}
\usepackage{adjustbox}
 \usepackage{float}

\hypersetup{
 unicode=false,          
 pdftoolbar=true,        
 pdfmenubar=true,        
 pdffitwindow=false,     
 pdfstartview={FitH},    
 pdftitle={Constants of motion and fundamental frequencies for elliptic orbits \\ at fourth post-Newtonian order: aligned-spin contributions},    
 pdfauthor={David Trestini, Jan Steinhoff},     
 pdfsubject={Subject},   
 pdfcreator={Creator},   
 pdfproducer={Producer}, 
 pdfkeywords={keyword1} {key2} {key3}, 
 pdfnewwindow=true,      
 colorlinks=true,       
 linkcolor=red,          
 citecolor=cyan,        
 filecolor=magenta,      
 urlcolor=darkgreen,           
 linktocpage=true
}

\allowdisplaybreaks

\DeclareSymbolFontAlphabet{\mathrsfs}{rsfs}
\DeclareMathAlphabet{\mathcal}{OMS}{cmsy}{m}{n}

\newcommand{\be}{\begin{equation}}
\newcommand{\ee}{\end{equation}}
\newcommand{\bse}{\begin{subequations}}
\newcommand{\ese}{\end{subequations}}
\newcommand{\ba}{\begin{align}}
\newcommand{\ea}{\end{align}}
\newcommand{\nn}{\nonumber}

\renewcommand{\dd}{\mathrm{d}}
\newcommand{\di}{\mathrm{i}}

\newcommand{\chip}{\chi_{+}}
\newcommand{\chim}{\chi_{-}}

\makeatletter
\g@addto@macro\bfseries{\boldmath}
\makeatother

\defcitealias{BBT22}{paper~I}
\defcitealias{T24_QK}{paper~II}
\begin{document}
\interfootnotelinepenalty=10000
\title{Constants of motion and fundamental frequencies for elliptic orbits \\ at fourth post-Newtonian order: aligned-spin contributions}

\author{David \textsc{Trestini}\,\orcidlink{0000-0002-4140-0591}}\email{david.trestini@southampton.ac.uk}

\affiliation{School of Mathematical Sciences and STAG Research Centre, University of Southampton, Southampton SO17 1BJ, United Kingdom}

\author{Jan \textsc{Steinhoff}\,\orcidlink{0000-0002-1614-0214}}\email{jan.steinhoff@aei.mpg.de}

\affiliation{Max-Planck-Institute for Gravitational Physics (Albert-Einstein-Institute),
Am Mühlenberg 1, 14476 Potsdam-Golm, Germany}

\date{\today}

\begin{abstract}
For eccentric systems with aligned spins, we compute all spin contributions to the gauge-invariant conservative map between the constants of motion (energy and angular momentum) and the fundamental (radial and azimuthal) frequencies at the fourth post-Newtonian order. This extends the nonspinning map derived in~\mbox{\href{https://doi.org/10.1088/1361-6382/ae60dc}{[Class. Quantum Grav. 43, 095009 (2026)]}} to include terms that are linear, quadratic, cubic, and quartic in spin, as well as the associated spin-deformability parameters. We also obtain derived quantities such as the redshift and gyroscopic invariants, as well as circular links. We notice that the Blanchet-Iyer-Favata  
post-Newtonian stability criterion that determines the dimensionless frequency of the innermost stable circular orbit (ISCO) is in fact intricately related to the periastron advance for circular orbits. Finally, in the case of unbound orbits, we complete the 4PN scattering angle with all spin contributions. 
\end{abstract}

\maketitle
\tableofcontents

\section{Introduction}
\label{sec:intro}
 
Future gravitational-wave detectors will observe binary systems throughout a much wider range of parameter space than what is currently being observed with LIGO, Virgo, and KAGRA. In particular, large eccentricities and spins are expected to be common, especially in systems with more disparate mass ratios. Gravitational-wave detection, source parameter estimation and LISA's global fit will all require accurate waveforms throughout the observed parameter space. This work is a step towards improving these waveforms in the case of eccentric, spin-aligned systems, including spin-induced deformability (SID): these are the most general nonprecessing compact binary systems (tidal effects can be added separately). Eccentric systems are described by a radial phase $\ell$, often called mean anomaly, and an azimuthal phase $\lambda$, as well as a set of complex amplitudes $h_{\ell m}$ associated with each spherical harmonic mode. The phases are the quantities which need to be modeled with the most accuracy in order for the waveform template to remain in phase with the signal over the whole duration of the detected signal. In the absence of dissipation, these phases are in fact angle variables in the sense of the action-angle formalism, and their time derivatives are, respectively:\footnote{In the context of gravitational self-force, one usually denotes the frequencies $\Omega_r=n$ and $\Omega_\phi = \omega$ and the associated phases $\varphi_r=\ell$ and $\varphi_\phi=\lambda$.} (i) the radial frequency or the mean motion $n=\dot{\ell} = 2\pi/P$ where $P$ is the radial period, i.e. the time elapsed between two periastron passages; and (ii) the azimuthal frequency $\omega=\dot{\lambda}$ at which the separation vector rotates by $2\pi$. When dissipative effects are included, the frequencies $(n,\omega)$ evolve slowly and the phases $(\ell,\lambda)$ will ``chirp''. In order to model this chirp with the best possible accuracy, one usually starts from the flux-balance laws associated with energy and angular momentum in the center-of-mass frame:
\begin{align}\label{eq:flux_balance_EJ}
    \frac{\dd E}{\dd t}&= - \mathcal{F}(E,J) &&\text{and}&\frac{\dd J}{\dd t} &= \mathcal{G}(E,J)\,.
\end{align}
Using the chain rule, this becomes
\begin{subequations}\label{eq:orbital_element_evolution}\begin{align}
    \frac{\dd n}{\dd t}&= - \frac{\partial n(E,J)}{\partial E} \ \mathcal{F}\left[E(n,\omega),J(n,\omega)\right] - \frac{\partial n(E,J)}{\partial J} \ \mathcal{G}\left[E(n,\omega),J(n,\omega)\right] \,,\\
    \frac{\dd \omega}{\dd t}&= - \frac{\partial \omega(E,J)}{\partial E} \ \mathcal{F}\left[E(n,\omega),J(n,\omega)\right] - \frac{\partial \omega(E,J)}{\partial J} \ \mathcal{G}\left[E(n,\omega),J(n,\omega)\right] \,. 
\end{align}\end{subequations}
It thus becomes apparent that two important ingredients are needed to obtain this chirp: (i) the expression of the fluxes in terms of either $(n,\omega)$ or $(E,J)$ ; and (ii) the map between $(n,\omega)$ and $(E,J)$. The latter ingredient is what we address in this work. Note that at fourth post-Newtonian order, this map differs depending on whether it is established in the conservative sector or for the full dissipative dynamics: the small difference between the two is called a Schott term~\cite{Trestini:2025nzr}. We will neglect such terms here and focus only on the conservative relation, but recall that the Schott term will have to be added separately later on.

Specifically, we complete the \textit{conservative} gauge-invariant map between constants of motion (energy and angular momentum) and fundamental (radial and azimuthal) frequencies at fourth post-Newtonian order (4PN). This map was obtained in the nonspinning case in Ref.~\cite{Trestini:2025yyc} for any eccentricity and includes hereditary effects; see also previous works~\cite{Bini:2020wpo, Bini:2020nsb} which perform a small eccentricity expansion of the tail contributions. We complete this map with all spin-dependent terms, including the contributions of SID in the case of neutron stars. Our results heavily rely on the Hamiltonian obtained in Ref.~\cite{Levi:2016ofk} in some modified harmonic coordinates, which we will call EFT coordinates. Since the results of Ref.~\cite{Trestini:2025yyc} were obtained using ADM coordinates, we found it useful to combine the spinning and nonspinning results at the level of gauge-invariant relations.

In Sec.~\ref{sec:hamiltonian}, we summarize the 4PN spinning dynamics. We then rewrite these dynamics in action-angle form in Sec.~\ref{sec:action_angle}, allowing us to combine spinning and nonspinning results at the level of the radial action. We obtain in particular the Hamiltonian in terms of the action variables only. Using this, we obtain in Sec.~\ref{sec:map} the invariant map between constants of motion and fundamental frequencies. From the first law of binary black hole mechanics~\cite{LeTiec:2011ab,Blanchet:2012at,LeTiec:2015kgg}, we deduce in Sec.~\ref{sec:redshift_gyroscopic} the redshift and gyroscopic invariants. Setting the radial action to zero in Sec.~\ref{sec:circular} then leads us to find circular links between constants of motion (or, equivalently, between fundamental frequencies). In Sec.~\ref{sec:isco} we use the circular expression for the periastron advance to reproduce  the Blanchet-Iyer-Favata stability criterion for the ISCO recently computed at 4PN~\cite{Blanchet:2025agj,Blanchet:2026wwa}. Finally, in Sec.~\ref{sec:scattering}, we go back to the level of the dynamics, and this time in the case of unbound orbits, we compute the spin contributions to the 4PN scattering angle. We conclude by summarizing our results in Sec.~\ref{sec:conclusion}.

\section{Spin contributions to the Hamiltonian and equations of motion}
\label{sec:hamiltonian}

We consider two bodies within the pole-dipole approximation, which we label by $1$ and $2$. They are endowed with masses $(m_1,m_2)$, where we choose the labels such that $m_1\ge m_2$, and spins $(\bm{S}_1,\bm{S}_2)$. The two particles are separated by the vector $\bm{x}=\bm{y}_1-\bm{y}_2$, and we define the relative velocity $\bm{v}=\dd \bm{x}/\dd t$ and acceleration $\bm{a}=\dd \bm{v}/\dd t$. The conservative 4PN dynamics of the bodies are described by the following integro-differential equations with respect to coordinate time:
\begin{subequations}\label{eq:eom_precession_equations}\begin{align} 
    \frac{\dd^2 \bm{x}}{\dd t^2} &=\frak{f}_\text{loc}(\bm{x},\bm{v},\bm{S}_1,\bm{S}_2)+\frak{f}_\text{tail}[\bm{x},\bm{v}] \,, \\
    \frac{\dd \bm{S}_A}{\dd t}&= \bm{\Omega}^s_A(\bm{x},\bm{v},\bm{S}_1,\bm{S}_2)\times \bm{S}_A \,,
\end{align}\end{subequations}
where $A\in\{1,2\}$ and where at 4PN, there is a hereditary tail contribution $\frak{f}_\text{tail}[\bm{x},\bm{v}]$   which depends on the whole history of $(\bm{x},\bm{v})$. The tail contribution was extensively studied in Ref.~\cite{Trestini:2025yyc}; since it is decoupled from the spin sector at 4PN order, we can ignore it here. Restricting now to the local sector, there exist coordinate systems which allow these dynamics to be described in terms of an ordinary\footnote{
This is to be contrasted with generalized Hamiltonians, which depend not only on the coordinates $\bm{x}$ and their conjugate momenta  $\bm{p}$, but also on derivatives of the latter:  $\dot{\bm{p}}$, $\ddot{\bm{p}}$, etc.}
Hamiltonian $H_\text{loc}(\bm{x},\bm{p},\bm{S}_1,\bm{S}_2)$. The dynamics then read
\begin{subequations}\label{eq:local_Hamilton_equations}
    \begin{align} 
    \frac{\dd \bm{x}}{\dd t} &= \frac{\partial H_\text{loc}}{\partial \bm{p}}\,, &&&
    \frac{\dd \bm{p}}{\dd t} &= -\frac{\partial H_\text{loc}}{\partial \bm{x}} \,,\\
    \frac{\dd \bm{S}_A}{\dd t} &= \bm{\Omega}^s_A \times \bm{S}_A\,,&&& \bm{\Omega}^s_A&= \frac{\partial H_\text{loc}}{\partial \bm{S}_A}\,.
\end{align}
\end{subequations}
In this work, we decompose the local Hamiltonian as follows:
\begin{align} \label{eq:local_Hamiltonian_decomposition_pp_spin_SID}
    H_\text{loc} = H_\text{loc}^\text{pp} + H_\text{spin} + H_\text{SID}\,.
\end{align}
Here, $H_\text{loc}^\text{pp}$ are the local point-particle contributions in the absence of spin, $H_\text{spin}$ are spin contributions in the case where the two objects are black holes, and $H_\text{SID}$ are spin-induced deformability (SID) corrections, e.g., in the case where the compact objects are neutron stars.
The spinning sector (in the case of two black holes) is decomposed in powers of spin:
\begin{align} \label{eq:spin_Hamiltonian_decomposition_S1234}
    H_\text{spin} = H_{S^1}+H_{S^2}+H_{S^3}+H_{S^4} \,.
\end{align}
The SID piece depends on parameters associated with each particle: at 4PN, we need the  quadrupolar ($\kappa_A$), octupolar  ($\lambda_A$) and hexadecapolar ($\mu_A$) SID
parameters.\footnote{
We relate our notation to the notations used elsewhere in the literature~\cite{Levi:2015msa}: $\kappa_1=C_{1(ES^2)}$, $\kappa_2=C_{2(ES^2)}$,  $\lambda_1=C_{1(BS^3)}$, $\lambda_2=C_{2(BS^3)}$, $\mu_1=C_{1(ES^4)}$, $\mu_2=C_{2(ES^4)}$. We also note that Ref.~\cite{Blanchet:2026wwa} uses $\kappa_A$ and $\lambda_A$ like us, but denotes by $\iota_A$ what we call $\mu_A$.}
For black holes, $\kappa_1 =\kappa_2 = \lambda_1 = \lambda_2= \mu_1=\mu_2  =1$ and $H_\text{SID}$ vanishes. For neutron stars, one typically has $\kappa_A \sim 2 - 14$~\cite{Rahman:2026qho,Laarakkers:1997hb,Pappas:2012qg, Harry:2018hke}, $\lambda_A \sim 4-30$ \cite{Saini:2023gaw} and $\mu_A \sim 8 - 500$ \cite{Yagi:2014bxa}. Finally, for white dwarfs, typical values are $\kappa_A \sim 10^3 - 10^5$ \cite{Rahman:2026qho, Taylor:2019hle}. 
The full nonlocal dynamics can also be described by a Hamiltonian, but the setup is more subtle: we write this full Hamiltonian
\begin{align} \label{eq:Hamiltonian_decomposition_loc_tail}
    H = H_\text{loc} + H_\text{tail}
\end{align}
and will refer to Ref.~\cite{Trestini:2025yyc} for the description of the tail contributions, which will not acquire any spin corrections at 4PN order.  

The point-particle local Hamiltonian was obtained in ADM coordinates in Ref.~\cite{Jaranowski:2015lha}, whereas the nonlocal pieces were studied in~\cite{Damour:2014jta,Damour:2015isa}. The spin contributions were obtained at 4PN in Refs.~\cite{Levi:2016ofk,Perrodin:2010dy,Porto:2010tr,Levi:2010zu,Levi:2015uxa,Porto:2006bt,Porto:2008tb,Levi:2008nh,Porto:2008jj,Levi:2011eq,Levi:2015ixa,Levi:2014gsa,Cho:2022syn} in EFT coordinates, alongside the 2PN nonspinning contributions in these same coordinates (important previous results and checks in other formalisms can be found in Refs.~\cite{Damour:2007nc,Hartung:2011te,Hartung:2013dza,Marsat:2012fn,Bohe:2012mr,Steinhoff:2007mb,Steinhoff:2008ji,Hergt:2010pa,Hergt:2011ik,Hartung:2011ea,Hartung:2013dza,Levi:2014sba,Jakobsen:2022fcj,Hergt:2007ha,Hergt:2008jn,Marsat:2014xea,Vaidya:2014kza}). This result includes all relevant contributions: linear, quadratic, cubic, and quartic in spin, including the SID parameters to account for the case where the two bodies are not black holes.

In this work, we specialize to the case of aligned (or antialigned) spin vectors, which automatically removes any precession of the orbital plane. The spin vectors  are then constant and can be considered as parameters rather than dynamical degrees of freedom: we denote their constant norms as $S_A = |\bm{S}_A|$. 

The Hamilton equations are then typically written in polar coordinates defined by $\bm{x}=(r \cos\phi,r\sin\phi,0)$, where the $z$-axis was chosen to be orthogonal to the orbital plane. We then have the phase-space variables $(r,\phi,p_r,p_\phi)$ which satisfy $|\bm{p}|^2=p_r^2+p_\phi^2/r^2$ and $\bm{p} \cdot \bm{x}/r = p_r$. The Hamiltonian $H_\text{loc}(r,p_r,p_\phi,S_1,S_2)$ does not depend on $\phi$, so the angular momentum $p_\phi$ is conserved and  Hamilton's equations read
\begin{align} \label{eq:local_Hamilton_equations_polar}
    \frac{\dd r}{\dd t} &= \frac{\partial H_\text{loc}}{\partial p_r}\,, &
    \frac{\dd p_r}{\dd t} &= -\frac{\partial H_\text{loc}}{\partial r} \,,&
    \frac{\dd \phi}{\dd t} &= \frac{\partial H_\text{loc}}{\partial p_\phi}\,, &
    \frac{\dd p_\phi}{\dd t} &= -\frac{\partial H_\text{loc}}{\partial \phi} =0\,.
\end{align}

We also introduce the very convenient dimensionless spin parameters $\chi_A= S_A/(G m_A^2)$ which are defined to  correspond to the reduced Kerr parameter: $\chi_A=0$ for a Schwarzschild black hole and $\chi_A=\pm 1$ for a maximally rotating Kerr black hole. The symmetrized and antisymmetrized versions are defined as $\chi_\pm=(\chi_1\pm\chi_2)/2$. Similarly, we define\footnote{The tilde in these definitions is to distinguish these quantities from the common definition $\kappa_\pm=\kappa_1 \pm \kappa_2$, etc.; see e.g. Ref.~\cite{Blanchet:2026wwa}.} the following SID parameters:
\begin{align}
    \tilde{\kappa}_+ &= \frac{(\kappa_1-1)+(\kappa_2-1)}{2}\,, &&& \tilde{\kappa}_- &= \frac{(\kappa_1-1)-(\kappa_2-1)}{2} \,,\nn\\
    \tilde{\lambda}_+ &= \frac{(\lambda_1-1)+(\lambda_2-1)}{2} \,,&&& \tilde{\lambda}_- &= \frac{(\lambda_1-1)-(\lambda_2-1)}{2} \,, \\
    \tilde{\mu}_+ &= \frac{(\mu_1-1)+(\mu_2-1)}{2}\,, &&& \tilde{\mu}_- &= \frac{(\mu_1-1)-(\mu_2-1)}{2} \,.\nn
\end{align}
We choose this definition such that, in the case of two black holes, we have $ \tilde{\kappa}_+= \tilde{\kappa}_-=  \tilde{\lambda}_+ =  \tilde{\lambda}_- = \tilde{\mu}_+ = \tilde{\mu}_-= 0$.

\section{Action-angle formulation of the spinning dynamics}
\label{sec:action_angle}

We wish to construct a set of action-angle variables associated with our problem following the steps described in Sec.~II.C of \cite{Trestini:2025yyc}. 
First, we need to address the following issue related to coordinate systems. On the one hand, the point-particle Hamiltonian $H_\text{loc}^\text{pp}$ is known to 4PN accuracy in ADM coordinates~\cite{Jaranowski:2015lha}. On the other hand, spinning contributions are known to 4PN order in EFT coordinates from Ref.~\cite{Levi:2016ofk}, but the accompanying point-particle contributions are only provided\footnote{While this work was in preparation, Ref.~\cite{Blanchet:2026wwa} obtained the local nonspinning Hamiltonian up to 4PN in these EFT coordinates.} to 2.5PN in that reference (in fact the 2.5PN coefficient is vanishing, so the highest nonvanishing coefficient is 2PN). The lowest-order spin effects enter at 1.5PN: when they couple with uncontrolled 3PN terms, they would contribute to 4.5PN order spin effects, which is beyond the precision sought in this work. To check this in practice, we have completed the 3PN and 4PN point-particle Hamiltonian in EFT coordinates using some agnostic parametrization. Using this parametrized Hamiltonian, we follow the steps described hereafter to obtain the gauge-invariant radial action. We then check that (i) the point-particle radial action obtained here agrees at 2PN with the one obtained in \cite{Trestini:2025yyc} using ADM coordinates; and (ii) the spinning contributions to the radial action do not depend on any of the arbitrary parameters introduced, which means that there are no couplings between the 3PN point-particle contributions and the spin sector. Since the result is gauge invariant, we then keep only the spinning contributions and add them to the complete 4PN nonspinning contributions obtained in Ref.~\cite{Trestini:2025yyc}. This also allows us to only focus on the local piece.

We first solve for $(p_r^\text{loc},p_\phi^\text{loc})$ in Hamilton's equations \eqref{eq:local_Hamilton_equations_polar}, which can be done iteratively up to 4PN order. We are left with expressions of the type
\begin{subequations}\label{eq:pr_pphi_of_rDot_phiDot}
    \begin{align} \label{seq:pr_of_rDot_phiDot}
    p_r^\text{loc}&= m \nu \dot{r}   \Biggl\{1+\frac{1}{c^2}\Biggl[\frac{\dot{r}^2+r^2 \dot{\phi}^2}{2} \left(1-3 \nu \right)  +\frac{G m}{r} (3+2\nu )  \Biggr] + \dots \Biggr\}\,,\\
    \label{seq:pphi_of_rDot_phiDot}
    p_\phi^\text{loc}=J^\text{loc}&= m \nu r^2 \dot{\phi}\Biggl\{1+\frac{1}{c^2}\Biggl[\frac{\dot{r}^2+r^2 \dot{\phi}^2}{2} \left(1-3 \nu \right)  +\frac{G m}{r} (3+\nu )   \Biggr] + \cdots \Biggr\} \,.
\end{align}
\end{subequations}
We can inject these expressions into the Hamiltonian to obtain the conserved energy $E^\text{loc}=H^\text{loc}(r,\dot{r},\dot{\phi})$, which has the structure 
\begin{align}\label{eq:E_of_rDot_phiDot}
    E^\text{loc} & =m \nu  \, \Biggl\{\frac{\dot{r}^2}{2}+\frac{r^2 \dot{\phi}^2}{2}-\frac{G m}{r}+ \cdots \Biggr\} \,.
\end{align}
We have now constructed the constants of motion $(E,J)$ in terms of $(r,\dot{r},\dot{\phi})$. We can invert these expressions iteratively. Omitting the ``loc'' labels for legibility, we find that the solution reads
\begin{align} \label{eq:rDot2_phiDot_of_r}
    \dot{r}^2 &= \mathcal{R}(1/r) &\text{and}&& \dot{\phi} &= \mathcal{S}(1/r)\,,
\end{align}
where $\mathcal{R}(s)$ and $\mathcal{S}(s)$ are polynomials in $s$. They are given by
\begin{subequations}
   \label{eq:R_S_of_s}
\begin{align}   \label{seq:R_of_s}
   \mathcal{R}(s) &= A + 2B s + C s^2 + D_1 s^3 + D_2 s^4 + D_3 s^5 + D_4 s^6  + D_5 s^7 + D_6 s^8 + D_7 s^9 + \mathcal{O}(9)   \,,  \\  \label{seq:S_of_s}
   \mathcal{S}(s) &= F s^2 + I_1 s^3 + I_2 s^4 + I_3 s^5 + I_4 s^6  + I_5 s^7 + I_6 s^8 + I_7 s^9 + \mathcal{O}(9)  \,,
\end{align} 
\end{subequations}
where $\mathcal{O}(9)$ denotes ignorable 4.5PN terms. The coefficients $(A,B,C,F,D_n,I_n)$ are expressed as functions of energy and angular momentum. In practice, we express them in terms of the reduced variables
\begin{align}   \label{eq:def_varepsilon_j}
    \varepsilon= - \frac{2E}{m\nu c^2} &&\text{and}&& j = -\frac{2 J^2 E}{G^2 m^5 \nu^3} \,,
\end{align}
which are defined such that $\varepsilon=\mathcal{O}(2)$, $j=\mathcal{O}(1)$, $\varepsilon>0$, and $j>0$. 

We then inject the  expressions for $(\dot{r}^2,\dot{\phi})$ obtained in~\eqref{eq:rDot2_phiDot_of_r}  into~\eqref{seq:pr_of_rDot_phiDot}. We then find that the radial momentum reads
\begin{align}   \label{eq:pr2_of_r}
    p_r^2&=\mathcal{I}(1/r) \,,
\end{align}
where $\mathcal{I}(s)$ is also a polynomial which reads
\begin{align} \label{eq:I_of_s}
   \mathcal{I}(s) &= \mathcal{A} + 2\mathcal{B} s + \mathcal{C} s^2 + \mathcal{D}_1 s^3 + \mathcal{D}_2 s^4 + \mathcal{D}_3 s^5 + \mathcal{D}_4 s^6  + \mathcal{D}_5 s^7 + \mathcal{D}_6 s^8 + \mathcal{D}_7 s^9 + \mathcal{O}(9)  \,,
\end{align}
and where the coefficients  $(\mathcal{A},\mathcal{B},\mathcal{C},\mathcal{D}_n)$ are expressed as functions of energy and angular momentum.

We now have everything set up to compute the gauge-invariant radial action, which is given by
\begin{align} \label{eq:Ir_loop_integral}
    I_r &= \frac{1}{2\pi} \oint \dd r \,p_r(r)= \frac{1}{2\pi} \oint \frac{\dd s}{s^2} \sqrt{\mathcal{I}(s)} \,.
\end{align}
This integral is performed \textit{à la Sommerfeld} as described in detail in Sec. III.A. of \cite{Trestini:2025yyc}; see Eq.~(3.7) of that reference for the explicit result for $I_r(\mathcal{A},\mathcal{B},\mathcal{C},\mathcal{D}_n)$. Finally, we can replace the coefficients by their expressions in terms of energy and angular momentum.

Since the relation between the radial action and the energy and angular momentum is gauge-invariant, there is no issue with combining the point-particle contribution obtained through ADM coordinates with the spinning contribution obtained using EFT coordinates. In particular, this means that our local treatment will only inform the spinning sector, whereas the point-particle contributions are taken from Ref.~\cite{Trestini:2025yyc}, where both local and nonlocal effects were carefully accounted for.
Introducing the reduced radial action $i_r=I_r/(G m^2 \nu)$, we find that the radial action decomposes into point-particle, spinning and deformability contributions:
\begin{align}\label{eq:ir_decomposition_pp_spin_SID}
    i_r = i_r^\text{pp}(\varepsilon,j) + i_r^\text{spin}(\varepsilon,j,\chi_A) + i_r^\text{SID}(\varepsilon,j,\chi_A,\kappa_A,\lambda_A,\mu_A)\,.
\end{align}
 The point-particle contribution $i_r^\text{pp}(\varepsilon,j)$ is provided in (4.49b) of \cite{Trestini:2025yyc}; it contains local and nonlocal contributions that are provided, respectively,  in Eqs.~(3.9) and  (4.48) of that reference. The BH spin radial action is decomposed into 
\begin{align} \label{eq:spin_ir_decomposition_S1234}
    i_r^\text{spin} &= i_r^{S^1} + i_r^{S^2} + i_r^{S^3}  +i_r^{S^4}   \,. 
\end{align}
The contributions read
\begin{subequations}\label{eq:ir_S1234_of_varpesilon_j_expression}
\begin{align} \label{seq:irS1_of_varpesilon_j_expression}
    i_r^{S^1}&= \frac{1}{c\sqrt{\varepsilon}} \Bigg\{\frac{\varepsilon^{3/2}}{j}\Bigg[\chip(-2+\nu) - 2 \delta \chim\Bigg] \nn\\
    & \qquad + \varepsilon^{5/2}\Bigg[\frac{1}{j}\Bigg(\chip(6-8\nu+\nu^2) + \delta \chim(6-2\nu)\Bigg)+\frac{1}{j^2}\Bigg(\chip(-21+\frac{147}{8}\nu- \frac{3}{4}\nu^2) + \delta \chim(-21+\frac{21}{8}\nu)\Bigg)\Bigg] \nn\\
    & \qquad + \varepsilon^{7/2}\Bigg[
    \frac{1}{j}\Bigg(\chip \bigg(-\frac{15}{4}+\frac{21}{2}\nu-\frac{27}{4}\nu^2+\frac{3}{4}\nu^3 \bigg) + \delta \chim \bigg(-\frac{15}{4}+ \frac{39}{8}\nu- \frac{3}{2}\nu^2\bigg)\Bigg)\nn\\
   &\qquad\qquad + \frac{1}{j^2}\Bigg(\chip \bigg(105-\frac{717}{4}\nu+\frac{207}{4}\nu^2-\frac{3}{2}\nu^3 \bigg) + \delta \chim \bigg(105- \frac{297}{4}\nu+ \frac{21}{4}\nu^2\bigg)\Bigg) \nn\\
   &\qquad\qquad + \frac{1}{j^3}\Bigg(\chip \bigg(-\frac{495}{2}+\frac{5055}{16}\nu-\frac{115}{2}\nu^2+\frac{5}{8}\nu^3 \bigg) + \delta \chim \bigg(-\frac{495}{2}+ \frac{1755}{16}\nu- \frac{25}{8}\nu^2 \bigg)\Bigg)\Bigg]
    +\mathcal{O}(\varepsilon^{9/2})\Bigg\} \,, \\
    \label{seq:irS2_of_varpesilon_j_expression}
    i_r^{S^2} &= \frac{1}{c \sqrt{\varepsilon }}\Bigg\{\frac{\varepsilon ^2}{j^{3/2}} \Bigg[\left(\frac{1}{2}-2 \nu \right) \chi_{-}^2+\delta  \chi_{-} \chi_{+}+\frac{\chi_{+}^2}{2}\Bigg]\nn\\
    &\qquad+\varepsilon ^3
   \Bigg[\frac{1}{j^{3/2}}\Bigg(\left(-\frac{11}{2}+23 \nu -3 \nu ^2\right) \chi_{-}^2+\delta  (-11+9 \nu ) \chi_{-} \chi_{+}+\left(-\frac{11}{2}+8 \nu -2 \nu ^2\right) \chi_{+}^2\Bigg) \nn\\
   &\qquad\qquad +\frac{1}{j^{5/2}}\Bigg(\left(21-\frac{171 \nu }{2}+3 \nu ^2\right) \chi_{-}^2+\delta  (42-24 \nu ) \chi_{-} \chi_{+}+\left(21-\frac{45 \nu }{2}+6 \nu ^2\right) \chi_{+}^2\Bigg)\Bigg]\nn\\
   &\qquad +\varepsilon ^4 \Bigg[\frac{1}{j^{7/2}}\Bigg(\left(\frac{7425}{16}-\frac{3945 \nu }{2}+\frac{6705 \nu ^2}{16}-\frac{15 \nu ^3}{4}\right) \chi_{-}^2+\delta \left(\frac{7425}{8}-\frac{3405 \nu }{4}+\frac{645 \nu ^2}{8}\right) \chi_{-} \chi_{+} \nn\\
   &\qquad\qquad\qquad\qquad +\left(\frac{7425}{16}-735 \nu +\frac{4725 \nu ^2}{16}-15 \nu ^3\right) \chi_{+}^2\Bigg)\nn\\
   &\qquad\qquad+\frac{1}{j^{3/2}}\Bigg(\left(\frac{117}{16}-37 \nu +\frac{489 \nu ^2}{16}-3 \nu ^3\right) \chi_{-}^2+\delta  \left(\frac{117}{8}-32 \nu +\frac{97 \nu ^2}{8}\right) \chi_{-}
   \chi_{+} \nn\\
   &\qquad\qquad\qquad\qquad +\left(\frac{117}{16}-\frac{97 \nu }{4}+\frac{321 \nu ^2}{16}-3 \nu ^3\right) \chi_{+}^2\Bigg)\nn\\
   &\qquad\qquad  +\frac{1}{j^{5/2}}\Bigg(\left(-\frac{777}{4}+\frac{3453 \nu }{4}-\frac{2577 \nu
   ^2}{8}+\frac{15 \nu ^3}{2}\right) \chi_{-}^2+\delta  \left(-\frac{777}{2}+477 \nu -\frac{321 \nu ^2}{4}\right) \chi_{-} \chi_{+} \nn\\
   &\qquad\qquad\qquad\qquad +\left(-\frac{777}{4}+\frac{1563 \nu
   }{4}-\frac{1581 \nu ^2}{8}+18 \nu ^3\right) \chi_{+}^2\Bigg)\Bigg]\Bigg\} \,,\\
   \label{seq:irS3_of_varpesilon_j_expression}
   i_r^{S^3} &= \frac{\varepsilon ^{7/2}}{c\sqrt{\varepsilon}} \Bigg\{\frac{1}{j^2}\Bigg[\delta  (3-12 \nu ) \chi_{-}^3+\left(9-39 \nu +12 \nu ^2\right) \chi_{-}^2 \chi_{+}+\delta  (9-6 \nu ) \chi_{-} \chi_{+}^2+(3-3 \nu
   ) \chi_{+}^3\Bigg]\nn\\*
   &\qquad +\frac{1}{j^3}\Bigg[\delta  (-10+40 \nu ) \chi_{-}^3+\left(-30+\frac{255 \nu }{2}-30 \nu ^2\right) \chi_{-}^2 \chi_{+}+\delta  (-30+15 \nu ) \chi_{-}
   \chi_{+}^2+\left(-10+\frac{15 \nu }{2}\right) \chi_{+}^3 \Bigg]\Bigg\} \,,\\
   \label{seq:irS4_of_varpesilon_j_expression}
   i_r^{S^4} &= \frac{\varepsilon^{4}}{c\sqrt{\varepsilon}}\Bigg\{  \frac{1}{j^{5/2}}\Bigg[\left(-\frac{3}{4}+6 \nu -12 \nu ^2\right) \chi_{-}^4+\delta  (-3+12 \nu ) \chi_{-}^3 \chi_{+}+\left(-\frac{9}{2}+18 \nu \right) \chi
   _-^2 \chi_{+}^2-3 \delta  \chi_{-} \chi_{+}^3-\frac{3 \chi_{+}^4}{4}\Bigg] \nn\\*
   &\qquad +\frac{1}{j^{7/2}}\Bigg[\left(\frac{15}{8}-15 \nu +30 \nu ^2\right) \chi_{-}^4+\delta 
   \left(\frac{15}{2}-30 \nu \right) \chi_{-}^3 \chi_{+}+\left(\frac{45}{4}-45 \nu \right) \chi_{-}^2 \chi_{+}^2+\frac{15}{2} \delta  \chi_{-} \chi
   _+^3+\frac{15 \chi_{+}^4}{8}\Bigg] \Bigg\} \,.
\end{align}
\end{subequations}
We relegate the SID contribution to the Supplemental Material~\cite{Supplemental}. 

We now wish to invert this relation to obtain the Hamiltonian in action-angle form. We introduce 
$i_\phi = J/(G m^2 \nu) =\sqrt{j/(c^2\varepsilon)}$ and $i_{r\phi} = i_r + i_\phi$. Inverting the expression for the radial action order by order and identifying the energy to the Hamiltonian, we obtain:  
\begin{align}\label{eq:Delaunay_Hamiltonian_decomposition_pp_spin_SID}
    H(i_{r\phi},i_\phi,\chi_A,\kappa_A,\lambda_A,\mu_A)&= H_\text{pp}(i_{r\phi},i_\phi)+ H_\text{spin}(i_{r\phi},i_\phi,\chi_A)+H_\text{SID}(i_{r\phi},i_\phi,\chi_A,\kappa_A,\lambda_A,\mu_A)\,.
\end{align}
The point-particle Hamiltonian is given by (4.47) of \cite{Trestini:2025yyc}; in that reference, see (3.13) for the local part and (4.46) for the tail part. The spin part (decomposed into powers of spin) and the SID contributions are given in the Supplemental Material~\cite{Supplemental}. An interesting direction for future work would be to extend this Hamiltonian to include dissipative effects as in Ref.~\cite{Blanco:2026prj}, which follows the Magnusian framework~\cite{Kim:2025gis}. 

\section{Map between constants of motion and fundamental frequencies}
\label{sec:map}

Now that we have the action-angle formulation of the dynamics, the fundamental frequencies of motion follow immediately. The radial frequency $n$ and the azimuthal frequency $\omega$ are obtained through
\begin{align} \label{eq:n_omega_from_Delaunay_Hamiltonian}
    n &= \frac{1}{G m^2  \nu} \frac{\partial H(i_{r\phi},i_\phi)}{\partial i_{r\phi}} &&\text{and}& \omega &= n+\frac{1}{G m^2 \nu} \frac{\partial H(i_{r\phi},i_\phi)}{\partial i_{\phi}} \,.
\end{align}
This yields expressions in terms of $(i_{r\phi},i_\phi)$, which are then replaced in terms of $(\varepsilon,j)$ using \eqref{eq:ir_decomposition_pp_spin_SID}. Alternatively, we can obtain the radial period $P = 2\pi/n$ and periastron advance $K=\omega/n$ directly from the radial action using
\begin{align}
    \label{eq:P_K_from_Ir}
    P &= 2\pi \,\frac{\partial I_r(E,J)}{\partial E} &&\text{and}&  K &= - \frac{\partial I_r(E,J)}{\partial J} \,,
\end{align}
which yields a result directly expressed in terms of energy and angular momentum.  We then define the Blanchet frequency parameters $x=(Gm\omega/c^3)^{2/3}$ and $\iota = 3x/(K-1)$. All these results are naturally decomposed as 
\begin{subequations}\label{eq:n_omega_etc_decomposition_pp_spin_SID}
    \begin{align} 
    \label{seq:n_decomposition_pp_spin_SID}
    n&=n_\text{pp}(\varepsilon,j)+n_\text{spin}(\varepsilon,j,\chi_A)+n_\text{SID}(\varepsilon,j,\chi_A,\kappa_A,\lambda_A,\mu_A)\,,\\
    \label{seq:omega_decomposition_pp_spin_SID}
    \omega&=\omega_\text{pp}(\varepsilon,j)+\omega_\text{spin}(\varepsilon,j,\chi_A)+\omega_\text{SID}(\varepsilon,j,\chi_A,\kappa_A,\lambda_A,\mu_A)\,,\\
    \label{seq:P_decomposition_pp_spin_SID}
    P&=P_\text{pp}(\varepsilon,j)+P_\text{spin}(\varepsilon,j,\chi_A)+P_\text{SID}(\varepsilon,j,\chi_A,\kappa_A,\lambda_A,\mu_A)\,,\\
    \label{seq:K_decomposition_pp_spin_SID}
    K&=K_\text{pp}(\varepsilon,j)+K_\text{spin}(\varepsilon,j,\chi_A)+K_\text{SID}(\varepsilon,j,\chi_A,\kappa_A,\lambda_A,\mu_A)\,,\\
    \label{seq:x_decomposition_pp_spin_SID}
    x&=x_\text{pp}(\varepsilon,j)+x_\text{spin}(\varepsilon,j,\chi_A)+x_\text{SID}(\varepsilon,j,\chi_A,\kappa_A,\lambda_A,\mu_A)\,,\\
    \label{seq:iota_decomposition_pp_spin_SID}
    \iota&=\iota_\text{pp}(\varepsilon,j)+\iota_\text{spin}(\varepsilon,j,\chi_A)+\iota_\text{SID}(\varepsilon,j,\chi_A,\kappa_A,\lambda_A,\mu_A) \,.
\end{align}
\end{subequations}
The point-particle contributions can all be found in Ref.~\cite{Trestini:2025yyc}; see Table I of that reference for details. Note that due to cross terms, $\iota(\varepsilon,j)$ includes $S^5$ and $S^6$ contributions.  For brevity, we relegate the new spin and SID contributions to the Supplemental Material~\cite{Supplemental}. We have checked that the radial frequency and the periastron advance are in agreement with the 3PN results of Ref.~\cite{Henry:2023tka}. 

Finally, this map can be inverted order-by-order: we find that the constants of motion can be expressed in terms of the Blanchet frequency variables as 
\begin{subequations}\label{eq:varepsilon_j_pp_spin_SID}
    \begin{align}
    \label{seq:varepsilon_decomposition_pp_spin_SID}
    \varepsilon&=\varepsilon_\text{pp}(x,\iota)+\varepsilon_\text{spin}(x,\iota,\chi_A)+\varepsilon_\text{SID}(x,\iota,\chi_A,\kappa_A,\lambda_A,\mu_A) \,,\\
    \label{eq:j_decomposition_pp_spin_SID}
    j&=j_\text{pp}(x,\iota)+j_\text{spin}(x,\iota,\chi_A)+j_\text{SID}(x,\iota,\chi_A,\kappa_A,\lambda_A,\mu_A) \,.
\end{align}
\end{subequations}
The point-particle contributions can be found in (5.11) and (5.12) of Ref.~\cite{Trestini:2025yyc}.
The spin contributions are decomposed as 
\begin{align}
    \varepsilon_\text{spin} &=  \varepsilon_{S^1}+\varepsilon_{S^2}+\varepsilon_{S^3}+\varepsilon_{S^4} &\text{and}&&
    j_\text{spin} &=  j_{S^1}+j_{S^2}+j_{S^3}+j_{S^4}+j_{S^5}+j_{S^6}\,.
\end{align}
The individual pieces associated with the energy read
\begin{subequations}
\label{eq:varepsilon_S1234_of_x_iota_expression}
\begin{align}
\label{seq:varepsilon_S1_of_x_iota_expression}
    \varepsilon_{S^1} &= \frac{x^{7/2}}{\iota }\Bigg\{\delta  \left(-\frac{14}{3}+\frac{4 \nu }{3}\right) \chi_{-}+\left(-\frac{14}{3}+\frac{23 \nu }{3}-\frac{2 \nu ^2}{3}\right) \chi_{+}\Bigg\}
    \nn\\
    &\quad +x^{9/2}
   \Bigg\{\frac{1}{\iota ^2}\Bigg[\delta  \left(\frac{35}{6}+\left(-\frac{887}{36}+\frac{41 \pi ^2}{48}\right) \nu +\frac{19 \nu ^2}{6}\right) \chi_{-}\nn\\
   &\qquad\qquad\quad \qquad+\left(\frac{35}{6}+\left(-\frac{541}{18}+\frac{41 \pi
   ^2}{48}\right) \nu +\left(\frac{163}{9}-\frac{41 \pi ^2}{96}\right) \nu ^2-\frac{7 \nu ^3}{3}\right) \chi_{+}\Bigg]\nn\\
   &\qquad\qquad +\frac{1}{\iota }\Bigg[\delta  \left(-15+\frac{7 \nu }{9}+\frac{10 \nu ^2}{9}\right)
   \chi_{-}+\left(-15+\frac{329 \nu }{18}+\frac{32 \nu ^2}{9}-\frac{5 \nu ^3}{9}\right) \chi_{+}\Bigg]\Bigg\} \,, \\
   \label{seq:varepsilon_S2_of_x_iota_expression}
   \varepsilon_{S^2} &= \frac{x^4}{\iota ^{3/2}} \Bigg\{\left(\frac{35}{36}-\frac{61 \nu }{18}-\frac{10 \nu ^2}{3}\right) \chi_{-}^2+\delta  \left(\frac{35}{18}+\frac{19 \nu }{9}-\frac{4 \nu ^2}{3}\right) \chi_{-} \chi_{+}+\left(\frac{35}{36}+\frac{29 \nu }{18}-\frac{47 \nu ^2}{18}+\frac{\nu ^3}{3}\right) \chi_{+}^2\Bigg\} \nn\\
   &\quad +x^5 \Bigg\{\frac{1}{\iota^{3/2}}\Bigg[\left(\frac{235}{32}-\frac{2977 \nu
   }{108}-\frac{2365 \nu ^2}{216}-\frac{65 \nu ^3}{18}\right) \chi_{-}^2+\delta  \left(\frac{235}{16}+\frac{173 \nu }{54}-\frac{125 \nu ^2}{108}-\frac{13 \nu ^3}{9}\right) \chi_{-}
   \chi_{+} \nn\\
   &\qquad\qquad\qquad+\left(\frac{235}{32}+\frac{301 \nu }{216}-\frac{2129 \nu ^2}{432}-\frac{181 \nu ^3}{108}+\frac{13 \nu ^4}{36}\right) \chi_{+}^2\Bigg] \nn\\
   &\qquad +\frac{1}{\iota ^{5/2}}\Bigg[\left(-\frac{1295}{288}+\left(-\frac{5683}{216}+\frac{1927 \pi ^2}{1152}\right) \nu +\left(\frac{39799}{216}-\frac{1927 \pi ^2}{288}\right) \nu ^2-\frac{425 \nu ^3}{18}\right)
   \chi_{-}^2 \nn\\
   &\qquad\qquad\qquad+\delta  \left(-\frac{1295}{144}+\left(-\frac{10105}{108}+\frac{1927 \pi ^2}{576}\right) \nu +\left(\frac{2138}{27}-\frac{287 \pi ^2}{144}\right) \nu ^2-\frac{185 \nu
   ^3}{18}\right) \chi_{-} \chi_{+} \nn\\
   &\qquad\qquad\qquad +\left(-\frac{1295}{288}+\left(-\frac{5321}{108}+\frac{1927 \pi ^2}{1152}\right) \nu +\left(\frac{33095}{432}-\frac{287 \pi ^2}{144}\right) \nu
   ^2+\left(-\frac{1621}{54}+\frac{287 \pi ^2}{576}\right) \nu ^3+\frac{115 \nu ^4}{36}\right) \chi_{+}^2\Bigg]\Bigg\} \,,\\
   \label{seq:varepsilon_S3_of_x_iota_expression}
    \varepsilon_{S^3} &= \frac{x^{9/2}}{\iota ^2}  \Bigg\{\delta  \left(\frac{5}{27}-\frac{10 \nu }{27}-\frac{112 \nu ^2}{27}\right) \chi_{-}^3+\left(\frac{5}{9}+\frac{5 \nu }{18}-\frac{137 \nu ^2}{9}+\frac{104 \nu ^3}{9}\right)
   \chi_{-}^2 \chi_{+} \nn \\
   &\qquad\qquad  +\delta  \left(\frac{5}{9}+\frac{35 \nu }{9}-\frac{74 \nu ^2}{9}+\frac{16 \nu ^3}{9}\right) \chi_{-} \chi_{+}^2+\left(\frac{5}{27}+\frac{95 \nu
   }{54}-\frac{17 \nu ^2}{3}+\frac{80 \nu ^3}{27}-\frac{8 \nu ^4}{27}\right) \chi_{+}^3\Bigg\} \,,\\
   \label{seq:varepsilon_S4_of_x_iota_expression}
    \varepsilon_{S^4} &= \frac{x^5}{\iota ^{5/2}}\Bigg\{\left(\frac{49}{864}-\frac{65 \nu }{432}-\frac{182 \nu ^2}{27}+\frac{695 \nu ^3}{27}\right) \chi_{-}^4+\delta  \left(\frac{49}{216}+\frac{263 \nu }{108}-\frac{776 \nu
   ^2}{27}+\frac{700 \nu ^3}{27}\right) \chi_{-}^3 \chi_{+} \nn \\
   &\qquad\quad +\left(\frac{49}{144}+\frac{493 \nu }{72}-\frac{479 \nu ^2}{8}+\frac{1199 \nu ^3}{12}-\frac{245 \nu ^4}{9}\right) \chi_{-}^2 \chi_{+}^2+\delta  \left(\frac{49}{216}+\frac{821 \nu }{108}-\frac{1039 \nu ^2}{36}+\frac{1015 \nu ^3}{54}-\frac{70 \nu ^4}{27}\right) \chi_{-} \chi_{+}^3 \nn \\
   &\qquad\quad +\left(\frac{49}{864}+\frac{1051 \nu }{432}-\frac{2585 \nu ^2}{216}+\frac{1337 \nu ^3}{108}-\frac{847 \nu ^4}{216}+\frac{35 \nu ^5}{108}\right) \chi_{+}^4\Bigg\} \,.
\end{align}
\end{subequations}
The individual pieces of the angular momentum read
\begin{subequations}
    \label{eq:j_S1234_of_x_iota_expression}
\begin{align}
   \label{seq:j_S1_of_x_iota_expression}
    j_{S^1} &= \sqrt{x} \sqrt{\iota } \Bigg\{-\frac{4 \delta  \chi_{-}}{3}+\left(-\frac{4}{3}+\frac{2 \nu }{3}\right) \chi_{+}\Bigg\}\nn\\
    &\quad +x^{3/2} \Bigg\{\sqrt{\iota } \Bigg[\delta  \left(\frac{3}{2}-\frac{8 \nu }{9}\right) \chi_{-}+\left(\frac{3}{2}-\frac{131 \nu }{36}+\frac{4 \nu ^2}{9}\right) \chi_{+}\Bigg] +\frac{1}{\sqrt{\iota }}\Bigg[\delta  \left(-\frac{47}{6}-\frac{3 \nu }{2}\right) \chi_{-} +\left(-\frac{47}{6}+\frac{113 \nu }{12}+\frac{3 \nu ^2}{2}\right) \chi_{+} \Bigg]\Bigg\} \nn\\
   &\quad +x^{5/2} \Bigg\{\delta  \left(-\frac{34}{3}+4 \nu \right) \chi_{-}+\left(-\frac{34}{3}+\frac{41 \nu }{3}-2 \nu ^2\right) \chi_{+}\nn\\
   &\qquad +\sqrt{\iota } \Bigg[\delta  \left(\frac{625}{96}-\frac{13 \nu }{36}-\frac{29 \nu ^2}{72}\right) \chi_{-}+\left(\frac{625}{96}-\frac{4243 \nu }{576}-\frac{91 \nu ^2}{72}+\frac{29 \nu ^3}{144}\right) \chi_{+}\Bigg] \nn\\
   &\qquad +\frac{1}{\sqrt{\iota }}\Bigg[\delta  \left(\frac{289}{16}+\left(-\frac{175}{144}-\frac{41 \pi
   ^2}{64}\right) \nu -\frac{7 \nu ^2}{4}\right) \chi_{-}+\left(\frac{289}{16}+\left(-\frac{6947}{288}-\frac{41 \pi ^2}{64}\right) \nu +\left(\frac{53}{144}+\frac{41 \pi ^2}{128}\right) \nu
   ^2+\frac{7 \nu ^3}{4}\right) \chi_{+}\Bigg] \nn\\
   &\qquad +\frac{1}{\iota ^{3/2}}\Bigg[\delta  \left(\frac{581}{96}+\left(-\frac{17453}{144}+\frac{1025 \pi ^2}{192}\right) \nu -\frac{25 \nu ^2}{4}\right) \chi_{-} \nn\\
   &\quad\qquad\qquad+\left(\frac{581}{96}+\left(-\frac{72779}{576}+\frac{1025 \pi ^2}{192}\right) \nu +\left(\frac{11683}{144}-\frac{1025 \pi ^2}{384}\right) \nu ^2+\frac{95 \nu ^3}{16}\right) \chi_{+}\Bigg]\Bigg\} \,,\\
      \label{seq:j_S2_of_x_iota_expression}
    j_{S^2} &= x \Bigg\{\left(-\frac{7}{18}+\frac{14 \nu }{9}\right) \chi_{-}^2+\delta  \left(-\frac{7}{9}+\frac{8 \nu }{9}\right) \chi_{-} \chi_{+}+\left(-\frac{7}{18}+\frac{8 \nu }{9}-\frac{2
   \nu ^2}{9}\right) \chi_{+}^2\Bigg\} \nn\\
   & +x^2 \Bigg\{\left(-\frac{9}{4}+\frac{1921 \nu }{216}+\frac{77 \nu ^2}{54}\right) \chi_{-}^2+\delta  \left(-\frac{9}{2}+\frac{229 \nu }{108}+\frac{22
   \nu ^2}{27}\right) \chi_{-} \chi_{+}+\left(-\frac{9}{4}+\frac{481 \nu }{216}+\frac{35 \nu ^2}{54}-\frac{11 \nu ^3}{54}\right) \chi_{+}^2\nn\\
   &\qquad+\frac{1}{\iota }\Bigg[\left(-\frac{35}{36}-\frac{\nu
   }{9}+11 \nu ^2\right) \chi_{-}^2+\delta  \left(-\frac{35}{18}+\frac{17 \nu }{9}+6 \nu ^2\right) \chi_{-} \chi_{+}+\left(-\frac{35}{36}+\frac{53 \nu }{9}+\frac{\nu ^2}{9}-2 \nu
   ^3\right) \chi_{+}^2\Bigg]\Bigg\} \nn\\
   & +x^3 \Bigg\{\left(-\frac{749}{144}+\frac{16799 \nu }{864}+\frac{1649 \nu ^2}{216}+\frac{91 \nu ^3}{108}\right) \chi_{-}^2+\delta 
   \left(-\frac{749}{72}-\frac{253 \nu }{432}+\frac{8 \nu ^2}{3}+\frac{13 \nu ^3}{27}\right) \chi_{-} \chi_{+}\nn\\
   &\qquad\qquad\quad+\left(-\frac{749}{144}+\frac{671 \nu }{864}+\frac{1001 \nu
   ^2}{216}-\frac{\nu ^3}{27}-\frac{13 \nu ^4}{108}\right) \chi_{+}^2 \nn\\
   &\qquad + \frac{1}{\sqrt{\iota
   }}\Bigg[\left(\frac{21}{4}-\frac{43 \nu }{2}+\frac{2 \nu ^2}{3}\right) \chi_{-}^2+\delta  \left(\frac{21}{2}-7 \nu
   -\frac{4 \nu ^2}{3}\right) \chi_{-} \chi_{+}+\left(\frac{21}{4}-\frac{13 \nu }{2}+\frac{\nu ^2}{2}+\frac{\nu ^3}{3}\right) \chi_{+}^2\Bigg]\nn\\
   &\qquad + \frac{1}{\iota ^2}\Bigg[\left(-\frac{5}{72}+\left(-\frac{7027}{24}+\frac{11275 \pi ^2}{768}\right) \nu +\left(\frac{84701}{72}-\frac{11275 \pi ^2}{192}\right) \nu ^2+\frac{147 \nu ^3}{2}\right) \chi_{-}^2\nn\\
   &\qquad\qquad\quad +\delta  \left(-\frac{5}{36}+\left(-\frac{21997}{36}+\frac{11275 \pi ^2}{384}\right) \nu +\left(\frac{7289}{18}-\frac{205 \pi ^2}{12}\right) \nu ^2+38 \nu ^3\right) \chi_{-}
   \chi_{+} \nn\\
   &\qquad\qquad\quad  +\left(-\frac{5}{72}+\left(-\frac{7631}{24}+\frac{11275 \pi ^2}{768}\right) \nu +\left(\frac{3317}{8}-\frac{205 \pi ^2}{12}\right) \nu ^2+\left(-\frac{2045}{18}+\frac{205 \pi
   ^2}{48}\right) \nu ^3-14 \nu ^4\right) \chi_{+}^2\Bigg]\nn\\
   &\quad +\frac{1}{\iota }\Bigg[\left(-\frac{109}{12}+\left(\frac{27223}{432}-\frac{533 \pi ^2}{384}\right) \nu +\left(-\frac{23117}{216}+\frac{533 \pi^2}{96}\right) \nu ^2+\frac{187 \nu ^3}{12}\right) \chi_{-}^2 \nn\\
   &\qquad\qquad\quad  +\delta  \left(-\frac{109}{6}+\left(\frac{17449}{216}-\frac{533 \pi ^2}{192}\right) \nu +\left(-\frac{2417}{54}+\frac{41 \pi
   ^2}{24}\right) \nu ^2+\frac{17 \nu ^3}{2}\right) \chi_{-} \chi_{+}  \nn\\
   &\qquad\qquad\quad +\left(-\frac{109}{12}+\left(\frac{23371}{432}-\frac{533 \pi ^2}{384}\right) \nu +\left(-\frac{12197}{216}+\frac{41
   \pi ^2}{24}\right) \nu ^2+\left(\frac{1885}{108}-\frac{41 \pi ^2}{96}\right) \nu ^3-\frac{17 \nu ^4}{6}\right) \chi_{+}^2\Bigg]\Bigg\} \,, \\
      \label{seq:j_S3_of_x_iota_expression}
   j_{S^3} &= \frac{x^{3/2}}{\sqrt{\iota }} \Bigg\{\delta  \left(-\frac{13}{27}+\frac{52 \nu }{27}\right) \chi_{-}^3+\left(-\frac{13}{9}+\frac{15 \nu }{2}-\frac{62 \nu ^2}{9}\right) \chi_{-}^2 \chi_{+} \nn\\
   &\qquad\qquad  +\delta 
   \left(-\frac{13}{9}+\frac{31 \nu }{9}-\frac{10 \nu ^2}{9}\right) \chi_{-} \chi_{+}^2+\left(-\frac{13}{27}+\frac{31 \nu }{18}-\frac{10 \nu ^2}{9}+\frac{5 \nu ^3}{27}\right)
   \chi_{+}^3\Bigg\} \nn\\
   & +x^{5/2} \Bigg\{\frac{1}{\sqrt{\iota }} \Bigg[\delta  \left(-\frac{5}{8}+\frac{395 \nu }{162}+\frac{182 \nu ^2}{81}\right) \chi_{-}^3+\left(-\frac{15}{8}+\frac{4087 \nu
   }{432}-\frac{49 \nu ^2}{12}-\frac{217 \nu ^3}{27}\right) \chi_{-}^2 \chi_{+} \nn\\ 
   &\qquad +\delta  \left(-\frac{15}{8}+\frac{887 \nu }{216}+\frac{317 \nu ^2}{108}-\frac{35 \nu ^3}{27}\right)
   \chi_{-} \chi_{+}^2+\left(-\frac{5}{8}+\frac{2701 \nu }{1296}+\frac{127 \nu ^2}{108}-\frac{343 \nu ^3}{216}+\frac{35 \nu ^4}{162}\right) \chi_{+}^3\Bigg] \nn\\
   &\quad +\frac{1}{\iota ^{3/2}}\Bigg[\delta  \left(-\frac{17}{216}-\frac{2047 \nu }{216}+\frac{45 \nu ^2}{2}\right) \chi_{-}^3+\left(-\frac{17}{72}-\frac{2255 \nu }{144}+\frac{1709 \nu ^2}{24}-\frac{155 \nu
   ^3}{2}\right) \chi_{-}^2 \chi_{+} \nn\\
   &\qquad\qquad  +\delta  \left(-\frac{17}{72}-\frac{23 \nu }{6}+\frac{190 \nu ^2}{9}-\frac{175 \nu ^3}{12}\right) \chi_{-} \chi_{+}^2+\left(-\frac{17}{216}+\frac{143 \nu }{48}-\frac{25 \nu ^2}{72}-\frac{1685 \nu ^3}{216}+\frac{35 \nu ^4}{12}\right) \chi_{+}^3\Bigg]\Bigg\} \,,\\
    \label{seq:j_S4_of_x_iota_expression}
   j_{S^4} &= \frac{x^2 }{\iota }\Bigg\{\left(-\frac{241}{324}+\frac{482 \nu }{81}-\frac{964 \nu ^2}{81}\right) \chi_{-}^4+\delta  \left(-\frac{241}{81}+\frac{140 \nu }{9}-\frac{1184 \nu ^2}{81}\right) \chi_{-}^3 \chi_{+} \nn\\
   &\qquad\quad +\left(-\frac{241}{54}+\frac{778 \nu }{27}-\frac{1294 \nu ^2}{27}+\frac{440 \nu ^3}{27}\right) \chi_{-}^2 \chi_{+}^2+\delta  \left(-\frac{241}{81}+\frac{296 \nu
   }{27}-\frac{220 \nu ^2}{27}+\frac{128 \nu ^3}{81}\right) \chi_{-} \chi_{+}^3 \nn\\
   &\qquad\quad +\left(-\frac{241}{324}+\frac{296 \nu }{81}-\frac{110 \nu ^2}{27}+\frac{128 \nu ^3}{81}-\frac{16 \nu
   ^4}{81}\right) \chi_{+}^4 \Bigg\}\nn\\
   &+x^3 \Bigg\{\frac{1}{\iota ^2}\Bigg[\left(\frac{259}{162}-\frac{2944 \nu }{81}+\frac{27247 \nu ^2}{162}-\frac{1742 \nu ^3}{9}\right) \chi_{-}^4+\delta 
   \left(\frac{518}{81}-\frac{5629 \nu }{54}+\frac{23803 \nu ^2}{81}-\frac{2072 \nu ^3}{9}\right) \chi_{-}^3 \chi_{+}\nn\\
   &\qquad\qquad +\left(\frac{259}{27}-\frac{7183 \nu }{54}+\frac{26209 \nu
   ^2}{54}-\frac{17038 \nu ^3}{27}+\frac{848 \nu ^4}{3}\right) \chi_{-}^2 \chi_{+}^2 \nn\\
   &\qquad\qquad +\delta  \left(\frac{518}{81}-\frac{2567 \nu }{54}+\frac{2498 \nu ^2}{27}-\frac{8284 \nu
   ^3}{81}+\frac{280 \nu ^4}{9}\right) \chi_{-} \chi_{+}^3 \nn\\
   &\qquad\qquad +\left(\frac{259}{162}-\frac{1295 \nu }{162}+\frac{292 \nu ^2}{27}-\frac{2290 \nu ^3}{81}+\frac{1772 \nu ^4}{81}-\frac{40 \nu
   ^5}{9}\right) \chi_{+}^4\Bigg]\nn\\
   &\qquad  +\frac{1}{\iota }\Bigg[\left(-\frac{7}{9}+\frac{24307 \nu }{3888}-\frac{4057 \nu ^2}{486}-\frac{4097 \nu ^3}{243}\right) \chi_{-}^4+\delta 
   \left(-\frac{28}{9}+\frac{15703 \nu }{972}-\frac{7 \nu ^2}{3}-\frac{5032 \nu ^3}{243}\right) \chi_{-}^3 \chi_{+} \nn\\
   &\qquad\qquad  +\left(-\frac{14}{3}+\frac{19195 \nu }{648}-\frac{2431 \nu
   ^2}{81}-\frac{7885 \nu ^3}{162}+\frac{1870 \nu ^4}{81}\right) \chi_{-}^2 \chi_{+}^2 \nn\\
   &\qquad\qquad  +\delta  \left(-\frac{28}{9}+\frac{10591 \nu }{972}+\frac{601 \nu ^2}{81}-\frac{989 \nu
   ^3}{81}+\frac{544 \nu ^4}{243}\right) \chi_{-} \chi_{+}^3  \nn\\
   &\qquad\qquad  +\left(-\frac{7}{9}+\frac{14083 \nu }{3888}+\frac{869 \nu ^2}{486}-\frac{1097 \nu ^3}{162}+\frac{688 \nu ^4}{243}-\frac{68
   \nu ^5}{243}\right) \chi_{+}^4\Bigg]\Bigg\} \,, \\
   j_{S^5} &= \frac{x^{5/2}}{\iota^{3/2}} \Bigg\{
\delta\left(-\frac{833}{648} + \frac{833}{81}\,\nu - \frac{1666}{81}\,\nu^{2}\right) \chi_-^{5} + \left(-\frac{4165}{648} + \frac{77105}{1296}\,\nu - \frac{27125}{162}\,\nu^{2} + \frac{10465}{81}\,\nu^{3}\right) \chi_-^{4}\,\chi_+ \nonumber \\
&\qquad\quad  + \delta\left(-\frac{4165}{324} + \frac{27125}{324}\,\nu - \frac{23065}{162}\,\nu^{2} + \frac{4270}{81}\,\nu^{3}\right) \chi_-^{3}\,\chi_+^{2} \nonumber \\
&\qquad\quad + \left(-\frac{4165}{324} + \frac{64715}{648}\,\nu - \frac{700}{3}\,\nu^{2} + \frac{54005}{324}\,\nu^{3} - \frac{2765}{81}\,\nu^{4}\right) \chi_-^{2}\,\chi_+^{3} \nonumber \\
&\qquad\quad  + \delta\left(-\frac{4165}{648} + \frac{10465}{324}\,\nu - \frac{2135}{54}\,\nu^{2} + \frac{2765}{162}\,\nu^{3} - \frac{385}{162}\,\nu^{4}\right) \chi_-\,\chi_+^{4} \nonumber \\
&\qquad\quad + \left(-\frac{833}{648} + \frac{10465}{1296}\,\nu - \frac{2135}{162}\,\nu^{2} + \frac{2765}{324}\,\nu^{3} - \frac{385}{162}\,\nu^{4} + \frac{77}{324}\,\nu^{5}\right) \chi_+^{5} \Bigg\} \,,\\
j_{S^6} &= \frac{x^{3}}{\iota^{2}} \Bigg\{
\left(-\frac{6935}{2916} + \frac{6935}{243}\,\nu - \frac{27740}{243}\,\nu^{2} + \frac{110960}{729}\,\nu^{3}\right) \chi_-^{6} + \delta\left(-\frac{6935}{486} + \frac{32176}{243}\,\nu - \frac{90968}{243}\,\nu^{2} + \frac{70976}{243}\,\nu^{3}\right) \chi_-^{5}\,\chi_+ \nonumber \\
&\qquad\quad + \left(-\frac{34675}{972} + \frac{1130}{3}\,\nu - \frac{325825}{243}\,\nu^{2} + \frac{48040}{27}\,\nu^{3} - \frac{154960}{243}\,\nu^{4}\right) \chi_-^{4}\,\chi_+^{2} \nonumber \\
&\qquad\quad + \delta\left(-\frac{34675}{729} + \frac{271780}{729}\,\nu - \frac{24020}{27}\,\nu^{2} + \frac{492080}{729}\,\nu^{3} - \frac{108800}{729}\,\nu^{4}\right) \chi_-^{3}\,\chi_+^{3} \nonumber \\
&\qquad\quad + \left(-\frac{34675}{972} + \frac{26345}{81}\,\nu - \frac{235550}{243}\,\nu^{2} + \frac{259640}{243}\,\nu^{3} - \frac{37640}{81}\,\nu^{4} + \frac{16480}{243}\,\nu^{5}\right) \chi_-^{2}\,\chi_+^{4} \nonumber \\
&\qquad\quad + \delta\left(-\frac{6935}{486} + \frac{22180}{243}\,\nu - \frac{38740}{243}\,\nu^{2} + \frac{27200}{243}\,\nu^{3} - \frac{8240}{243}\,\nu^{4} + \frac{896}{243}\,\nu^{5}\right) \chi_-\,\chi_+^{5} \nonumber \\
&\qquad\quad + \left(-\frac{6935}{2916} + \frac{4436}{243}\,\nu - \frac{9685}{243}\,\nu^{2} + \frac{27200}{729}\,\nu^{3} - \frac{4120}{243}\,\nu^{4} + \frac{896}{243}\,\nu^{5} - \frac{224}{729}\,\nu^{6}\right) \chi_+^{6} \Bigg\} \,.
\end{align}
\end{subequations}
The SID contributions for the energy and angular momentum are relegated to the Supplemental Material~\cite{Supplemental}.

\section{Redshift and gyroscopic invariants}
\label{sec:redshift_gyroscopic}
\subsection{Overview of the first law of binary black hole mechanics}
\label{subsec:first_law}

The first law of binary black hole mechanics is a `thermodynamic' differential relation linking various gauge-invariant quantities. It was first established in the context of nonspinning black holes on circular orbits~\cite{LeTiec:2011ab} and reads
\begin{align}
    \delta M - \omega\, \delta J = z_1 \delta m_1 + z_2 \delta m_2 \,,
\end{align}
where $M=m_1+m_2+E/c^2$ is the ADM mass of the spacetime and $J$ is the total angular momentum, which here coincides with the orbital angular momentum $L$. We have introduced the invariant redshift variable which corresponds to the ratio between the orbital frequency in coordinate time (for an asymptotic observer) and the orbital frequency in the proper time of body $A$, denoted $\tau_A$; it can also be understood as the redshift undergone by light emitted at the particle and reaching an asymptotic observer~\cite{Detweiler:2008ft}. In self-force theory, it is computed directly from the regularized metric and serves, using the first law, as a potential to compute the constants of motion. In post-Newtonian theory, it is computed from the metric---which is regularized with dimensional regularization~\cite{Blanchet:2009sd}---using the formula
\begin{align}
    z_1 = \frac{\dd \tau_1}{\dd t}=\sqrt{-(g_{\alpha\beta})_1 \frac{v_1^\alpha v_1^\beta}{c^2}} \,,
\end{align}
where $(g_{\alpha\beta})_1$ is the Hadamard- or dimensionally-regularized metric at the location of particle $1$, and where $v_1^\alpha=(c,\bm{v}_1)$. Of course, $z_2$ is defined by switching labels $1 \leftrightarrow 2$ in all of these expressions.

A consequence of the first law is that the redshift invariant can in fact be computed through a much simpler route, as  it is simply given by
\begin{align}
    z_1 = \frac{\partial M(m_1,m_2,J)}{\partial m_1} \,.
\end{align}
The first law was shown to still be true even in the presence of hereditary tails at 4PN order~\cite{Blanchet:2017rcn}. The first law was also generalized to the case of spinning, nonprecessing compact binaries, i.e., binaries with (anti)aligned spin. However, this analysis was restricted to the level of linearized spin~\cite{Blanchet:2012at,Ramond:2022ctc}, in which case the first law can be extended to 
\begin{align}
    \delta M- \omega \delta L = z_1 \delta m_1 + z_2 \delta m_2  + \Omega_1^s\, \delta S_1 + \Omega_2^s \, \delta S_2 + \mathcal{O}(S^2) \,,
\end{align}
where $L \equiv J - S_1 -S_2$ is the \textit{orbital} angular momentum. It can be rewritten in terms of the \textit{total} angular momentum~$J$: 
\begin{align}
    \delta M- \omega \delta J = z_1 \delta m_1 + z_2 \delta m_2 + (\Omega_1^s-\omega) \delta S_1  + (\Omega_2^s-\omega) \delta S_2 + \mathcal{O}(S^2) \,.
\end{align}
Consequently, the redshift can be obtained using the same formula as in the nonspinning case, provided that the dimensionful spin magnitudes $(S_1,S_2)$ are kept fixed:
\begin{align}\label{eq:z1_first_law_linear_in_spin}
    z_1 &= \frac{\partial M(m_1, m_2,J,S_1,S_2)}{\partial m_1} + \mathcal{O}(S^2) \,.
\end{align}
Note that the relations between the reduced spin variables $(\chi_1,\chi_2)$ and  the dimensionful spin variables $(S_1,S_2)$ involve mass-dependent terms. Therefore, varying with respect to the mass while keeping the \textit{reduced} spin variables fixed would lead to a different, and hence incorrect, expression of the redshift.

In parallel to these works for circular orbits, Ref.~\cite{LeTiec:2015kgg} extended the first law to eccentric nonspinning systems. For such systems, it is convenient to use action-angle variables and introduce the radial action $I_r$. It was then shown that the first law becomes 
\begin{align}
    \delta M= n \, \delta I_r + \omega\,\delta J+\langle z_1\rangle \delta m_1 +\langle z_2\rangle \delta m_2\,,
\end{align}
where $\langle z_A\rangle$ is the \textit{orbit average} of the instantaneous redshift. We recall that $n$ and $\omega$ are, respectively, the radial and azimuthal frequencies. The redshift is then obtained using the formula
\begin{align}
    \langle z_1 \rangle = \frac{\partial M(m_2,I_r,J)}{\partial m_1} \,,
\end{align}
where both the dimensionful angular momentum and radial action are kept fixed. It was furthermore proven in Ref.~\cite{Blanchet:2017rcn} that the first law holds even in the presence of hereditary tail effects: this was used for the computation of the 4PN tail contribution to the redshift over eccentric orbits in Ref.~\cite{Trestini:2025yyc}.

In order to extend the first law to include spin-spin interactions, it was initially assumed that Eq.~\eqref{eq:z1_first_law_linear_in_spin} should still hold for quadratic-in-spin interactions in the case of a black hole binary~\cite{Bini:2019lcd}. However, when trying to extend this argument to the case of neutron stars, the question of the treatment of the SID coefficients arose~\cite{Bini:2020zqy}. This question was ultimately resolved to any order in spin in Ref.~\cite{Antonelli:2020ybz}, which used the general form of the Lagrangian---provided in (4.16) of~\cite{Levi:2015msa}---to argue that the first law holds to any order provided the spin-induced coupling constants are rescaled by a power of the mass related to their multipolarity.
Structurally, the point-particle actions for $A\in\{1,2\}$ can be written as
\begin{equation}
\mathcal{S}^\text{pp}_A = - m_A \int \dd \tau_A + \mathcal{S}^\text{pp,spin-orbit}_A + \mathcal{S}^\text{pp,nonminimal}_A \,,
\end{equation}
with the linear-in-spin part $\mathcal{S}^\text{pp,spin-orbit}_A$ following from a minimal coupling and $\mathcal{S}^\text{pp,nonminimal}_A$ containing nonminimal couplings to the curvature tensor. The latter includes the SID couplings and parameters---see (4.16) in Ref.~\cite{Levi:2015msa}---and at higher orders also contains tidal and mixed spin-tidal couplings (with corresponding parameters). Now, $\mathcal{S}^\text{pp,spin-orbit}_A$ is independent of the masses $m_A$. If $\mathcal{S}^\text{pp,nonminimal}_A$ is also independent of $m_A$, then
\begin{equation}\label{eq:Lredshift}
 z_A = \frac{\dd \tau_A}{\dd t} = - \frac{\partial L}{\partial m_A} \,,
\end{equation}
with the coordinate-time binary Lagrangian $L$ defined by $\int \dd t \, L = \mathcal{S} = \mathcal{S}^\text{grav} + \mathcal{S}^\text{pp}_1 + \mathcal{S}^\text{pp}_2$ and $\mathcal{S}^\text{grav}$ being the gauge-fixed Einstein-Hilbert action. Although it is customary to make the parameters in front of the couplings in $\mathcal{S}^\text{pp,nonminimal}_A$ dimensionless by rescaling them with an appropriate power of $m_A$, this choice is not preferred here as it would spoil Eq.~\eqref{eq:Lredshift}. Instead, we introduce the \textit{dimensionful} SID parameters $\bar{C}_{A(ES^{2n})}={C}_{A(ES^{2n})}/m_A^{2n-1}$ and $\bar{C}_{A(BS^{2n+1})}={C}_{A(BS^{2n+1})}/m_A^{2n}$, following the notation\footnote{In our case, since we work at 4PN order, the maximum required order in spin is quartic, so we use the following notation for the coupling constants: $\bar{\kappa}_A = \kappa_A/m_A$, $\bar{\lambda}_A = \lambda_A/m_A^2$, and $\bar{\mu}_A = \mu_A/m_A^3$.} of Ref.~\cite{Levi:2015msa}. This choice makes $\mathcal{S}^\text{pp,nonminimal}_A$ independent of $m_A$, and hence Eq.~\eqref{eq:Lredshift} remains valid to all orders in spin. Following the arguments of Ref.~\cite{Blanchet:2012at}, it was further shown in Ref.~\cite{Antonelli:2020ybz} that Eq.~\eqref{eq:Lredshift} remains valid in time-averaged form (and to all orders in spin) under rather general transformations of the dynamical variables. It further remains valid upon elimination of the gravitational field---see Eq.~(4.6) therein---and also under replacement of the Lagrangian $L$ by $-E$, i.e., minus the Hamiltonian or ADM energy. Consequently, the redshift can be computed for nonprecessing eccentric orbits and to all orders in spin (including SID parameters for extended bodies)  through 
 \begin{align}
    \langle z_1 \rangle &= \frac{\partial M(m_2,J,I_r, S_1,S_2,\bar{C}_{1(ES^{2n})},\bar{C}_{2(ES^{2n})},\bar{C}_{1(BS^{2n+1})},\bar{C}_{2(BS^{2n+1})})}{\partial m_1} \,.
\end{align}
This result can immediately be specialized to the circular case by eliminating the dependence on the radial action~$I_r$. 
These conclusions appear to invalidate the scheme laid out in~Ref.~\cite{Bini:2019lcd}: indeed, Eqs.~(62) and (63) of Ref.~\cite{Bini:2019lcd} disagree with Eq.~(6.4)~of~\cite{Bini:2020zqy}. Thus, \textit{even in the case of two black holes}, the SID coefficients need to be kept arbitrary and adequately rescaled before varying with respect to the mass; only after variation can they be set to $1$.

In the presence of spin, another invariant typically obtained by variation of the ADM mass is the gyroscopic (or spin-precession) invariant. First, we recall that the (average) precession frequency of the spinning object $1$ is obtained from the ADM mass using~\cite{Bini:2019lkm}  
\begin{align}
    \Omega_1^s = \frac{\partial M\Big(m_1,m_2,J,I_r, S_2,\bar{C}_{1(ES^{2n})},\bar{C}_{2(ES^{2n})},\bar{C}_{1(BS^{2n+1})},\bar{C}_{2(BS^{2n+1})} \Big)}{\partial S_1}  \,.
\end{align}
Note that here, the rest-mass contribution to the ADM mass does not contribute, so $M$ can be replaced with the action-angle Hamiltonian $H$. Moreover,  it actually does not matter in this case whether the variation is taken with the original or rescaled spin-deformability parameters kept fixed. The gyroscopic invariant is then defined as~\cite{Bini:2019lkm}
\begin{align}
    \psi_1&= \frac{\Omega_1^s}{\omega} \,.
\end{align}

\subsection{Results for the redshift}
\label{subsec:redshift}

We have obtained the redshift invariant at 4PN using the prescription laid out previously. We have then restored our usual dimensionless variables and SID parameters. The redshift in terms of energy and angular momentum is decomposed as
    \begin{align}
    \langle z_1\rangle&=\langle z_1^\text{pp} \rangle(\varepsilon,j)+\langle z_1^\text{spin}\rangle (\varepsilon,j,\chi_A)+\langle  z_1^\text{SID} \rangle (\varepsilon,j,\chi_A,\kappa_A,\lambda_A,\mu_A) \,,
\end{align}
where $\langle z_1^\text{pp}\rangle (\varepsilon,j)$ is given in (6.5) and (6.6) of \cite{Trestini:2025yyc}, including the nonlocal tail contribution. The spinning contributions are decomposed as 
\begin{align}
    \langle z_1^\text{spin}  \rangle (\varepsilon,j,\chi_A) &= \langle z_1^{S^1}  \rangle(\varepsilon,j,\chi_A) + \langle z_1^{S^2} \rangle (\varepsilon,j,\chi_A) + \langle z_1^{S^3} \rangle (\varepsilon,j,\chi_A) + \langle z_1^{S^4} \rangle (\varepsilon,j,\chi_A) \,.
\end{align}
The expressions are lengthy, so we relegate the spin results to Appendix~\ref{app:redshift} and the SID piece to the Supplemental Material~\cite{Supplemental}.
We then reexpress the redshift in terms of the frequency variables thanks to  Eq.~\eqref{eq:varepsilon_j_pp_spin_SID}. We find that 
\begin{align}
    \langle z_1 \rangle &=\langle z_1^\text{pp} \rangle (x,\iota)+\langle z_1^\text{spin} \rangle (x,\iota,\chi_A)+\langle z_1^\text{SID} \rangle (x,\iota,\chi_A,\kappa_A,\lambda_A,\mu_A) \,,
\end{align}
where $\langle z_1^\text{pp}(x,\iota) \rangle$ is given in (6.8) and (6.9) of \cite{Trestini:2025yyc}, including the nonlocal tail contribution. Due to this reexpansion, we introduce extra couplings which generate higher-order terms in the spin expansion, namely up to $S^6$:
\begin{align}
    \langle z_1^\text{spin} \rangle (x,\iota,\chi_A) &= \langle z_1^{S^1} \rangle (x,\iota,\chi_A)
    + \langle z_1^{S^2} \rangle (x,\iota,\chi_A) 
    + \langle z_1^{S^3} \rangle (x,\iota,\chi_A) 
    + \langle z_1^{S^4} \rangle (x,\iota,\chi_A) 
    + \langle z_1^{S^5} \rangle (x,\iota,\chi_A) 
    + \langle  z_1^{S^6} \rangle (x,\iota,\chi_A) \,.
\end{align}
All these spin contributions, as well as the  SID piece, are provided in the Supplemental Material~\cite{Supplemental}.

As discussed in Sec.~\ref{subsec:first_law}, for two black holes, the linear-in-spin contribution agrees at 4PN with (50) of \cite{Bini:2019lcd}, but the quadratic-in-spin piece disagrees already at leading 2PN order with (51) of that same reference. Instead, the quadratic-in-spin piece perfectly agrees at 4PN with (6.4) of \cite{Bini:2020zqy}, including the associated SID terms. We have also checked that, in the case of a binary composed of one spinning black hole and one nonspinning black hole, we find perfect agreement with the result presented in the Supplemental Material of \cite{Bautista:2024agp} at 4PN, to all orders in the mass ratio and to \textit{all orders in spin} (i.e. $S^1$ to $S^6$).

\subsection{Gyroscopic invariant}
\label{subsec:gyroscopic}

Similarly, we have obtained the gyroscopic invariant at 4PN. We deviate slightly from our usual nomenclature: although there is a spin-independent contribution, it arises from the linear-in-spin piece of the Hamiltonian and cannot be interpreted as a `point-particle' contribution. Indeed, the latter has been nullified by the derivative with respect to spin. We thus refer to this spin-independent contribution as an $S^0$ piece. Moreover, there is no $S^4$ piece, as this would arise from an $S^5$ piece in the Hamiltonian, which enters at higher PN order. With these conventions in mind, the gyroscopic invariant is decomposed as 
\begin{align}
    \psi_1(\varepsilon,j)&= \psi_1^{S^0}(\varepsilon,j)+\psi_1^{S^1}(\varepsilon,j,\chi_A)+\psi_1^{S^2}(\varepsilon,j,\chi_A)+\psi_1^{S^3}(\varepsilon,j,\chi_A)+\psi_1^\text{SID}(\varepsilon,j,\chi_A,\kappa_A,\lambda_A,\mu_A) \,.
\end{align}
The expressions are rather lengthy, so we relegate the black-hole results to Appendix~\ref{app:gyroscopic} and the SID piece to the Supplemental Material~\cite{Supplemental}. Note that the leading order of the gyroscopic invariant arises from the 1.5PN spin-orbit coupling in the Hamiltonian: the 4PN gyroscopic invariant is therefore only 2.5PN orders higher relative to the leading 1.5PN order.

We then reexpress the gyroscopic invariant in terms of the frequency variables thanks to  Eq.~\eqref{eq:varepsilon_j_pp_spin_SID}. Due to couplings, the highest power in spin is $S^5$:
\begin{align}
    \psi_1(x,\iota) &= \psi_1^{S^0}(x,\iota) + \psi_1^{S^1}(x,\iota,\chi_A)   +\psi_1^{S^2}(x,\iota,\chi_A) + \psi_1^{S^3}(x,\iota,\chi_A) \nn\\
    &\qquad+ \psi_1^{S^4}(x,\iota,\chi_A) + \psi_1^{S^5}(x,\iota,\chi_A)  + \psi_1^\text{SID}(x,\iota,\chi_A,\kappa_A,\lambda_A,\mu_A)\,.
\end{align}
All these contributions are provided in the Supplemental Material~\cite{Supplemental}. We find that our 4PN result perfectly reproduces the 3PN results for the $S^0$, $S^1$ and $S^2$ sectors obtained in (82)--(85) of Ref.~\cite{Bini:2019lkm} (note that those expressions are only valid in the case of two black holes).  In the case of a binary composed of one spinning black hole and one nonspinning black hole, we have also found perfect overlap with the results in the Supplemental Material of Ref.~\cite{Bautista:2024agp}.

\section{Circular orbits}
\label{sec:circular}

In order to recover the case of circular orbits, we first need to obtain the associated relations satisfied by the constants of motion (or by the frequencies), which are called ``circular links''. These are obtained by imposing that the radial action vanish: $i_r =0$. Solving this equation iteratively yields the circular links  (i) between the energy and angular momentum; and (ii) between the two fundamental frequencies. These links are decomposed as
\begin{subequations}
\begin{align}
    j_\text{circ}(\varepsilon) &= j_\text{circ}^\text{pp}(\varepsilon)+j_\text{circ}^\text{spin}(\varepsilon)+j_\text{circ}^\text{SID}(\varepsilon)\,,\\
    K_\text{circ}(x) &= K_\text{circ}^\text{pp}(x)+K_\text{circ}^\text{spin}(x)+K_\text{circ}^\text{SID}(x) \,.
\end{align}
\end{subequations}
The point-particle contributions $j_\text{circ}^\text{pp}(\varepsilon)$ and $K_\text{circ}^\text{pp}(x)$ are given, respectively, in (7.2) and (7.4b) of \cite{Trestini:2025yyc}. The spinning contributions read
\begin{subequations}
\begin{align}
j_{\text{circ}}^{S^1}(\varepsilon) &= \varepsilon^{3/2}\Bigg\{\chi_{+}\Bigg[-4 + 2 \nu\Bigg] - 4 \delta\chi_{-}\Bigg\}  + \varepsilon^{5/2}\Bigg\{\chi_{+}\Bigg[-\frac{27}{2} + 13 \nu + \frac{1}{4} \nu^{2}\Bigg] + \delta\chi_{-}\Bigg[-\frac{27}{2} + \frac{7}{4} \nu\Bigg]\Bigg\} \nn \\
& + \varepsilon^{7/2}\Bigg\{\chi_{+}\Bigg[-\frac{1863}{32} + \frac{6281}{64} \nu - \frac{587}{32} \nu^{2} + \frac{3}{64} \nu^{3}\Bigg] + \delta\chi_{-}\Bigg[-\frac{1863}{32} + \frac{1399}{32} \nu - \frac{3}{16} \nu^{2}\Bigg]\Bigg\} \\
j_{\text{circ}}^{S^2}(\varepsilon) &= \varepsilon^{2}\Bigg\{\chi_{+}^2 + 2 \delta\chi_{+}\chi_{-} + \chi_{-}^2\Bigg[1 - 4 \nu\Bigg]\Bigg\} \nn \\
& + \varepsilon^{3}\Bigg\{\chi_{+}^2\Bigg[\frac{55}{4} - \frac{69}{4} \nu + 5 \nu^{2}\Bigg] + \delta\chi_{+}\chi_{-}\Bigg[\frac{55}{2} - \frac{37}{2} \nu\Bigg] + \chi_{-}^2\Bigg[\frac{55}{4} - \frac{225}{4} \nu + \nu^{2}\Bigg]\Bigg\} \nn \\
& + \varepsilon^{4}\Bigg\{\chi_{+}^2\Bigg[99 - \frac{1517}{8} \nu + \frac{311}{4} \nu^{2} - 2 \nu^{3}\Bigg] + \delta\chi_{+}\chi_{-}\Bigg[198 - \frac{905}{4} \nu + \frac{85}{4} \nu^{2}\Bigg] + \chi_{-}^2\Bigg[99 - \frac{3461}{8} \nu + 135 \nu^{2}\Bigg]\Bigg\} \\
j_{\text{circ}}^{S^3}(\varepsilon) &= \varepsilon^{7/2}\Bigg\{\chi_{+}^3\Bigg[-6 + 5 \nu\Bigg] + \delta\chi_{+}^2\chi_{-}\Bigg[-18 + 10 \nu\Bigg] + \chi_{+}\chi_{-}^2\Bigg[-18 + 77 \nu - 20 \nu^{2}\Bigg] + \delta\chi_{-}^3\Bigg[-6 + 24 \nu\Bigg]\Bigg\} \\
j_{\text{circ}}^{S^4}(\varepsilon) &= \varepsilon^{4}\Bigg\{\chi_{+}^4 + 4 \delta\chi_{+}^3\chi_{-} + \chi_{+}^2\chi_{-}^2\Bigg[6 - 24 \nu\Bigg] + \delta\chi_{+}\chi_{-}^3\Bigg[4 - 16 \nu\Bigg] + \chi_{-}^4\Bigg[1 - 8 \nu + 16 \nu^{2}\Bigg]\Bigg\}
\end{align}
\end{subequations}
and
\begin{subequations}
    \begin{align}
K_{\text{circ}}^{S^1}(x) &= x^{3/2}\Bigg\{\chi_{+}\Bigg[-4 + 2 \nu\Bigg] - 4 \delta\chi_{-}\Bigg\} + x^{5/2}\Bigg\{\chi_{+}\Bigg[-34 + \frac{81}{2} \nu - 2 \nu^{2}\Bigg] + \delta\chi_{-}\Bigg[-34 + \frac{17}{2} \nu\Bigg]\Bigg\} \nn \\
& + x^{7/2}\Bigg\{\chi_{+}\Bigg[-252 + \frac{11581}{24} \nu - \frac{733}{6} \nu^{2} + \frac{2}{3} \nu^{3}\Bigg] + \delta\chi_{-}\Bigg[-252 + \frac{5317}{24} \nu - \frac{22}{3} \nu^{2}\Bigg]\Bigg\} \\
K_{\text{circ}}^{S^2}(x) &= x^{2}\Bigg\{\frac{3}{2}\chi_{+}^2 + 3 \delta\chi_{+}\chi_{-} + \chi_{-}^2\Bigg[\frac{3}{2} - 6 \nu\Bigg]\Bigg\} \nn \\
&+ x^{3}\Bigg\{\chi_{+}^2\Bigg[\frac{67}{2} - \frac{99}{2} \nu + 14 \nu^{2}\Bigg] + \delta\chi_{+}\chi_{-}\Bigg[67 - 55 \nu\Bigg] + \chi_{-}^2\Bigg[\frac{67}{2} - \frac{279}{2} \nu + 10 \nu^{2}\Bigg]\Bigg\} \nn \\
& + x^{4}\Bigg\{\chi_{+}^2\Bigg[\frac{4835}{12} - \frac{6995}{8} \nu + \frac{1298}{3} \nu^{2} - \frac{88}{3} \nu^{3}\Bigg]  + \delta\chi_{+}\chi_{-}\Bigg[\frac{4835}{6} - \frac{12913}{12} \nu + \frac{497}{3} \nu^{2}\Bigg] \nn\\
& \qquad\quad  + \chi_{-}^2\Bigg[\frac{4835}{12} - \frac{14507}{8} \nu + \frac{1501}{2} \nu^{2} - \frac{20}{3} \nu^{3}\Bigg]\Bigg\}
\\
K_{\text{circ}}^{S^3}(x) &= x^{7/2}\Bigg\{\chi_{+}^3\Bigg[-16 + 17 \nu\Bigg] + \delta\chi_{+}^2\chi_{-}\Bigg[-48 + 34 \nu\Bigg] + \chi_{+}\chi_{-}^2\Bigg[-48 + 209 \nu - 68 \nu^{2}\Bigg] + \delta\chi_{-}^3\Bigg[-16 + 64 \nu\Bigg]\Bigg\}
\\
K_{\text{circ}}^{S^4}(x) &= x^{4}\Bigg\{\frac{27}{8}\chi_{+}^4 + \frac{27}{2} \delta\chi_{+}^3\chi_{-} + \chi_{+}^2\chi_{-}^2\Bigg[\frac{81}{4} - 81 \nu\Bigg] + \delta\chi_{+}\chi_{-}^3\Bigg[\frac{27}{2} - 54 \nu\Bigg] + \chi_{-}^4\Bigg[\frac{27}{8} - 27 \nu + 54 \nu^{2}\Bigg]\Bigg\}
\end{align}
\end{subequations}
The SID contributions  $j_{\text{circ}}^{\text{SID}}(\varepsilon)$ and $K_{\text{circ}}^{\text{SID}}(x)$ are relegated to the Supplemental Material~\cite{Supplemental}.
Note that we do not present the link $\iota_\text{circ}(x)$ in the text because the reexpansion leads to terms up to $S^6$, namely
\begin{align}
    \iota(x) = \iota_\text{pp}(x)+ \iota_{S^1}(x)+ \iota_{S^2}(x)+ \iota_{S^3}(x)+ \iota_{S^4}(x)+ \iota_{S^5}(x)+ \iota_{S^6}(x)+ \iota_\text{SID}(x) \,.
\end{align}
These contributions are provided in the Supplemental Material~\cite{Supplemental}.
Using these links, we can express the following invariants in terms of the frequency:
\begin{subequations}
    \begin{align}
    \varepsilon(x)&=  \varepsilon_\text{circ}^\text{pp}(x)+\varepsilon_\text{circ}^\text{spin}(x)+\varepsilon_\text{circ}^\text{SID}(x) \\
    z_\text{1\,circ}(x) &= z_\text{1\,circ}^\text{pp}(x)+z_\text{1\,circ}^\text{spin}(x)+z_\text{1\,circ}^\text{SID}(x) \\
    \psi_\text{1\,circ}(x) &=  \psi_\text{1\,circ}^\text{spin}(x)+\psi_\text{1\,circ}^\text{SID}(x)
\end{align}
\end{subequations}
The point-particle contributions for the energy and redshift are given, respectively, in (4.11) of~\cite{Bernard:2016wrg} and in (7.5b) of~\cite{Trestini:2025yyc}. Remarkably, the quartic-in-spin contribution to the energy (expressed in terms of frequency) vanishes at 4PN for black holes: $\varepsilon_{\text{circ}}^{S^4}(x) =0$. This curiosity was already noticed in \cite{Siemonsen:2017yux}. We provide here the spinning contributions of all three invariants, including the nonspinning contribution of the gyroscopic invariant, in the case of black holes. The SID contributions are relegated to the Supplemental Material~\cite{Supplemental}.

The spinning contributions to the energy $\varepsilon_\text{circ}^\text{spin}(x) = \varepsilon_{\text{circ}}^{S^1}(x) +\varepsilon_{\text{circ}}^{S^2}(x) +\varepsilon_{\text{circ}}^{S^3}(x) +\varepsilon_{\text{circ}}^{S^4}(x) $ read
\begin{subequations}
\begin{align}
\varepsilon_{\text{circ}}^{S^1}(x) &= x^{5/2}\Bigg\{\chi_{+}\Bigg[\frac{8}{3} - \frac{4}{3} \nu\Bigg] + \frac{8}{3} \delta\chi_{-}\Bigg\} \nn \\
& + x^{7/2}\Bigg\{\chi_{+}\Bigg[8 - \frac{121}{9} \nu + \frac{2}{9} \nu^{2}\Bigg] + \delta\chi_{-}\Bigg[8 - \frac{31}{9} \nu\Bigg]\Bigg\} \nn \\
& + x^{9/2}\Bigg\{\chi_{+}\Bigg[27 - \frac{373}{4} \nu + \frac{86}{3} \nu^{2} + \frac{1}{6} \nu^{3}\Bigg] + \delta\chi_{-}\Bigg[27 - \frac{211}{4} \nu + \frac{7}{6} \nu^{2}\Bigg]\Bigg\}\\
\varepsilon_{\text{circ}}^{S^2}(x) &= x^{3}\Bigg\{-\chi_{+}^2 - 2 \delta\chi_{+}\chi_{-} + \chi_{-}^2\Bigg[-1 + 4 \nu\Bigg]\Bigg\} \nn \\
& + x^{4}\Bigg\{\chi_{+}^2\Bigg[-\frac{65}{18} + \frac{245}{18} \nu - \frac{40}{9} \nu^{2}\Bigg] + \delta\chi_{+}\chi_{-}\Bigg[-\frac{65}{9} + \frac{145}{9} \nu\Bigg] + \chi_{-}^2\Bigg[-\frac{65}{18} + \frac{305}{18} \nu - \frac{10}{3} \nu^{2}\Bigg]\Bigg\} \nn \\
& + x^{5}\Bigg\{\chi_{+}^2\Bigg[-\frac{469}{24} + \frac{19775}{216} \nu - \frac{15113}{216} \nu^{2} + \frac{196}{27} \nu^{3}\Bigg] + \delta\chi_{+}\chi_{-}\Bigg[-\frac{469}{12} + \frac{13223}{108} \nu - \frac{4249}{108} \nu^{2}\Bigg] \nn\\
&\qquad + \chi_{-}^2\Bigg[-\frac{469}{24} + \frac{23555}{216} \nu - \frac{27125}{216} \nu^{2} + \frac{7}{18} \nu^{3}\Bigg]\Bigg\}\\
\varepsilon_{\text{circ}}^{S^3}(x) &= x^{9/2}\Bigg\{-4 \nu\chi_{+}^3 - 8 \delta \nu\chi_{+}^2\chi_{-} + \chi_{+}\chi_{-}^2\Bigg[-4 \nu + 16 \nu^{2}\Bigg]\Bigg\}\\
\varepsilon_{\text{circ}}^{S^4}(x) &= 0
\end{align}
\end{subequations}
We have checked that the linear and quadratic-in-spin contributions agree with (12) of \cite{Cho:2022syn} (including the SID contributions) and that the cubic-in-spin contributions agree with (6.16) and (6.17) of \cite{Marsat:2014xea} (including the SID contributions). The vanishing of the quartic-in-spin contribution in the case of two black holes confirms the prediction of \cite{Siemonsen:2017yux} --- note that the SID piece does indeed have a nonvanishing quartic contribution; see the Supplemental Material~\cite{Supplemental}.

The spinning contributions to the redshift  $z_1^{\text{circ}}(x)=z_1^{\text{circ},S^1}(x) + z_1^{\text{circ},S^2}(x) + z_1^{\text{circ},S^3}(x) +z_1^{\text{circ},S^4}(x)$ read
\begin{subequations}
\begin{align}
z_1^{\text{circ},S^1}(x) &= x^{5/2}\Bigg\{\chi_{+}\Bigg[1 - \frac{19}{6} \nu + \frac{4}{3} \nu^{2} + \delta\Bigg(-1 + \frac{1}{2} \nu\Bigg)\Bigg] + \chi_{-}\Bigg[-1 + \frac{5}{2} \nu + \delta\Bigg(1 - \frac{7}{6} \nu\Bigg)\Bigg]\Bigg\} \nn \\
& + x^{7/2}\Bigg\{\chi_{+}\Bigg[\frac{3}{2} - \frac{31}{6} \nu + \frac{217}{36} \nu^{2} - \frac{2}{9} \nu^{3} + \delta\Bigg(-\frac{3}{2} + \frac{19}{6} \nu - \frac{1}{12} \nu^{2}\Bigg)\Bigg] \nn \\
&\qquad + \chi_{-}\Bigg[-\frac{3}{2} + \frac{37}{6} \nu - \frac{47}{12} \nu^{2} + \delta\Bigg(\frac{3}{2} - \frac{13}{6} \nu + \frac{79}{36} \nu^{2}\Bigg)\Bigg]\Bigg\} \nn \\
& + x^{9/2}\Bigg\{\chi_{+}\Bigg[\frac{27}{8} - \frac{73}{4} \nu + \frac{539}{16} \nu^{2} - \frac{713}{48} \nu^{3} - \frac{1}{6} \nu^{4} + \delta\Bigg(-\frac{27}{8} + \frac{73}{4} \nu - \frac{117}{16} \nu^{2} - \frac{1}{16} \nu^{3}\Bigg)\Bigg] \nn \\
&\qquad + \chi_{-}\Bigg[-\frac{27}{8} + 25 \nu - \frac{695}{16} \nu^{2} + \frac{23}{16} \nu^{3} + \delta\Bigg(\frac{27}{8} - \frac{23}{2} \nu + \frac{273}{16} \nu^{2} - \frac{41}{48} \nu^{3}\Bigg)\Bigg]\Bigg\} \\
z_1^{\text{circ},S^2}(x) &= x^{3}\Bigg\{\chi_{+}^2\Bigg[-\frac{1}{4} + \frac{5}{4} \nu + \delta\Bigg(\frac{1}{4} - \frac{1}{4} \nu\Bigg)\Bigg] + \chi_{+}\chi_{-}\Bigg[\frac{1}{2} - \frac{3}{2} \nu + \delta\Bigg(-\frac{1}{2} + \frac{3}{2} \nu\Bigg)\Bigg] \nn \\
&\qquad + \chi_{-}^2\Bigg[-\frac{1}{4} + \frac{5}{4} \nu - 2 \nu^{2} + \delta\Bigg(\frac{1}{4} - \frac{1}{4} \nu\Bigg)\Bigg]\Bigg\} \nn \\
& + x^{4}\Bigg\{\chi_{+}^2\Bigg[-\frac{13}{24} + \frac{293}{72} \nu - \frac{761}{72} \nu^{2} + \frac{28}{9} \nu^{3} + \delta\Bigg(\frac{13}{24} - \frac{79}{24} \nu + \frac{19}{24} \nu^{2}\Bigg)\Bigg] \nn \\
& \qquad + \chi_{+}\chi_{-}\Bigg[\frac{13}{12} - \frac{35}{4} \nu + \frac{113}{12} \nu^{2} + \delta\Bigg(-\frac{13}{12} + \frac{215}{36} \nu - \frac{307}{36} \nu^{2}\Bigg)\Bigg] \nn \\
& \qquad + \chi_{-}^2\Bigg[-\frac{13}{24} + \frac{293}{72} \nu - \frac{773}{72} \nu^{2} + \frac{7}{3} \nu^{3} + \delta\Bigg(\frac{13}{24} - \frac{79}{24} \nu + \frac{5}{8} \nu^{2}\Bigg)\Bigg]\Bigg\} \nn \\
& + x^{5}\Bigg\{\chi_{+}^2\Bigg[-\frac{67}{32} + \frac{35467}{2016} \nu - \frac{159031}{3024} \nu^{2} + \frac{36773}{864} \nu^{3} - \frac{154}{27} \nu^{4} + \delta\Bigg(\frac{67}{32} - \frac{31495}{2016} \nu + \frac{13553}{1008} \nu^{2} - \frac{451}{288} \nu^{3}\Bigg)\Bigg] \nn \\
&\qquad  + \chi_{+}\chi_{-}\Bigg[\frac{67}{16} - \frac{39937}{1008} \nu + \frac{2309}{28} \nu^{2} - \frac{4121}{144} \nu^{3} + \delta\Bigg(-\frac{67}{16} + \frac{27025}{1008} \nu - \frac{45467}{756} \nu^{2} + \frac{10411}{432} \nu^{3}\Bigg)\Bigg] \nn \\
&\qquad + \chi_{-}^2\Bigg[-\frac{67}{32} + \frac{35467}{2016} \nu - \frac{173269}{3024} \nu^{2} + \frac{52661}{864} \nu^{3} - \frac{11}{36} \nu^{4} + \delta\Bigg(\frac{67}{32} - \frac{31495}{2016} \nu + \frac{19559}{1008} \nu^{2} - \frac{3}{32} \nu^{3}\Bigg)\Bigg]\Bigg\} \\
z_1^{\text{circ},S^3}(x) &= x^{9/2}\Bigg\{\chi_{+}^3\Bigg[\frac{9}{2} \nu^{2} + \delta\Bigg(\nu - \frac{1}{2} \nu^{2}\Bigg)\Bigg] + \chi_{+}^2\chi_{-}\Bigg[3 \nu - 4 \nu^{2} + 5 \delta \nu^{2}\Bigg] \nn \\
&\qquad + \chi_{+}\chi_{-}^2\Bigg[\frac{5}{2} \nu^{2} - 8 \nu^{3} + \delta\Bigg(3 \nu - \frac{1}{2} \nu^{2}\Bigg)\Bigg] + \chi_{-}^3\Bigg[\nu - 5 \nu^{2} + 2 \delta \nu^{2}\Bigg]\Bigg\} \\
z_1^{\text{circ},S^4}(x) &= x^{5}\Bigg\{\chi_{+}^4\Bigg[-\frac{3}{16} \nu - \frac{15}{16} \nu^{2} + \delta\Bigg(-\frac{3}{16} \nu + \frac{3}{16} \nu^{2}\Bigg)\Bigg] + \chi_{+}^3\chi_{-}\Bigg[-\frac{3}{4} \nu + \frac{3}{4} \nu^{2} + \delta\Bigg(-\frac{3}{4} \nu - \frac{3}{4} \nu^{2}\Bigg)\Bigg] \nn \\
&\qquad + \chi_{+}^2\chi_{-}^2\Bigg[-\frac{9}{8} \nu + \frac{39}{8} \nu^{2} + \delta\Bigg(-\frac{9}{8} \nu - \frac{3}{8} \nu^{2}\Bigg)\Bigg] + \chi_{+}\chi_{-}^3\Bigg[-\frac{3}{4} \nu + \frac{15}{4} \nu^{2} + \delta\Bigg(-\frac{3}{4} \nu + \frac{9}{4} \nu^{2}\Bigg)\Bigg] \nn \\
& \qquad  + \chi_{-}^4\Bigg[-\frac{3}{16} \nu + \frac{9}{16} \nu^{2} + \delta\Bigg(-\frac{3}{16} \nu + \frac{27}{16} \nu^{2}\Bigg)\Bigg]\Bigg\}
\end{align}
\end{subequations}
We have checked that 
the spin-orbit contribution to the redshift agrees at 4PN with  (61) of \cite{Bini:2019lcd} and that the quadratic-in-spin contribution agrees at 4PN  with (6.4) of \cite{Bini:2020zqy}, including the SID terms. However, the quadratic-in-spin contribution disagrees already at 2PN with (62) and (63) of \cite{Bini:2019lcd}, as discussed previously in Sec.~\ref{subsec:first_law}. We have also checked that, in the case of a binary composed of one spinning black hole and one nonspinning black hole, our redshift perfectly agrees with the result presented in the Supplemental Material of \cite{Bautista:2024agp} at 4PN, to all orders in the mass ratio and to all orders in spin. In the test-mass limit $\nu\rightarrow 0$, our full result also agrees at 4PN with the exact geodesic redshift of the secondary black hole in a Kerr background~\cite{Siemonsen:2019dsu}:
\begin{align}
    z_2^\text{circ}(x) = \sqrt{\Big(1-x^{3/2}\chi_1\Big)\Big(1+x^{3/2}\chi_1-3x(1-x^{3/2} \chi_1)^{1/3}\Big)} + \mathcal{O}(\nu)\,.
\end{align}

The spinning contributions to the gyroscopic invariant $\psi_{1\,\text{circ}}^\text{spin}(x) = \psi_{1\,\text{circ}}^{S^0}(x) + \psi_{1\,\text{circ}}^{S^1}(x) + \psi_{1\,\text{circ}}^{S^2}(x) + \psi_{1\,\text{circ}}^{S^3}(x)$ read
\begin{subequations}
\begin{align}
\psi_{1\,\text{circ}}^{S^0}(x) &= x\Bigg\{\frac{3}{4} + \frac{1}{2} \nu - \frac{3}{4} \delta\Bigg\}  + x^{2}\Bigg\{\frac{9}{16} + \frac{5}{4} \nu - \frac{1}{24} \nu^{2} + \delta\Bigg(-\frac{9}{16} + \frac{5}{8} \nu\Bigg)\Bigg\} \nn \\
& + x^{3}\Bigg\{\frac{27}{32} + \frac{3}{16} \nu - \frac{105}{32} \nu^{2} - \frac{1}{48} \nu^{3} + \delta\Bigg(-\frac{27}{32} + \frac{39}{8} \nu - \frac{5}{32} \nu^{2}\Bigg)\Bigg\} \\
\psi_{1\,\text{circ}}^{S^1}(x) &= x^{3/2}\Bigg\{\chi_{+}\Bigg[-\frac{1}{2} + \frac{1}{2} \delta\Bigg] + \chi_{-}\Bigg[\frac{1}{2} - 2 \nu - \frac{1}{2} \delta\Bigg]\Bigg\} \nn \\
& + x^{5/2}\Bigg\{\chi_{+}\Bigg[-\frac{1}{4} + \frac{1}{12} \nu + \frac{4}{3} \nu^{2} + \delta\Bigg(\frac{1}{4} - \frac{7}{4} \nu\Bigg)\Bigg] + \chi_{-}\Bigg[\frac{1}{4} - \frac{3}{4} \nu + \nu^{2} + \delta\Bigg(-\frac{1}{4} - \frac{23}{12} \nu\Bigg)\Bigg]\Bigg\} \nn \\
& + x^{7/2}\Bigg\{\chi_{+}\Bigg[-\frac{15}{16} + \frac{59}{48} \nu + \frac{1057}{144} \nu^{2} - \frac{14}{9} \nu^{3} + \delta\Bigg(\frac{15}{16} - \frac{371}{48} \nu + \frac{55}{16} \nu^{2}\Bigg)\Bigg] \nn \\
&\qquad + \chi_{-}\Bigg[\frac{15}{16} - \frac{407}{48} \nu + \frac{889}{48} \nu^{2} - \frac{1}{12} \nu^{3} + \delta\Bigg(-\frac{15}{16} - \frac{85}{48} \nu + \frac{601}{144} \nu^{2}\Bigg)\Bigg]\Bigg\} \\
\psi_{1\,\text{circ}}^{S^2}(x) &= x^{3}\Bigg\{\chi_{+}^2\Bigg[-\frac{1}{4} - \frac{1}{2} \nu + \delta\Bigg(\frac{1}{4} + \nu\Bigg)\Bigg] + \chi_{+}\chi_{-}\Bigg[\frac{1}{2} - \nu - 4 \nu^{2} - \frac{1}{2} \delta\Bigg] + \chi_{-}^2\Bigg[-\frac{1}{4} + \frac{3}{2} \nu - 2 \nu^{2} + \delta\Bigg(\frac{1}{4} - \nu\Bigg)\Bigg]\Bigg\} \\
\psi_{1\,\text{circ}}^{S^3}(x) &= 0
\end{align}
\end{subequations}
Notice the vanishing of the $S^3$ piece in the case of two black holes, whereas the SID piece does indeed have a nonvanishing $S^3$ contribution; see the Supplemental Material~\cite{Supplemental}.
We have checked that the gyroscopic invariant perfectly agrees with the results presented in (92) and (93) of \cite{Bini:2019lkm}.
In the case of a binary composed of one spinning black hole and one nonspinning black hole, we have also found perfect overlap with the results in the Supplemental Material of Ref.~\cite{Bautista:2024agp}. In the test-mass limit $\nu\rightarrow 0$, our full result also agrees at 4PN with the exact geodesic gyroscopic invariant of the secondary black hole in a Kerr background~\cite{Siemonsen:2019dsu}:
\begin{align}
    \psi_2^\text{circ}(x) = 1-\sqrt{ \frac{1+x^{3/2}\chi_1-3x(1-x^{3/2} \chi_1)^{1/3}}{1- \chi_1 x^{3/2}} } + \mathcal{O}(\nu)\,.
\end{align}

Finally, for completeness, we provide in the Supplemental Material~\cite{Supplemental} the expressions of the redshift and gyroscopic invariants in terms of the energy:
\begin{subequations}
    \begin{align}
    z_\text{1\,circ}(\varepsilon) &= z_\text{1\,circ}^\text{pp}(\varepsilon)+z_\text{1\,circ}^\text{spin}(\varepsilon)+z_\text{1\,circ}^\text{SID}(\varepsilon) \,,\\
    \psi_\text{1\,circ}(\varepsilon) &=  \psi_\text{1\,circ}^\text{spin}(\varepsilon)+\psi_\text{1\,circ}^\text{SID}(\varepsilon) \,;
\end{align}
\end{subequations}
see also (7.5a) of Ref.~\cite{Trestini:2025yyc} for the expression of $z_\text{1\,circ}^\text{pp}(\varepsilon)$.

\section{Innermost stable circular orbit (ISCO)}
\label{sec:isco}

The separatrix between stable bound orbits and unstable plunging orbits is determined by analyzing the stability of Hamilton's equations. When restricting to circular orbits, the separatrix corresponds to the innermost stable circular orbit (ISCO), and it is possible to determine its (azimuthal) frequency. In gravitational self-force theory, the stability analysis is performed at the level of the effective potential~\cite{Barack:2009ey,Isoyama:2014mja}. In post-Newtonian theory, as well as in the EOB approach, different criteria have been considered~\cite{Damour:2000we, Buonanno:2002ft,Favata:2010yd}, but the one that stood out as most robust when compared against self-force results was the Blanchet-Iyer-Favata criterion~\cite{Blanchet:2002mb, Favata:2010ic}, which is based on an explicit stability analysis of the Hamiltonian. This criterion has recently been extended to fourth post-Newtonian order, including the contributions of tails~\cite{Blanchet:2025agj} and spins~\cite{Blanchet:2026wwa}. 

In gravitational self-force theory, the separatrix between bound and unbound orbits is equivalently determined by the criterion $\mathcal{C} = K^{-2}=(\Omega_r/\Omega_\phi)^2 = 0$. If one additionally restricts to circular orbits, this same criterion yields the ISCO frequency. Of course, this criterion is equivalent to the simpler condition $\Omega_r = 0$; however, as one approaches circular orbits, $\Omega_r$ becomes an extremely singular quantity that is not suitable for a perturbative study, be it the post-Newtonian or mass-ratio expansion; see, e.g., Sec. II.A of \cite{Barack:2011ed}. It was thus argued in Sec. II.A.2 of \cite{Barack:2010ny} that~$\mathcal{C}$~is a well-behaved quantity near the ISCO, such that $\mathcal{C}=0$ is the correct condition to perturbatively determine the ISCO frequency; see also Refs.~\cite{Barack:2010tm, Trestini:2026tky}.

Here, for the first time, we apply this alternative criterion $\mathcal{C} = K^{-2} = 0$ (which stems from the self-force and EOB approaches) to the case of post-Newtonian theory. The ISCO frequency is a solution to $\mathcal{C}(x)=0$ where
\begin{align}
    \mathcal{C}(x)&=\mathcal{C}_\text{pp}(x)+\mathcal{C}_{S^1}(x)+\mathcal{C}_{S^2}(x)+\mathcal{C}_{S^3}(x)+\mathcal{C}_{S^4}(x)+\mathcal{C}_\text{SID}(x)\,.
\end{align}
The point-particle criterion includes tail effects and reads
\begin{align}
    \mathcal{C}_\text{pp}(x) &= 1-6 x+ \nu \Bigg\{14 x^2   +x^3 \Bigg[\left(\frac{397}{2}-\frac{123 \pi ^2}{16}\right) -14 \nu \Bigg]\nn\\
    &\qquad +x^4 \Bigg[-\frac{215729}{180}+\frac{5024}{15}\gamma_E+\frac{58265}{1536}\pi
   ^2+\frac{1184}{15}\ln 2+\frac{2916}{5} \ln 3+\frac{2512}{15} \ln x \nn\\
   &\qquad\qquad + \nu\left(-\frac{4223}{6}+\frac{451 \pi
   ^2}{16}\right)  +\frac{196}{27}\nu^2  \Bigg] \Bigg\} \,.
\end{align}
This exactly reproduces the Blanchet-Iyer-Favata criterion, including the recent 4PN results obtained in Ref.~\cite{Blanchet:2025agj}, which accounts for the tail contributions. In that work, the stability analysis was performed both at the level of the equations of motion and at the level of the ADM Hamiltonian. Their derivation involved a detailed treatment of the tail term, and our result is a nice confirmation that their calculation can be directly reproduced from the expression of the periastron advance, see e.g. (5.10) of \cite{Bernard:2017ktp}. Moreover, notice that this criterion is exact at leading order in the mass ratio, namely it reproduces the criterion for the Schwarzschild ISCO:
\begin{align}
    \mathcal{C}_\text{Schwarzschild}(x)=1-6x
\end{align}
This was already noticed in, e.g., Eq. (16) of Ref.~\cite{Barack:2010ny}, which led to the ansatz~$\mathcal{C}(x)=1-6x+\nu \rho(x)+\mathcal{O}(\nu^2)$.

The spin contributions to the criterion then read
\begin{subequations}
\begin{align}
\mathcal{C}_{S^1}(x) &= x^{3/2}\Bigg\{\chi_{+}\Bigg[8 - 4 \nu\Bigg] + 8 \delta\chi_{-}\Bigg\} \nn \\
&\quad + x^{5/2}\Bigg\{\chi_{+}\Bigg[-4 - 45 \nu + 4 \nu^{2}\Bigg] + \delta\chi_{-}\Bigg[-4 - 17 \nu\Bigg]\Bigg\} \nn \\
&\quad + x^{7/2}\Bigg\{\chi_{+}\Bigg[-\frac{1465}{12} \nu + \frac{373}{3} \nu^{2} - \frac{4}{3} \nu^{3}\Bigg] + \delta\chi_{-}\Bigg[-\frac{1465}{12} \nu + \frac{44}{3} \nu^{2}\Bigg]\Bigg\} \\
\mathcal{C}_{S^2}(x) &= x^{2}\Bigg\{-3\chi_{+}^2 - 6 \delta\chi_{+}\chi_{-} + \chi_{-}^2\Bigg[-3 + 12 \nu\Bigg]\Bigg\} \nn \\
&\quad + x^{3}\Bigg\{\chi_{+}^2\Bigg[8 + 51 \nu - 16 \nu^{2}\Bigg] + \delta\chi_{+}\chi_{-}\Bigg[16 + 62 \nu\Bigg] + \chi_{-}^2\Bigg[8 - 21 \nu - 20 \nu^{2}\Bigg]\Bigg\} \nn \\
&\quad + x^{4}\Bigg\{\chi_{+}^2\Bigg[-\frac{10}{3} - \frac{37}{4} \nu - \frac{670}{3} \nu^{2} + \frac{104}{3} \nu^{3}\Bigg] + \delta\chi_{+}\chi_{-}\Bigg[-\frac{20}{3} + \frac{169}{6} \nu - \frac{544}{3} \nu^{2}\Bigg] \nn\\
&\qquad\qquad + \chi_{-}^2\Bigg[-\frac{10}{3} + \frac{203}{4} \nu - 253 \nu^{2} + \frac{40}{3} \nu^{3}\Bigg]\Bigg\} \\
\mathcal{C}_{S^3}(x) &= x^{7/2}\Bigg\{\chi_{+}^3\Bigg[-4 - 16 \nu\Bigg] + \delta\chi_{+}^2\chi_{-}\Bigg[-12 - 32 \nu\Bigg] + \chi_{+}\chi_{-}^2\Bigg[-12 + 32 \nu + 64 \nu^{2}\Bigg]  + \delta\chi_{-}^3\Bigg[-4 + 16 \nu\Bigg]\Bigg\} \\
\mathcal{C}_{S^4}(x) &= 0
\end{align}
\end{subequations}
The SID terms are relegated to the Supplemental Material~\cite{Supplemental}.
Remarkably, this expression exactly reproduces the very recent 4PN  stability criterion with spin \cite{Blanchet:2026wwa}, including the SID terms. We have also checked that our result agrees with  Eq.~(10) of \cite{LeTiec:2013uey} in the region of overlap (3.5PN and up to quartic-in-spin for black holes).
Taking our result in the $\nu \rightarrow 0$ limit (where we identify $\chi_1=a$ and $\chi_2$ does not appear at this order), we recover the 4PN expansion of the Kerr criterion,
\begin{align}
    \mathcal{C}_\text{Kerr}(x) &= 1-\frac{6}{p(x,a)} + \frac{8 a}{p^{3/2}(x,a)}- \frac{3a^2}{p^2(x,a)}
\end{align}
where $p(x,a) = x^{-1}(1-a x^{3/2})^{2/3} $. Unlike the Schwarzschild case, the PN expansion of the geodesic criterion does not truncate, so our 4PN result is not exact at leading order in the mass ratio. This suggests a `resummation' where the geodesic term is given by its exact expression, and higher-order terms of the small mass-ratio expansion are given by their 4PN expressions (note that $\delta=\sqrt{1-4\nu}$ needs to be reexpanded in small $\nu$ for this scheme).

We refer to Refs.~\cite{Blanchet:2025agj, Blanchet:2026wwa} for a detailed analysis of the accuracy of such a criterion.

\section{Scattering angle}
\label{sec:scattering}

Unlike in the rest of this paper, which was devoted to the bound case, we now consider the scattering case. We introduce $\bar\varepsilon=-\varepsilon$ and $\bar{\jmath}=-j$; in the scattering case, $\bar\varepsilon>0$ and $\bar{\jmath}>0$. Recalling the notations of Sec.~\ref{sec:action_angle}, we can write the local scattering angle as
\begin{align}
    \frac{\chi_\text{loc}}{2} = \int_0^{s_+}\frac{\dd s}{s^2} \frac{\mathcal{S}(s)}{\sqrt{\mathcal{R}(s)}}
\end{align}

To compute this explicitly, let $s_-$ and $s_+$ be the two roots of the polynomial $\mathcal{R}(s)$ that admit a nonzero limit as $c\rightarrow\infty$; we choose $s_-<0<s_+$. These are constructed perturbatively from the Newtonian roots
\begin{align}
    s_\pm=\frac{B\pm\sqrt{B^2-AC}}{-C} + \mathcal{O}(2) \,,
\end{align}
where one should remember that in the scattering case, $A>0$, $B>0$, and $C<0$. Up to neglected 4PN terms, one can then factorize the polynomial
\begin{align}
    \mathcal{R}(s)=(s-s_-)(s_+-s)\widetilde{\mathcal{R}}(s)\,,
\end{align}
where at 4PN, $\widetilde{\mathcal{R}}(s)$ is a polynomial of order 7. We then find that the scattering angle is given, after PN-expanding inside the integral~\cite{Damour:1988mr}, by
\begin{align}
     \frac{\chi_\text{loc}}{2} = \int_0^{s_+}\dd s \frac{\overline{\mathcal{Q}}(s)}{\sqrt{(s-s_-)(s_+-s)}} + \mathcal{O}(10)\,,
\end{align}
where $\overline{\mathcal{Q}}(s)$ is a polynomial of order 7 resulting from the PN-expansion of 
\begin{align}
    \mathcal{Q}(s)=\frac{\mathcal{S}(s)}{s^2\sqrt{\widetilde{\mathcal{R}}(s)}}\,.
\end{align}
Thus, we are only left with the master integrals
\begin{align}
    \mathcal{J}_n=\int_0^{s_+}\dd s \, \frac{s^n}{\sqrt{(s-s_-)(s_+-s)}} \,\,
\end{align}
As we show in App.~\ref{app:integrals}, these can be written in closed form as
\begin{align}
    \mathcal{J}_n &=      2 \arctan\left(\sqrt{-\frac{s_+}{s_-}}\right)  \frak{R}_n(s_+,s_-)
    + \sqrt{-s_+ s_-}  \sum_{k=1}^{n} \frac{1}{k}    \frak{R}_{k-1}(s_+,s_-)   \frak{R}_{n-k}(s_+,s_-) \,, 
\end{align}
where $\frak{R}_n(s_+,s_-)$ is a bivariate polynomial in $s_+$ and $s_-$ which reads explicitly
\begin{align}
    \frak{R}_n(s_+,s_-) =  \sum_{k=0}^{\lfloor n/2 \rfloor} \frac{1}{4^{n-k}}\begin{pmatrix}
        n\\k
    \end{pmatrix} \begin{pmatrix}
        2n-2k\\n
    \end{pmatrix}  (s_+ + s_-)^{n-2k}(-s_+ s_-)^{k} \,.
\end{align}
Substituting $s_+$ and $s_-$ by their expressions, one then first obtains the scattering angle in terms of $A$, $B$, $C$, etc. The generic expression we find extends Eq.~(3.7) of~Ref.~\cite{Usseglio:2025iwt}, but is too lengthy to present here. Instead, we relegate it to the Supplemental Material~\cite{Supplemental} and present here only its structure: 
\begin{align}
    \chi_\text{loc} &= - \pi + \frac{1}{\sqrt{-C}} \arctan\left(\frac{B+\sqrt{B^2-A C}}{\sqrt{-A C}}\right)\Bigg\{4F + \frac{\epsilon}{C^2}\Bigg[\dots\Bigg] + \frac{\epsilon^2}{C^4}\Bigg[\dots\Bigg]  + \frac{\epsilon^3}{C^6}\Bigg[\dots\Bigg]  + \frac{\epsilon^4}{C^8}\Bigg[\dots\Bigg] \Bigg\}\nn\\
    &\qquad\qquad\qquad + \sqrt{A}\Bigg\{\frac{\epsilon}{C^2(B^2-AC)}\Bigg[\dots\Bigg]+\frac{\epsilon^2}{C^4(B^2-AC)^2}\Bigg[\dots\Bigg]+\frac{\epsilon^3}{C^6(B^2-AC)^3}\Bigg[\dots\Bigg]\Bigg\} \,,
\end{align}
where the ellipses represent polynomials in the coefficients $(A,B,C,F,D_n,I_n)$ and $\epsilon$ is a bookkeeping parameter that counts the PN orders. 
One then substitutes the coefficients $(A,B,C,F,D_n,I_n)$ with their expressions in terms of $(\bar{\varepsilon},\bar{\jmath})$ and reexpands to 4PN order. We encounter the same difficulty as in the bound case: the expressions of these coefficients are known in the nonspinning sector to 4PN in ADM coordinates, whereas in EFT coordinates, the spin sector is known to 4PN but the nonspinning sector is only known to 2PN. We adopt the same strategy as for the radial action: (i) we compute the nonspinning sector using ADM coordinates; (ii) we compute the full spinning scattering angle using EFT coordinates, then discard the nonspinning sector arising from that computation; (iii) we add the two together, which is legitimate since the relation between the scattering angle and the constants of motion is gauge-invariant.

The full scattering angle then naturally splits into point-particle, spinning and SID contributions, namely
\begin{align}
    \chi&= \chi_\text{pp}+\chi_\text{spin} + \chi_\text{SID}\,.
\end{align}
The spin part is split into powers of spin, namely 
\begin{align}
    \chi_\text{spin} &= \chi_{S^1} + \chi_{S^2} + \chi_{S^3} +\chi_{S^4} \,,
\end{align}
whereas the point-particle piece itself splits into a local and a tail part, namely
\begin{align}
    \chi_\text{pp} &=  \chi_\text{pp loc}+\chi_\text{pp tail}\,.
\end{align}
The techniques described in this section can only help us obtain the local part, whereas the tail part\footnote{Note that the split between local and tail parts is different in Ref.~\cite{Bini:2017wfr}, since logarithmic contributions are included in the local part.} was treated in Ref.~\cite{Bini:2017wfr}. We obtain all local and spinning contributions, but these are very lengthy. In Appendix~\ref{app:scattering}, we present the following contributions: $\chi_\text{pp loc}$, $\chi_{S^1}$, $\chi_{S^2}$, $\chi_{S^3}$, and $\chi_{S^4}$. The $\chi_\text{SID}$ contribution is only given in the Supplemental Material~\cite{Supplemental}.

As a sanity check, we have verified that we can recover the local 4PN periastron advance from the scattering angle using the boundary-to-bound map, including all spin contributions~\cite{Kalin:2019inp}:
\begin{align}
    2\pi\Big[K_\text{loc}(E,J,\chi_1,\chi_2)-1\Big]=\chi_\text{loc}(E,J,\chi_1,\chi_2)+\chi_\text{loc}(E,-J,-\chi_1,-\chi_2) \,.
\end{align}
We have also verified that we exactly recover the 3PN nonspinning scattering angle of \cite{Bini:2017wfr} [we have omitted the comparison with the 4PN part due to the different conventions for the split between local and tail pieces]. We have checked that the linear-in-spin piece completely recovers (160) of \cite{Bini:2017wfr}. It also agrees with (3.12) of Ref.~\cite{Khalil:2021fpm}, which is limited to 5PM accuracy, in the region of overlap. Straightforwardly extending to 6PM the ansatz laid out in (3.12) of \cite{Khalil:2021fpm}, we find the following values for the 6PM coefficients: 
\begin{align}
{X}_{61} =0 \,, \  
X_{63} = 0\,,  \quad\mathsf{X}_{65} = -\frac{9405}{8}\,, \ 
{X}_{65}^{\delta} = \frac{2475}{8}\,, \ 
\mathsf{X}_{65}^{\nu} = \frac{9225}{16}\,, \ 
{X}_{65}^{\nu\delta} = -\frac{1305}{16}\,, \ 
{X}_{65}^{\nu^{2}} = -\frac{285}{16}\,, \  
{X}_{65}^{\nu^{2}\delta} = \frac{15}{16} \,.
\end{align}

Finally, we have compared our scattering angle with the results of Ref.~\cite{Bautista:2024agp} and found perfect agreement in the overlap. In that work, the quantity $\theta$ is provided in the Supplemental Material, manifestly as a post-Minkowskian (PM) expansion: for each $S^n$ contribution with $n\in[\![0,4]\!]$, the leading PM order scales with the impact parameter $b$ as $\sim b^{-n-1}$, and the PM series is controlled at next-to-next-to-leading order, namely up to $\sim b^{-n-3}$. We find that our $\chi$ corresponds in that work to the quantity $\Gamma \theta$, where $\Gamma\equiv1+\bar\varepsilon \nu/2$. We then  
convert\footnote{This comparison requires the map between our variables and the EOB variables of  Ref.~\cite{Bautista:2024agp}, referred to as BKSKV. As usual~\cite{Trestini:2025yyc}: $\gamma_\text{\tiny BKSKV} = 1+\bar\varepsilon/2+ \bar\varepsilon^2\nu/8$, $\Gamma_\text{\tiny BKSKV}=1+\bar\varepsilon \nu/2$, and $L_\text{\tiny BKSKV}=(G m^2 \nu/c)\sqrt{\bar\jmath/\bar\varepsilon}$. From (15) of \cite{Bautista:2024agp} we further learn that
$$b_\text{\tiny BKSKV} = \frac{G m \sqrt{\bar\jmath}}{\bar\varepsilon} \cdot \frac{ 8+4\bar\varepsilon\nu -   \nu\bar\varepsilon \sqrt{\bar\varepsilon/\bar\jmath}\Big[(8+\bar\varepsilon)\chi_+ + \delta \bar\varepsilon \chi_-\Big]}{ \sqrt{4+ \bar\varepsilon \nu} \sqrt{16+4 \bar\varepsilon + \bar\varepsilon^2 \nu}}\,. $$
We have checked that the PN expansion of this exact relation in the nonspinning case recovers (D1) of \cite{Usseglio:2025iwt}. }
their expression in terms of our variables and PM-expand our result in large $\bar\jmath$ (since $b \sim {G m \sqrt{\bar\jmath}}/{\bar\varepsilon}$). For each $S^n$ contribution with $n\in[\![0,4]\!]$, we find that we exactly recover the next-to-next-to-leading PM truncation at 4PN order for the spinning terms and at 3PN order for the nonspinning terms. We also note that Ref.~\cite{Bautista:2024agp} also provides $S^5$ and $S^6$ terms: we have checked, after conversion to our variables, that these vanish at 4PN order, in agreement with our findings. Despite this perfect overlap at 4PN, we would like to stress that our results are nonperturbative in the PM sense, as can be seen from the arc-tangent contributions, whereas  Ref.~\cite{Bautista:2024agp} is truncated in the PM sense.

\section{Conclusion}
\label{sec:conclusion}

In this work, using the complete 4PN Hamiltonian to all orders in spin \cite{Levi:2016ofk},  we have entirely completed the map between constants of motion and fundamental frequencies at 4PN for aligned-spin eccentric systems. This includes all relevant spin contributions in the standard PN counting, i.e., up to $S^4$. Using the first law of binary black hole mechanics, we have then obtained the redshift and gyroscopic invariants at the same order. We have then reduced all these results to circular orbits. Based on arguments stemming from gravitational self-force, we have noticed that we could easily recover the Blanchet-Iyer-Favata criterion for the ISCO frequency which was recently computed at 4PN, including spin contributions, in Refs.~\cite{Blanchet:2025agj,Blanchet:2026wwa}: their criterion $\mathcal{C}(x)=0$ is in fact equivalent to the PN expansion of   $K^{-2}(x)=0$, where $K(x)$ is the periastron advance for circular orbits. Finally, we have completed the 4PN scattering angle with all spinning contributions. To ensure their robustness, we have successfully validated our results against large swaths of the literature~\cite{Henry:2023tka,
Bini:2019lcd,
Bini:2020zqy,
Bautista:2024agp,
Cho:2022syn,
Marsat:2014xea,
Siemonsen:2017yux,
Bini:2017wfr,
Khalil:2021fpm,
Blanchet:2025agj,
Blanchet:2026wwa,
LeTiec:2013uey} and checked explicitly the  boundary-to-bound map between the scattering angle and the periastron advance~\cite{Kalin:2019inp}. Some of our results will be included in the public repository \texttt{PNpedia}~\cite{PNpedia}.

This work is a first step towards obtaining a complete 4PN quasi-Keplerian parametrization~\cite{DamourDeruelle86, Schafer:1993pkg,Wex:1995pjg,Memmesheimer:2004cv, Tessmer:2012xr, Cho:2021oai,Henry:2023tka} for spin-aligned quasi-elliptic orbits:  the main difficulty would lie in the treatment of the tail term, whereas the spin terms should be straightforward to compute. This quasi-Keplerian parametrization is necessary to specialize the 4PN flux of energy~\cite{Blanchet:2023bwj,Blanchet:2023sbv,Cho:2022syn,Marsat:2014xea,Siemonsen:2017yux} to the case of spin-aligned elliptic orbits. One would then be in a position to obtain the spin-aligned eccentric phasing at 4PN, which is an important ingredient for waveform templates. We also believe that this work will be useful in synergy with gravitational self-force and post-Minkowskian theory, e.g., in the context of hybrid models~\cite{Honet:2025gge,Honet:2025lmk,Warburton:2025ymy}.

A natural extension of this work would be the treatment of the case of unaligned spin, which exhibits precession. This problem is much more difficult as it is not separable, i.e., we cannot obtain the radial action using $I_r = \oint p_r(r) \dd r$, but instead need to integrate over a loop on the torus in phase space. This issue has been mitigated at lower PN orders by the existence of an extra approximate constant of motion~\cite{Racine:2008qv}, but little progress has been achieved beyond the 1.5PN and 2PN orders~\cite{Klein:2021jtd,Samanta:2022yfe,Witzany:2024ttz,Morras:2025nlp,Colin:2026iry}.

\acknowledgments
DT thanks Luc Blanchet, David Langlois, and \'{E}tienne Ligout for sharing their results  before publication, as well as
Riccardo Gonzo, Paul Ramond, and Tanja Hinderer for interesting discussions. DT acknowledges the use of  PNpedia~\cite{PNpedia}. \texttt{Wolfram Mathematica} was used for most computations. \texttt{Claude AI}  was used to: (i) explore the literature; (ii)  format long equations; (iii) check consistency between the raw results, the Supplemental Material, the displayed equations, and equations in the literature; (iv) assist in the computation of the master integral of App.~\ref{app:integrals}; (v) flag typos in the draft. 
DT acknowledges the support of the ERC Consolidator/UKRI Frontier Research Grant GWModels (selected by the ERC and funded by UKRI [grant number EP/Y008251/1]). 
\appendix

\section{Master integrals for scattering}
\label{app:integrals}
Our goal is to perform the integral 
\begin{align}
    \mathcal{J}_n=\int_0^{s_+}\dd s \, \frac{s^n}{\sqrt{(s-s_-)(s_+-s)}} \,.
\end{align}
where $s_-<0<s_+$.
For the special case $n=0$,
we perform the change of variables
$s=\frac{s_++s_-}{2}+\frac{s_+-s_-}{2}x$, and find that
\begin{align}
    \mathcal{J}_0&=\int_{-\frac{s_++s_-}{s_+-s_-}}^{1}\frac{\dd x}{\sqrt{1-x^2}} \,.
\end{align}
We then recognize, for $-1<\alpha<0$, the identities
\begin{align}
   \int_{\alpha}^{1}\frac{\dd x}{\sqrt{1-x^2}}&= \frac{\pi}{2}-\arcsin\left(\alpha\right)=\arccos(\alpha) = 2 \arctan\left(\sqrt{\frac{1-\alpha}{1+\alpha}}\right) \,,
\end{align}
which immediately leads to 
\begin{align}\label{eq:J0_expression}
    \mathcal{J}_0 = 2 \arctan\left(\sqrt{-\frac{s_+}{s_-}}\right)\,.
\end{align}
In order to obtain the expression for $n\ge 1$, we use 
\begin{align}
    \frac{\dd}{\dd s} \Big[s^{n-1}\sqrt{(s-s_-)(s_+-s)}\Big] = \frac{-ns^n + \frac{2n-1}{2}(s_-+s_+) s^{n-1} - (n-1)s_-s_+s^{n-2}}{\sqrt{(s-s_-)(s_+-s) }} \,.
\end{align}
Integrating this identity over $s\in[0,s_+]$, we find
\begin{align}
    \Big[s^{n-1}\sqrt{(s-s_-)(s_+-s)}\Big]_0^{s_+} = - n \mathcal{J}_n + \frac{2n-1}{2}(s_-+s_+) \mathcal{J}_{n-1} - (n-1)s_-s_+ \mathcal{J}_{n-2} \,.
\end{align}
For the special case $n=1$, we obtain
\begin{align}
    -\sqrt{-s_- s_+}= - \mathcal{J}_1 +\frac{1}{2}(s_-+s_+)\mathcal{J}_0 \,,
\end{align}
therefore
\begin{align}
    \mathcal{J}_1= (s_-+s_+) \arctan\left(\sqrt{-\frac{s_+}{s_-}}\right) + \sqrt{-s_-s_+} \,.
\end{align}
For the general case $n\ge2$, we instead find that $\Big[s^{n-1}\sqrt{(s-s_-)(s_+-s)}\Big]_0^{s_+}=0$, hence the recurrence relation becomes
\begin{align}
    \mathcal{J}_n =  \frac{2n-1}{2n}(s_-+s_+) \mathcal{J}_{n-1} - \frac{(n-1)}{n}s_-s_+ \mathcal{J}_{n-2}\,.
\end{align}

Substituting $\mathcal{J}_n = \left(\di \sqrt{-s_- s_+}\right)^n \mathcal{K}_n$ in the previous equation, we recognize that $\mathcal{K}_n$ satisfies Bonnet's recursion formula~\cite{Bonnet:these},
\begin{align}
    (n+1) \mathcal{K}_{n+1}= (2n+1) z \mathcal{K}_n - n \mathcal{K}_{n-1} \,,
\end{align}
where $z = \frac{s_+ + s_-}{2\di \sqrt{-s_- s_+}}$. Two independent solutions are the Legendre functions of the first and second kind, $P_n(z)$ and $Q_n(z)$, such that 
\begin{align}
    \mathcal{K}_n  = A \,P_n(z) + B \, Q_n(z) \,,
\end{align}
where $A$ and $B$ are complex-valued and are to be determined from the initial conditions. The imaginary character of the argument $z$ of the Legendre functions can be mitigated by introducing the functions
\begin{align}
    \frak{P}_n(x) &= \di^{-n} P_n(\di x) = \frac{1}{2^n} \sum_{k=0}^{\lfloor n/2 \rfloor} \begin{pmatrix}
        n\\k
    \end{pmatrix} \begin{pmatrix}
        2n-2k\\n
    \end{pmatrix}
    x^{n-2k} \,,\\
    \frak{Q}_n(x) &= \di^{-n-1} Q_n(ix) = \frak{P}_n(x) \arctan(x) + \sum_{k=1}^{n} \frac{1}{k} \frak{P}_{k-1}(x) \frak{P}_{n-k}(x) \,,
\end{align}
which allows us to write the general solution in a basis of functions that are explicitly real-valued:
\begin{align}\label{eq:cal_Jn_of_frak_Pn_Qn}
    \mathcal{J}_n &= (-1)^n\left( \sqrt{-s_- s_+}\right)^n\ \bigg[A \,    \frak{P}_n\left(- \frac{s_+ + s_-}{2 \sqrt{-s_- s_+}} \right)+ \di B \,   \frak{Q}_n\left(- \frac{s_+ + s_-}{2 \sqrt{-s_- s_+}} \right)\bigg] \,.
\end{align}
Evaluating this expression for $n=0$, comparing it to Eq.~\eqref{eq:J0_expression}, and using the identity
\begin{align}\label{eq:arctan_identity}
    \arctan\left(-\frac{(s_+ + s_-)}{2\sqrt{-s_+ s_-}}\right) = \frac{\pi}{2} - 2 \arctan\left(\sqrt{\frac{s_+}{-s_-}}\right) \,,
\end{align}
we find that 
$A = \pi/2$ and $B=\di $. To further simplify the expression, we then write
\begin{align}\label{eq:frak_Pn_of_frak_Rn}
    \frak{P}_n\left(- \frac{s_+ + s_-}{2\sqrt{-s_+ s_-}}\right) & = \frac{(-1)^n}{(-s_+ s_-)^{n/2}}  \frak{R}_n(s_+,s_-) \,,
\end{align}
where   
\begin{align}\label{eq:frak_Rn_def}
    \frak{R}_n(s_+,s_-) =  \sum_{k=0}^{\lfloor n/2 \rfloor} \frac{1}{4^{n-k}}\begin{pmatrix}
        n\\k
    \end{pmatrix} \begin{pmatrix}
        2n-2k\\n
    \end{pmatrix}  (s_+ + s_-)^{n-2k}(-s_+ s_-)^{k}
\end{align}
is a bivariate polynomial in $s_+$ and $s_-$. Similarly,
\begin{align}\label{eq:frak_Qn_of_frak_Rn}
    \frak{Q}_n\!\left(\!- \frac{s_+ + s_-}{2\sqrt{-s_+ s_-}}\!\right) &= \frac{(-1)^n}{(-s_+ s_-)^{n/2}} \Bigg\{ \! \frak{R}_n(s_+,s_-)\arctan\left(\!- \frac{s_+ + s_-}{2\sqrt{-s_+ s_-}}\!\right) - 
    \sqrt{-s_+ s_-}  \sum_{k=1}^{n} \frac{1}{k}    \frak{R}_{k-1}(s_+,s_-)   \frak{R}_{n-k}(s_+,s_-)\! \Bigg\}\,.
\end{align}
Injecting \eqref{eq:frak_Pn_of_frak_Rn} and \eqref{eq:frak_Qn_of_frak_Rn}  into~\eqref{eq:cal_Jn_of_frak_Pn_Qn} and using the identity \eqref{eq:arctan_identity}, we finally find the closed-form expression
\begin{align}
    \mathcal{J}_n &=      2 \arctan\left(\sqrt{-\frac{s_+}{s_-}}\right)  \frak{R}_n(s_+,s_-)
    + \sqrt{-s_+ s_-}  \sum_{k=1}^{n} \frac{1}{k}    \frak{R}_{k-1}(s_+,s_-)   \frak{R}_{n-k}(s_+,s_-) \,.
\end{align}

Alternatively, one can compute the master integral in terms of hypergeometric functions. For this, we perform the variable change $x=s/s_+$ and define $\alpha=-s_-/s_+$, such that the integral reads
\begin{align}
    \mathcal{J}_n=\frac{s_+^{n}}{\sqrt{\alpha}}\int_0^1 \dd x \,x^n(1-x)^{-1/2}(1+x/\alpha)^{-1/2} \,.
\end{align}
We recognize a hypergeometric function:
\begin{align}
    {}_2F_1(a,b,c;z)= \frac{1}{B(b,c-b)}\int_0^1 \dd x \,x^{b-1}(1-x)^{c-b-1}(1-z x)^{-a} \,,
\end{align}
where we identify $a=1/2$, $b=n+1$, $c=n+3/2$ and $z=-1/\alpha$. The beta function simplifies to 
\begin{align}
    B(n+1,1/2) = \frac{\Gamma(n+1)\Gamma(1/2)}{\Gamma(n+3/2)} = \frac{2^{n+1}n!}{(2n+1)!!} \,.
\end{align}
The master integral then reads:
\begin{align}
    \mathcal{J}_n &= \frac{2^{n+1}n!}{(2n+1)!!} s_+^n\sqrt{-\frac{s_+}{s_-}} \ {}_2F_1 \!\left(\frac{1}{2},n+1;n+\frac{3}{2}; \frac{s_+}{s_-}\right) \,.
\end{align}
Using \href{https://dlmf.nist.gov/15.4#E3}{(15.4.3)} of \cite{NIST:DLMF}, we see that this formula recovers the case $n=0$. The other cases can be deduced recursively using Gauss's contiguous relations.
Note that a similar derivation in the bound case can be found in App.~A of \cite{Henry:2025uta}.

\section{Redshift invariant}
\label{app:redshift}

The linear-in-spin piece reads:
\begin{align}
    \langle z_1^{S^1} \rangle  &= \frac{\varepsilon ^{5/2}}{j} \Bigg\{\left(-3+\delta  \left(3-\frac{5 \nu }{2}\right)+\frac{21 \nu }{2}\right) \chi_{-}+\left(3-\frac{11 \nu }{2}+2 \nu ^2+\delta  \left(-3+\frac{3 \nu
   }{2}\right)\right) \chi_{+}\Bigg\} \nn\\
   &\quad +\varepsilon ^{7/2} \Bigg\{\frac{1}{j^2}\Bigg[\left(-\frac{105}{2}+\frac{819 \nu }{4}-\frac{117 \nu ^2}{4}+\delta  \left(\frac{105}{2}-\frac{147 \nu }{4}+\frac{33
   \nu ^2}{4}\right)\right) \chi_{-} \nn\\
   &\qquad\qquad +\left(\frac{105}{2}-\frac{357 \nu }{4}+\frac{177 \nu ^2}{4}-3 \nu ^3+\delta  \left(-\frac{105}{2}+\frac{189 \nu }{4}-\frac{9 \nu ^2}{4}\right)\right)
   \chi_{+}\Bigg]\nn\\
   &\qquad\quad +\frac{1}{j}\Bigg[\left(\frac{165}{8}-\frac{1329 \nu }{16}+\frac{381 \nu ^2}{16}+\delta  \left(-\frac{165}{8}+\frac{321 \nu }{16}-\frac{109 \nu ^2}{16}\right)\right) \chi_{-} \nn\\
   &\qquad\qquad  +\left(-\frac{165}{8}+\frac{699 \nu }{16}-\frac{439 \nu ^2}{16}+\frac{17 \nu ^3}{4}+\delta  \left(\frac{165}{8}-\frac{387 \nu }{16}+\frac{51 \nu ^2}{16}\right)\right) \chi_{+}\Bigg]\Bigg\}\nn\\
   &\quad + \varepsilon ^{9/2} \Bigg\{\frac{1}{j^{3/2}}\Bigg[\left(\frac{117}{2}-\frac{975 \nu }{4}+\frac{159 \nu ^2}{2}+\delta  \left(-\frac{117}{2}+\frac{159 \nu }{4}-\frac{15 \nu ^2}{2}\right)\right)
   \chi_{-}\nn\\
   &\qquad\qquad +\left(-\frac{117}{2}+\frac{393 \nu }{4}-\frac{75 \nu ^2}{2}+6 \nu ^3+\delta  \left(\frac{117}{2}-\frac{273 \nu }{4}+\frac{21 \nu ^2}{2}\right)\right) \chi_{+}\Bigg] \nn\\
   &\qquad +\frac{1}{j^2}\Bigg[\left(\frac{7455}{16}-\frac{67359 \nu }{32}+\frac{19857 \nu ^2}{16}-\frac{2997 \nu ^3}{32}+\delta  \left(-\frac{7455}{16}+\frac{16959 \nu }{32}-\frac{4635 \nu ^2}{16}+\frac{897
   \nu ^3}{32}\right)\right) \chi_{-}\nn\\
   &\qquad\qquad +\left(-\frac{7455}{16}+\frac{33549 \nu }{32}-\frac{3243 \nu ^2}{4}+\frac{7533 \nu ^3}{32}-\frac{75 \nu ^4}{8}+\delta  \left(\frac{7455}{16}-\frac{24309
   \nu }{32}+\frac{879 \nu ^2}{4}-\frac{225 \nu ^3}{32}\right)\right) \chi_{+}\Bigg] \nn\\
   &\qquad +\frac{1}{j^3}\Bigg[\left(-\frac{3465}{4}+\frac{30075 \nu }{8}-\frac{26985 \nu ^2}{16}+\frac{855 \nu ^3}{16}+\delta 
   \left(\frac{3465}{4}-\frac{6315 \nu }{8}+\frac{5925 \nu ^2}{16}-\frac{255 \nu ^3}{16}\right)\right) \chi_{-} \nn\\
   &\qquad\quad +\left(\frac{3465}{4}-\frac{13245 \nu }{8}+\frac{17325 \nu ^2}{16}-\frac{3825 \nu
   ^3}{16}+\frac{15 \nu ^4}{4}+\delta  \left(-\frac{3465}{4}+\frac{9285 \nu }{8}-\frac{3705 \nu ^2}{16}+\frac{45 \nu ^3}{16}\right)\right) \chi_{+}\Bigg]\nn\\
   &\qquad +\frac{1}{j}\Bigg[\left(-\frac{9825}{128}+\frac{85923 \nu }{256}-\frac{10887 \nu ^2}{64}+\frac{8343 \nu ^3}{256}+\delta  \left(\frac{9825}{128}-\frac{19443 \nu }{256}+\frac{2655 \nu
   ^2}{64}-\frac{2535 \nu ^3}{256}\right)\right) \chi_{-} \nn\\
   &\qquad +\left(\frac{9825}{128}-\frac{42453 \nu }{256}+\frac{7881 \nu ^2}{64}-\frac{12321 \nu ^3}{256}+\frac{363 \nu ^4}{64}+\delta 
   \left(-\frac{9825}{128}+\frac{30333 \nu }{256}-\frac{1365 \nu ^2}{32}+\frac{1089 \nu ^3}{256}\right)\right) \chi_{+}\Bigg]\Bigg\}
\end{align}
The quadratic-in-spin piece reads:
\begin{align}
    \langle z_1^{S^2} \rangle &= \frac{\varepsilon ^3}{j^{3/2}}\Bigg\{\left(-1+\delta  \left(1-\frac{13 \nu }{4}\right)+\frac{19 \nu }{4}-4 \nu ^2\right) \chi_{-}^2 \nn\\
    &\qquad \quad+\left(2-\frac{15 \nu }{2}+\delta  \left(-2+\frac{5 \nu
   }{2}\right)\right) \chi_{-} \chi_{+}+\left(-1+\delta  \left(1-\frac{\nu }{4}\right)+\frac{7 \nu }{4}\right) \chi_{+}^2 \Bigg\} \nn\\
   &  +\varepsilon ^4
   \Bigg\{\frac{1}{j^{3/2}}\Bigg[\left(\frac{147}{8}-\frac{2921 \nu }{32}+\frac{2567 \nu ^2}{32}-\frac{25 \nu ^3}{2}+\delta  \left(-\frac{147}{8}+\frac{2239 \nu }{32}-\frac{313 \nu ^2}{32}\right)\right) \chi_{-}^2 \nn\\
   &\qquad +\left(-\frac{147}{4}+\frac{2725 \nu }{16}-\frac{1707 \nu ^2}{16}+\delta  \left(\frac{147}{4}-\frac{1055 \nu }{16}+\frac{505 \nu ^2}{16}\right)\right) \chi_{-} \chi_{+} \nn\\
   &\qquad  +\left(\frac{147}{8}-\frac{1541 \nu }{32}+\frac{1259 \nu ^2}{32}-8 \nu ^3+\delta  \left(-\frac{147}{8}+\frac{859 \nu }{32}-\frac{205 \nu ^2}{32}\right)\right) \chi_{+}^2\Bigg]\nn\\
   &\qquad +\frac{1}{j^{5/2}}\Bigg[\left(-63+\frac{1221 \nu }{4}-\frac{1809 \nu ^2}{8}+12 \nu ^3+\delta  \left(63-\frac{921 \nu }{4}+\frac{75 \nu ^2}{8}\right)\right) \chi_{-}^2 \nn\\
   &\qquad\qquad +\left(126-555 \nu
   +\frac{1107 \nu ^2}{4}+\delta  \left(-126+201 \nu -\frac{321 \nu ^2}{4}\right)\right) \chi_{-} \chi_{+} \nn\\
   &\qquad \qquad +\left(-63+\frac{591 \nu }{4}-\frac{873 \nu ^2}{8}+24 \nu ^3+\delta 
   \left(63-\frac{291 \nu }{4}+\frac{147 \nu ^2}{8}\right)\right) \chi_{+}^2\Bigg]\Bigg\}\nn\\
   & +\varepsilon ^5 \Bigg\{\frac{1}{j^{7/2}}\Bigg[\Bigg(-\frac{7425}{4}+\frac{4085265 \nu }{448}-\frac{818655 \nu
   ^2}{112}+\frac{59205 \nu ^3}{32}-\frac{45 \nu ^4}{2} \nn\\
   &\qquad\qquad\qquad +\delta  \left(\frac{7425}{4}-\frac{3313455 \nu }{448}+\frac{388905 \nu ^2}{224}-\frac{555 \nu ^3}{32}\right)\Bigg) \chi_{-}^2\nn\\
   & \qquad\qquad +\Bigg(\frac{7425}{2}-\frac{4006455 \nu }{224}+\frac{779235 \nu ^2}{56}-\frac{22365 \nu ^3}{16} \nn\\
   &\qquad\qquad\qquad +\delta  \left(-\frac{7425}{2}+\frac{1451865 \nu }{224}-\frac{397755 \nu ^2}{112}+\frac{6735 \nu
   ^3}{16}\right)\Bigg) \chi_{-} \chi_{+} \nn\\
   & \qquad\qquad  +\Bigg(-\frac{7425}{4}+\frac{2144865 \nu }{448}-\frac{486015 \nu ^2}{112}+\frac{44985 \nu ^3}{32}-90 \nu ^4  \nn\\
   &\qquad\qquad\qquad +\delta 
   \left(\frac{7425}{4}-\frac{1373055 \nu }{448}+\frac{278025 \nu ^2}{224}-\frac{2175 \nu ^3}{32}\right)\Bigg) \chi_{+}^2\Bigg]\nn\\
   & \qquad +\frac{1}{j^{3/2}}\Bigg[\Bigg(-\frac{10899}{128}+\frac{223905 \nu
   }{512}-\frac{114525 \nu ^2}{256}+\frac{103737 \nu ^3}{512}-\frac{683 \nu ^4}{32}   \nn\\
   &\qquad\qquad\qquad +\delta  \left(\frac{10899}{128}-\frac{190287 \nu }{512}+\frac{44863 \nu ^2}{256}-\frac{8423 \nu
   ^3}{512}\right)\Bigg) \chi_{-}^2\nn\\
   &\qquad\qquad +\Bigg(\frac{10899}{64}-\frac{230325 \nu }{256}+\frac{122497 \nu ^2}{128}-\frac{61821 \nu ^3}{256} \nn\\
   &\qquad\qquad\qquad +\delta  \left(-\frac{10899}{64}+\frac{89559 \nu
   }{256}-\frac{33975 \nu ^2}{128}+\frac{19071 \nu ^3}{256}\right)\Bigg) \chi_{-} \chi_{+} \nn\\
   &\qquad\qquad +\Bigg(-\frac{10899}{128}+\frac{129597 \nu }{512}-\frac{73545 \nu ^2}{256}+\frac{72597 \nu
   ^3}{512}-20 \nu ^4 \nn\\
   &\qquad\qquad\qquad +\delta  \left(\frac{10899}{128}-\frac{95979 \nu }{512}+\frac{29251 \nu ^2}{256}-\frac{7907 \nu ^3}{512}\right)\Bigg) \chi_{+}^2\Bigg]\nn\\
   &\qquad +\frac{1}{j^2}\Bigg[\Bigg(\frac{153}{2}-\frac{2901 \nu }{8}+\frac{961 \nu ^2}{4}-52 \nu ^3+\delta  \left(-\frac{153}{2}+\frac{2427 \nu }{8}-\frac{279 \nu ^2}{4}\right)\Bigg) \chi_{-}^2 \nn\\
   &\qquad\qquad +\Bigg(-153+\frac{2913 \nu }{4}-\frac{1077 \nu ^2}{2}+45 \nu ^3+\delta  \left(153-\frac{939 \nu }{4}+\frac{199 \nu ^2}{2}-13 \nu ^3\right)\Bigg) \chi_{-} \chi_{+} \nn\\
   &\qquad\qquad +\Bigg(\frac{153}{2}-\frac{1425 \nu }{8}+\frac{541 \nu ^2}{4}-43 \nu ^3+4 \nu ^4+\delta  \left(-\frac{153}{2}+\frac{951 \nu }{8}-\frac{171 \nu ^2}{4}+3 \nu ^3\right)\Bigg) \chi_{+}^2 \Bigg]\nn\\
   &\qquad +\frac{1}{j^{5/2}}\Bigg[\Bigg(\frac{7161}{8}-\frac{256491 \nu }{56}+\frac{1985835 \nu ^2}{448}-\frac{97797 \nu ^3}{64}+\frac{93 \nu ^4}{2} \nn\\
   &\qquad\qquad\qquad +\delta  \left(-\frac{7161}{8}+\frac{418473 \nu
   }{112}-\frac{623745 \nu ^2}{448}+\frac{2295 \nu ^3}{64}\right)\Bigg) \chi_{-}^2 \nn\\
   &\qquad\qquad+\Bigg(-\frac{7161}{4}+\frac{1019667 \nu }{112}-\frac{1940115 \nu ^2}{224}+\frac{45927 \nu ^3}{32} \nn\\
   &\qquad\qquad\qquad +\delta 
   \left(\frac{7161}{4}-\frac{406653 \nu }{112}+\frac{540453 \nu ^2}{224}-\frac{13917 \nu ^3}{32}\right)\Bigg) \chi_{-} \chi_{+} \nn\\
   &\qquad\qquad +\Bigg(\frac{7161}{8}-\frac{294687 \nu
   }{112}+\frac{1232271 \nu ^2}{448}-\frac{69453 \nu ^3}{64}+111 \nu ^4 \nn\\
   &\qquad\qquad\qquad +\delta  \left(-\frac{7161}{8}+\frac{100089 \nu }{56}-\frac{399381 \nu ^2}{448}+\frac{5391 \nu ^3}{64}\right)\Bigg) \chi_{+}^2\Bigg]\Bigg\}
\end{align}
The cubic-in-spin piece reads

\begin{align}
    \langle z_1^{S^3} \rangle &=\varepsilon^{9/2}\Bigg\{
    \frac{1}{j^3}\Bigg[\left(-35+260\nu-\frac{965\nu^2}{2}+\delta\left(35-\frac{335\nu}{2}+135\nu^2\right)\right)\chi_{-}^3\nn\\
    &\qquad\qquad+\left(105-555\nu+\frac{2355\nu^2}{4}-120\nu^3+\delta\left(-105+\frac{825\nu}{2}-\frac{375\nu^2}{4}\right)\right)\chi_{-}^2\chi_{+}\nn\\
    &\qquad\qquad+ \left(-105+465\nu-225\nu^2+\delta\left(105-\frac{375\nu}{2}+\frac{135\nu^2}{2}\right)\right)\chi_{-}\chi_{+}^2\nn\\
    &\qquad\qquad+\left(35-80\nu+\frac{175\nu^2}{4}+\delta\left(-35+\frac{65\nu}{2}-\frac{15\nu^2}{4}\right)\right)\chi_{+}^3\Bigg]\nn\\
    &\qquad+\frac{1}{j^2}\Bigg[\left(\frac{21}{2}-\frac{315\nu}{4}+\frac{297\nu^2}{2}+\delta\left(-\frac{21}{2}+\frac{201\nu}{4}-42\nu^2\right)\right)\chi_{-}^3\nn\\
    &\qquad\qquad+\left(-\frac{63}{2}+\frac{675\nu}{4}-189\nu^2+48\nu^3+\delta\left(\frac{63}{2}-\frac{513\nu}{4}+\frac{75\nu^2}{2}\right)\right)\chi_{-}^2\chi_{+}\nn\\
    &\qquad\qquad+\left(\frac{63}{2}-\frac{585\nu}{4}+\frac{177\nu^2}{2}+\delta\left(-\frac{63}{2}+\frac{243\nu}{4}-27\nu^2\right)\right)\chi_{-}\chi_{+}^2\nn\\
    &\qquad\qquad+\left(-\frac{21}{2}+\frac{105\nu}{4}-18\nu^2+\delta\left(\frac{21}{2}-\frac{51\nu}{4}+\frac{3\nu^2}{2}\right)\right)\chi_{+}^3\Bigg]\Bigg\}
\end{align}
The quartic-in-spin piece reads
\begin{align}
\langle z_1^{S^4} \rangle  &= \varepsilon^5\Bigg\{\frac{1}{j^{5/2}}\Bigg[\left(3-\frac{831\nu}{32}+\frac{2181\nu^2}{32}-48\nu^3+\delta\left(-3+\frac{705\nu}{32}-\frac{1329\nu^2}{32}\right)\right)\chi_{-}^4\nn\\
&\qquad\qquad+\left(-12+\frac{729\nu}{8}-\frac{1389\nu^2}{8}+\delta\left(12-\frac{471\nu}{8}+\frac{405\nu^2}{8}\right)\right)\chi_{-}^3\chi_{+}\nn\\
&\qquad\qquad+\left(18-\frac{1485\nu}{16}+\frac{1371\nu^2}{16}+\delta\left(-18+\frac{1107\nu}{16}-\frac{39\nu^2}{16}\right)\right)\chi_{-}^2\chi_{+}^2\nn\\
&\qquad\qquad+\left(-12+\frac{393\nu}{8}-\frac{57\nu^2}{8}+\delta\left(12-\frac{135\nu}{8}+\frac{9\nu^2}{8}\right)\right)\chi_{-}\chi_{+}^3\nn\\
&\qquad\qquad+\left(3-\frac{159\nu}{32}+\frac{45\nu^2}{32}+\delta\left(-3+\frac{33\nu}{32}-\frac{9\nu^2}{32}\right)\right)\chi_{+}^4\Bigg]\nn\\
&\qquad+\frac{1}{j^{7/2}}\Bigg[\left(-\frac{15}{2}+\frac{2085\nu}{32}-\frac{5475\nu^2}{32}+120\nu^3+\delta\left(\frac{15}{2}-\frac{1755\nu}{32}+\frac{3255\nu^2}{32}\right)\right)\chi_{-}^4 \nn\\
&\qquad\qquad+\left(30-\frac{1815\nu}{8}+\frac{3435\nu^2}{8}+\delta\left(-30+\frac{1185\nu}{8}-\frac{1035\nu^2}{8}\right)\right)\chi_{-}^3\chi_{+}\nn\\
&\qquad\qquad+\left(-45+\frac{3735\nu}{16}-\frac{3525\nu^2}{16}+\delta\left(45-\frac{2745\nu}{16}+\frac{105\nu^2}{16}\right)\right)\chi_{-}^2\chi_{+}^2 \nn \\
&\qquad\qquad+\left(30-\frac{975\nu}{8}+\frac{135\nu^2}{8}+\delta\left(-30+\frac{345\nu}{8}-\frac{15\nu^2}{8}\right)\right)\chi_{-}\chi_{+}^3 \nn\\
&\qquad\qquad +\left(-\frac{15}{2}+\frac{405\nu}{32}-\frac{75\nu^2}{32}+\delta\left(\frac{15}{2}-\frac{75\nu}{32}+\frac{15\nu^2}{32}\right)\right)\chi_{+}^4\Bigg]\Bigg\}
\end{align}

\section{Gyroscopic invariant}
\label{app:gyroscopic}

The nonspinning contribution reads

\begin{align}
\psi_1^{S^0}(\varepsilon,j) &= \frac{\varepsilon}{j}\Bigg\{
\frac{3}{4} + \frac{1}{2}\nu - \frac{3}{4}\delta\Bigg\} + \frac{\varepsilon^{2}}{j}\Bigg\{
-\frac{3}{2} - \frac{9}{4}\nu + \frac{1}{2}\nu^{2} + \delta\bigg(\frac{3}{2} - \frac{3}{4}\nu\bigg) + \frac{1}{j}\Bigg[\frac{69}{16} + \frac{21}{4}\nu - \frac{3}{8}\nu^{2} + \delta\bigg(-\frac{69}{16} + \frac{9}{8}\nu\bigg)\Bigg]\Bigg\} \nn \\
&\quad + \frac{\varepsilon^{3}}{j}\Bigg\{
\frac{15}{32} + \frac{27}{16}\nu - \frac{33}{16}\nu^{2} + \frac{3}{8}\nu^{3} + \delta\bigg(-\frac{15}{32} + \frac{9}{8}\nu - \frac{9}{16}\nu^{2}\bigg) \nn \\
&\qquad + \frac{1}{j}\Bigg[-\frac{303}{16} - \frac{39}{2}\nu + \frac{75}{4}\nu^{2} - \frac{3}{4}\nu^{3} + \delta\bigg(\frac{303}{16} - \frac{177}{8}\nu + \frac{9}{4}\nu^{2}\bigg)\Bigg] \nn \\
&\qquad + \frac{1}{j^{2}}\Bigg[\frac{633}{16} + \frac{309}{8}\nu - \frac{669}{32}\nu^{2} + \frac{5}{16}\nu^{3} + \delta\bigg(-\frac{633}{16} + \frac{129}{4}\nu - \frac{45}{32}\nu^{2}\bigg)\Bigg]\Bigg\} 
\end{align}

The linear-in-spin contribution reads
\begin{align}
\psi_1^{S^1}(\varepsilon,j) &= \frac{\varepsilon^{3/2}}{j^{3/2}}\Bigg\{
\chi_{+}\Bigg(-\frac{1}{2} + \frac{1}{2}\delta\Bigg) + \chi_{-}\Bigg(\frac{1}{2} - 2 \nu - \frac{1}{2}\delta\Bigg)\Bigg\}   + \frac{\varepsilon^{5/2}}{j^{3/2}}\Bigg\{
\chi_{+}\Bigg(\frac{15}{4} - 2 \nu^{2} + \delta\bigg(-\frac{15}{4} + \frac{7}{2}\nu\bigg)\Bigg) \nn \\
&\qquad\qquad + \chi_{-}\Bigg(-\frac{15}{4} + \frac{31}{2}\nu - 3 \nu^{2} + \delta\bigg(\frac{15}{4} + 3 \nu\bigg)\Bigg) \nn \\
&\qquad + \frac{1}{j}\Bigg[\chi_{+}\Bigg(-\frac{45}{4} - \nu + 5 \nu^{2} + \delta\bigg(\frac{45}{4} - \frac{15}{2}\nu\bigg)\Bigg) + \chi_{-}\Bigg(\frac{45}{4} - 45 \nu + 3 \nu^{2} + \delta\bigg(-\frac{45}{4} - \frac{17}{2}\nu\bigg)\Bigg)\Bigg]\Bigg\} \nn \\
&\quad + \frac{\varepsilon^{7/2}}{j^{3/2}}\Bigg\{
\chi_{+}\Bigg(-\frac{51}{16} + \frac{3}{4}\nu + \frac{151}{16}\nu^{2} - 3 \nu^{3} + \delta\bigg(\frac{51}{16} - 9 \nu + \frac{73}{16}\nu^{2}\bigg)\Bigg) \nn \\
&\qquad\qquad + \chi_{-}\Bigg(\frac{51}{16} - \frac{69}{4}\nu + \frac{319}{16}\nu^{2} - 3 \nu^{3} + \delta\bigg(-\frac{51}{16} - \frac{15}{4}\nu + \frac{73}{16}\nu^{2}\bigg)\Bigg) \nn \\
&\qquad + \frac{1}{j}\Bigg[\chi_{+}\Bigg(90 - \frac{27}{4}\nu - \frac{859}{8}\nu^{2} + 16 \nu^{3} + \delta\bigg(-90 + \frac{261}{2}\nu - \frac{225}{8}\nu^{2}\bigg)\Bigg) \nn \\
&\qquad\qquad + \chi_{-}\Bigg(-90 + \frac{813}{2}\nu - \frac{1677}{8}\nu^{2} + \frac{15}{2}\nu^{3} + \delta\bigg(90 + \frac{309}{4}\nu - \frac{259}{8}\nu^{2}\bigg)\Bigg)\Bigg] \nn \\
&\qquad + \frac{1}{j^{2}}\Bigg[\chi_{+}\Bigg(-\frac{729}{4} - \frac{51}{4}\nu + \frac{2571}{16}\nu^{2} - \frac{51}{4}\nu^{3} + \delta\bigg(\frac{729}{4} - \frac{1629}{8}\nu + \frac{453}{16}\nu^{2}\bigg)\Bigg) \nn \\
&\qquad\qquad\qquad\quad + \chi_{-}\Bigg(\frac{729}{4} - 786 \nu + \frac{4269}{16}\nu^{2} - \frac{15}{4}\nu^{3} + \delta\bigg(-\frac{729}{4} - \frac{1275}{8}\nu + \frac{507}{16}\nu^{2}\bigg)\Bigg)\Bigg]\Bigg\} 
\end{align}

The quadratic-in-spin contribution reads

\begin{align}
\psi_1^{S^2}(\varepsilon,j) &= \frac{\varepsilon^{3}}{j^{2}}\Bigg\{
\chi_{+}^{2}\Bigg(-\frac{15}{4} + \frac{3}{2}\nu + \delta\bigg(\frac{15}{4} - 3 \nu\bigg)\Bigg)  + \chi_{+}\chi_{-}\Bigg(\frac{15}{2} - 33 \nu + 12 \nu^{2} - \frac{15}{2}\delta\Bigg) \nn \\
&\qquad\qquad + \chi_{-}^{2}\Bigg(-\frac{15}{4} + \frac{27}{2}\nu + 6 \nu^{2} + \delta\bigg(\frac{15}{4} - 15 \nu\bigg)\Bigg) \nn \\
&\qquad + \frac{1}{j}\Bigg[\chi_{+}^{2}\Bigg(10 - \frac{7}{2}\nu + \delta\bigg(-10 + \frac{13}{2}\nu\bigg)\Bigg)   + \chi_{+}\chi_{-}\Bigg(-20 + \frac{173}{2}\nu - 26 \nu^{2} + \delta\bigg(20 - \frac{1}{2}\nu\bigg)\Bigg) \nn \\
&\qquad\quad + \chi_{-}^{2}\Bigg(10 - 37 \nu - 12 \nu^{2} + \delta\bigg(-10 + 40 \nu\bigg)\Bigg)\Bigg]\Bigg\}
\end{align}

The cubic-in-spin contribution reads

\begin{align}
\psi_1^{S^3}(\varepsilon,j) &= \frac{\varepsilon^{7/2}}{j^{5/2}}\Bigg\{
\chi_{+}^{3}\Bigg(\frac{3}{2} - \frac{3}{2}\delta\Bigg)   + \chi_{+}^{2}\chi_{-}\Bigg(-\frac{9}{2} + 18 \nu + \frac{9}{2}\delta\Bigg) \nn \\
&\qquad\qquad + \chi_{+}\chi_{-}^{2}\Bigg(\frac{9}{2} - 18 \nu + \delta\bigg(-\frac{9}{2} + 18 \nu\bigg)\Bigg)  + \chi_{-}^{3}\Bigg(-\frac{3}{2} + 12 \nu - 24 \nu^{2} + \delta\bigg(\frac{3}{2} - 6 \nu\bigg)\Bigg) \nn \\
&\qquad + \frac{1}{j}\Bigg[\chi_{+}^{3}\Bigg(-3 + 3\delta\Bigg)  + \chi_{+}^{2}\chi_{-}\Bigg(9 - 36 \nu - 9\delta\Bigg) \nn \\
&\qquad\quad + \chi_{+}\chi_{-}^{2}\Bigg(-9 + 36 \nu + \delta\bigg(9 - 36 \nu\bigg)\Bigg)  + \chi_{-}^{3}\Bigg(3 - 24 \nu + 48 \nu^{2} + \delta\bigg(-3 + 12 \nu\bigg)\Bigg)\Bigg]\Bigg\}
\end{align}

\section{Scattering angle}
\label{app:scattering}

We present here the full expression for the local scattering angle. 

\begin{align}
\chi_\text{pp loc} &= 2\arctan\!\left(\frac{1}{\sqrt{\bar{\jmath}}}\right)\Bigg\{1 + \frac{3\bar{\varepsilon}}{\bar{\jmath}} + \bar{\varepsilon}^{2}\Bigg[\frac{1}{\bar{\jmath}}\Bigg(\frac{15}{4} - \frac{3}{2}\nu\Bigg) + \frac{1}{\bar{\jmath}^{2}}\Bigg(\frac{105}{4} - \frac{15}{2}\nu\Bigg)\Bigg] \nn \\
&\quad + \bar{\varepsilon}^{3}\Bigg[
\frac{1}{\bar{\jmath}}\Bigg(\frac{15}{16} - \frac{15}{16}\nu + \frac{3}{4}\nu^{2}\Bigg) + \frac{1}{\bar{\jmath}^{2}}\Bigg(\frac{315}{4} + \nu\bigg(-109 + \frac{123}{128}\pi^{2}\bigg) + \frac{45}{4}\nu^{2}\Bigg) \nn\\
&\qquad\qquad+ \frac{1}{\bar{\jmath}^{3}}\Bigg(\frac{1155}{4} + \nu\bigg(-\frac{625}{2} + \frac{615}{128}\pi^{2}\bigg) + \frac{105}{8}\nu^{2}\Bigg)\Bigg] \nn \\
&\quad + \bar{\varepsilon}^{4}\Bigg[
\frac{1}{\bar{\jmath}}\Bigg(\frac{15}{32}\nu^{2} - \frac{3}{8}\nu^{3}\Bigg) + \frac{1}{\bar{\jmath}^{2}}\Bigg(\frac{4725}{64} + \nu\bigg(-\frac{20323}{96} + \frac{35569}{8192}\pi^{2}\bigg) + \nu^{2}\bigg(\frac{4045}{32} - \frac{615}{512}\pi^{2}\bigg) - \frac{45}{4}\nu^{3}\Bigg) \nn \\
&\qquad + \frac{1}{\bar{\jmath}^{3}}\Bigg(\frac{45045}{32} + \nu\bigg(-\frac{293413}{96} + \frac{257195}{4096}\pi^{2}\bigg) + \nu^{2}\bigg(\frac{35065}{32} - \frac{615}{32}\pi^{2}\bigg) - \frac{525}{16}\nu^{3}\Bigg) \nn \\
&\qquad + \frac{1}{\bar{\jmath}^{4}}\Bigg(\frac{225225}{64} + \nu\bigg(-\frac{1736399}{288} + \frac{2975735}{24576}\pi^{2}\bigg) + \nu^{2}\bigg(\frac{132475}{96} - \frac{7175}{256}\pi^{2}\bigg) - \frac{315}{16}\nu^{3}\Bigg)\Bigg]\Bigg\} \nn \\
&+ \frac{\bar{\varepsilon}\,\bar{\jmath}^{1/2}}{\bar{\jmath}+1}\Bigg[
\frac{15}{4} - \frac{1}{4}\nu + \frac{3 \pi}{\bar{\jmath}^{1/2}} + \frac{6}{\bar{\jmath}} + \frac{3 \pi}{\bar{\jmath}^{3/2}}\Bigg] \nn \\
&+ \frac{\bar{\varepsilon}^{2}\,\bar{\jmath}^{3/2}}{(\bar{\jmath}+1)^{2}}\Bigg[
\frac{35}{64} + \frac{15}{32}\nu + \frac{3}{64}\nu^{2} + \frac{\pi}{\bar{\jmath}^{1/2}}\Bigg(\frac{15}{4} - \frac{3}{2}\nu\Bigg) + \frac{1}{\bar{\jmath}}\Bigg(\frac{2593}{64} - \frac{419}{32}\nu + \frac{1}{64}\nu^{2}\Bigg) + \frac{\pi}{\bar{\jmath}^{3/2}}\Bigg(\frac{135}{4} - \frac{21}{2}\nu\Bigg) \nn \\
&\qquad\quad + \frac{1}{\bar{\jmath}^{2}}\Bigg(95 - 28 \nu\Bigg)  + \frac{\pi}{\bar{\jmath}^{5/2}}\Bigg(\frac{225}{4} - \frac{33}{2}\nu\Bigg) + \frac{1}{\bar{\jmath}^{3}}\Bigg(\frac{105}{2} - 15 \nu\Bigg) + \frac{\pi}{\bar{\jmath}^{7/2}}\Bigg(\frac{105}{4} - \frac{15}{2}\nu\Bigg)\Bigg] \nn \\
&+ \frac{\bar{\varepsilon}^{3}\,\bar{\jmath}^{5/2}}{(\bar{\jmath}+1)^{3}}\Bigg[
-\frac{21}{512} + \frac{105}{512}\nu - \frac{15}{512}\nu^{2} - \frac{5}{512}\nu^{3} + \frac{\pi}{\bar{\jmath}^{1/2}}\Bigg(\frac{15}{16} - \frac{15}{16}\nu + \frac{3}{4}\nu^{2}\Bigg) \nn \\
&\qquad\quad + \frac{1}{\bar{\jmath}}\Bigg(\frac{12001}{256} - \frac{16805}{256}\nu + \frac{3227}{256}\nu^{2} - \frac{5}{768}\nu^{3}\Bigg)  + \frac{\pi}{\bar{\jmath}^{3/2}}\Bigg(\frac{1305}{16} + \nu\bigg(-\frac{1789}{16} + \frac{123}{128}\pi^{2}\bigg) + \frac{27}{2}\nu^{2}\Bigg) \nn \\
&\qquad\quad + \frac{1}{\bar{\jmath}^{2}}\Bigg(\frac{315135}{512} + \nu\bigg(-\frac{1176049}{1536} + \frac{533}{64}\pi^{2}\bigg) + \frac{33533}{512}\nu^{2} - \frac{1}{512}\nu^{3}\Bigg)  \nn \\
&\qquad\quad + \frac{\pi}{\bar{\jmath}^{5/2}}\Bigg(\frac{8445}{16} + \nu\bigg(-\frac{10277}{16} + \frac{123}{16}\pi^{2}\bigg) + \frac{393}{8}\nu^{2}\Bigg) + \frac{1}{\bar{\jmath}^{3}}\Bigg(\frac{13539}{8} + \nu\bigg(-\frac{46997}{24} + \frac{1681}{64}\pi^{2}\bigg) + \frac{477}{4}\nu^{2}\Bigg) \nn \\
&\qquad\quad + \frac{\pi}{\bar{\jmath}^{7/2}}\Bigg(\frac{17655}{16} + \nu\bigg(-\frac{20247}{16} + \frac{1107}{64}\pi^{2}\bigg) + \frac{591}{8}\nu^{2}\Bigg) + \frac{1}{\bar{\jmath}^{4}}\Bigg(\frac{3395}{2} + \nu\bigg(-\frac{5654}{3} + \frac{1763}{64}\pi^{2}\bigg) + \frac{185}{2}\nu^{2}\Bigg) \nn \\
&\qquad\quad + \frac{\pi}{\bar{\jmath}^{9/2}}\Bigg(945 + \nu\bigg(-\frac{2093}{2} + \frac{123}{8}\pi^{2}\bigg) + \frac{405}{8}\nu^{2}\Bigg) + \frac{1}{\bar{\jmath}^{5}}\Bigg(\frac{1155}{2} + \nu\bigg(-625 + \frac{615}{64}\pi^{2}\bigg) + \frac{105}{4}\nu^{2}\Bigg) \nn \\
&\qquad\quad + \frac{\pi}{\bar{\jmath}^{11/2}}\Bigg(\frac{1155}{4} + \nu\bigg(-\frac{625}{2} + \frac{615}{128}\pi^{2}\bigg) + \frac{105}{8}\nu^{2}\Bigg)\Bigg] \nn \\
&+ \frac{\bar{\varepsilon}^{4}\,\bar{\jmath}^{7/2}}{(\bar{\jmath}+1)^{4}}\Bigg[
\frac{99}{16384} - \frac{105}{4096}\nu + \frac{105}{8192}\nu^{2} + \frac{15}{4096}\nu^{3} + \frac{35}{16384}\nu^{4} + \frac{\pi}{\bar{\jmath}^{1/2}}\Bigg(\frac{15}{32}\nu^{2} - \frac{3}{8}\nu^{3}\Bigg) \nn \\
&\qquad\quad + \frac{1}{\bar{\jmath}}\Bigg(\frac{255651}{16384} - \frac{922597}{20480}\nu + \frac{368617}{8192}\nu^{2} - \frac{37481}{4096}\nu^{3} + \frac{35}{16384}\nu^{4}\Bigg) \nn \\
&\qquad\quad + \frac{\pi}{\bar{\jmath}^{3/2}}\Bigg(\frac{4725}{64} + \nu\bigg(-\frac{20323}{96} + \frac{35569}{8192}\pi^{2}\bigg) + \nu^{2}\bigg(\frac{4105}{32} - \frac{615}{512}\pi^{2}\bigg) - \frac{51}{4}\nu^{3}\Bigg) \nn \\
&\qquad\quad + \frac{1}{\bar{\jmath}^{2}}\Bigg(\frac{20911701}{16384} + \nu\bigg(-\frac{1735825289}{552960} + \frac{2322439}{36864}\pi^{2}\bigg) \nn \\
&\qquad\qquad\qquad + \nu^{2}\bigg(\frac{108333751}{73728} - \frac{15539}{768}\pi^{2}\bigg) - \frac{362983}{4096}\nu^{3} + \frac{21}{16384}\nu^{4}\Bigg) \nn \\
&\qquad\quad + \frac{\pi}{\bar{\jmath}^{5/2}}\Bigg(\frac{54495}{32} + \nu\bigg(-\frac{374705}{96} + \frac{328333}{4096}\pi^{2}\bigg) + \nu^{2}\bigg(\frac{51335}{32} - \frac{3075}{128}\pi^{2}\bigg) - \frac{1281}{16}\nu^{3}\Bigg) \nn \\
&\qquad\quad + \frac{1}{\bar{\jmath}^{3}}\Bigg(\frac{181195269}{16384} + \nu\bigg(-\frac{12821210033}{552960} + \frac{17442025}{36864}\pi^{2}\bigg) \nn\\
&\qquad\qquad\qquad + \nu^{2}\bigg(\frac{603251335}{73728} - \frac{105083}{768}\pi^{2}\bigg) - \frac{1239551}{4096}\nu^{3} + \frac{5}{16384}\nu^{4}\Bigg) \nn \\
&\qquad\quad + \frac{\pi}{\bar{\jmath}^{7/2}}\Bigg(\frac{613935}{64} + \nu\bigg(-\frac{5623169}{288} + \frac{9788657}{24576}\pi^{2}\bigg) + \nu^{2}\bigg(\frac{626245}{96} - \frac{7175}{64}\pi^{2}\bigg) - \frac{3519}{16}\nu^{3}\Bigg) \nn \\
&\qquad\quad + \frac{1}{\bar{\jmath}^{4}}\Bigg(\frac{527109}{16} + \nu\bigg(-\frac{9121243}{144} + \frac{7904153}{6144}\pi^{2}\bigg) + \nu^{2}\bigg(\frac{454247}{24} - \frac{88273}{256}\pi^{2}\bigg) - \frac{2029}{4}\nu^{3}\Bigg) \nn \\
&\qquad\quad + \frac{\pi}{\bar{\jmath}^{9/2}}\Bigg(\frac{365085}{16} + \nu\bigg(-\frac{6235453}{144} + \frac{5397197}{6144}\pi^{2}\bigg) + \nu^{2}\bigg(\frac{1209655}{96} - \frac{29725}{128}\pi^{2}\bigg) - 321 \nu^{3}\Bigg) \nn \\
&\qquad\quad + \frac{1}{\bar{\jmath}^{5}}\Bigg(\frac{715575}{16} + \nu\bigg(-\frac{176084807}{2160} + \frac{30370361}{18432}\pi^{2}\bigg) + \nu^{2}\bigg(\frac{3127685}{144} - \frac{319595}{768}\pi^{2}\bigg) - \frac{1819}{4}\nu^{3}\Bigg) \nn \\
&\qquad\quad + \frac{\pi}{\bar{\jmath}^{11/2}}\Bigg(\frac{1716435}{64} + \nu\bigg(-\frac{1555591}{32} + \frac{8044599}{8192}\pi^{2}\bigg) + \nu^{2}\bigg(\frac{409255}{32} - \frac{126075}{512}\pi^{2}\bigg) - \frac{2085}{8}\nu^{3}\Bigg) \nn \\
&\qquad\quad + \frac{1}{\bar{\jmath}^{6}}\Bigg(\frac{915915}{32} + \nu\bigg(-\frac{10870553}{216} + \frac{37362595}{36864}\pi^{2}\bigg) + \nu^{2}\bigg(\frac{886405}{72} - \frac{93685}{384}\pi^{2}\bigg) - 210 \nu^{3}\Bigg) \nn \\
&\qquad\quad + \frac{\pi}{\bar{\jmath}^{13/2}}\Bigg(\frac{495495}{32} + \nu\bigg(-\frac{7825835}{288} + \frac{6723055}{12288}\pi^{2}\bigg) + \nu^{2}\bigg(\frac{635095}{96} - \frac{8405}{64}\pi^{2}\bigg) - \frac{1785}{16}\nu^{3}\Bigg) \nn \\
&\qquad\quad + \frac{1}{\bar{\jmath}^{7}}\Bigg(\frac{225225}{32} + \nu\bigg(-\frac{1736399}{144} + \frac{2975735}{12288}\pi^{2}\bigg) + \nu^{2}\bigg(\frac{132475}{48} - \frac{7175}{128}\pi^{2}\bigg) - \frac{315}{8}\nu^{3}\Bigg) \nn \\
&\qquad\quad + \frac{\pi}{\bar{\jmath}^{15/2}}\Bigg(\frac{225225}{64} + \nu\bigg(-\frac{1736399}{288} + \frac{2975735}{24576}\pi^{2}\bigg) + \nu^{2}\bigg(\frac{132475}{96} - \frac{7175}{256}\pi^{2}\bigg) - \frac{315}{16}\nu^{3}\Bigg)\Bigg]
\end{align}
The linear-in-spin piece reads
\begin{align}
\chi_{S^1} &= 2\arctan\!\left(\frac{1}{\sqrt{\bar{\jmath}}}\right)\Bigg\{\bar{\varepsilon}^{3/2}\Bigg[
\frac{1}{\bar{\jmath}^{3/2}}\Bigg(\chi_{+}\bigg(-4 + 2 \nu\bigg) - 4\chi_{-}\delta\Bigg)\Bigg] \nn \\
&\qquad\qquad + \bar{\varepsilon}^{5/2}\Bigg[
\frac{1}{\bar{\jmath}^{3/2}}\Bigg(\chi_{+}\bigg(-12 + 16 \nu - 2 \nu^{2}\bigg) + \chi_{-}\delta\bigg(-12 + 4 \nu\bigg)\Bigg) \nn \\
&\qquad\qquad\qquad + \frac{1}{\bar{\jmath}^{5/2}}\Bigg(\chi_{+}\bigg(-84 + \frac{147}{2}\nu - 3 \nu^{2}\bigg) + \chi_{-}\delta\bigg(-84 + \frac{21}{2}\nu\bigg)\Bigg)\Bigg] \nn \\
&\qquad\qquad + \bar{\varepsilon}^{7/2}\Bigg[
\frac{1}{\bar{\jmath}^{3/2}}\Bigg(\chi_{+}\bigg(-\frac{15}{2} + 21 \nu - \frac{27}{2}\nu^{2} + \frac{3}{2}\nu^{3}\bigg) + \chi_{-}\delta\bigg(-\frac{15}{2} + \frac{39}{4}\nu - 3 \nu^{2}\bigg)\Bigg) \nn \\
&\qquad\qquad\qquad + \frac{1}{\bar{\jmath}^{5/2}}\Bigg(\chi_{+}\bigg(-420 + 717 \nu - 207 \nu^{2} + 6 \nu^{3}\bigg) + \chi_{-}\delta\bigg(-420 + 297 \nu - 21 \nu^{2}\bigg)\Bigg) \nn \\
&\qquad\qquad\qquad + \frac{1}{\bar{\jmath}^{7/2}}\Bigg(\chi_{+}\bigg(-1485 + \frac{15165}{8}\nu - 345 \nu^{2} + \frac{15}{4}\nu^{3}\bigg) + \chi_{-}\delta\bigg(-1485 + \frac{5265}{8}\nu - \frac{75}{4}\nu^{2}\bigg)\Bigg)\Bigg]\Bigg\} \nn \\
&+ \frac{\bar{\varepsilon}^{3/2}}{\bar{\jmath}+1}\Bigg[
\chi_{+}\bigg(-4 + 2 \nu\bigg) - 4\chi_{-}\delta + \frac{1}{\bar{\jmath}^{1/2}}\Bigg(\chi_{+}\pi\bigg(-4 + 2 \nu\bigg) - 4 \pi\chi_{-}\delta\Bigg) + \frac{1}{\bar{\jmath}}\Bigg(\chi_{+}\bigg(-8 + 4 \nu\bigg) - 8\chi_{-}\delta\Bigg) \nn \\
&\qquad\quad + \frac{1}{\bar{\jmath}^{3/2}}\Bigg(\chi_{+}\pi\bigg(-4 + 2 \nu\bigg) - 4 \pi\chi_{-}\delta\Bigg)\Bigg] \nn \\
&+ \frac{\bar{\varepsilon}^{5/2}\,\bar{\jmath}}{(\bar{\jmath}+1)^{2}}\Bigg[
\chi_{+}\bigg(-\frac{5}{2} + \frac{11}{2}\nu - \frac{5}{4}\nu^{2}\bigg) + \chi_{-}\delta\bigg(-\frac{5}{2} + \frac{7}{4}\nu\bigg) \nn \\
&\qquad\quad + \frac{1}{\bar{\jmath}^{1/2}}\Bigg(\chi_{+}\pi\bigg(-12 + 16 \nu - 2 \nu^{2}\bigg) + \chi_{-}\delta\pi\bigg(-12 + 4 \nu\bigg)\Bigg) \nn \\
&\qquad\quad + \frac{1}{\bar{\jmath}}\Bigg(\chi_{+}\bigg(-\frac{259}{2} + 132 \nu - \frac{39}{4}\nu^{2}\bigg) + \chi_{-}\delta\bigg(-\frac{259}{2} + \frac{99}{4}\nu\bigg)\Bigg) \nn \\
&\qquad\quad + \frac{1}{\bar{\jmath}^{3/2}}\Bigg(\chi_{+}\pi\bigg(-108 + \frac{211}{2}\nu - 7 \nu^{2}\bigg) + \chi_{-}\delta\pi\bigg(-108 + \frac{37}{2}\nu\bigg)\Bigg) \nn \\
&\qquad\quad + \frac{1}{\bar{\jmath}^{2}}\Bigg(\chi_{+}\bigg(-304 + 277 \nu - 14 \nu^{2}\bigg) + \chi_{-}\delta\bigg(-304 + 43 \nu\bigg)\Bigg) \nn \\
&\qquad\quad + \frac{1}{\bar{\jmath}^{5/2}}\Bigg(\chi_{+}\pi\bigg(-180 + 163 \nu - 8 \nu^{2}\bigg) + \chi_{-}\delta\pi\bigg(-180 + 25 \nu\bigg)\Bigg) \nn \\
&\qquad\quad + \frac{1}{\bar{\jmath}^{3}}\Bigg(\chi_{+}\bigg(-168 + 147 \nu - 6 \nu^{2}\bigg) + \chi_{-}\delta\bigg(-168 + 21 \nu\bigg)\Bigg) \nn \\
&\qquad\quad + \frac{1}{\bar{\jmath}^{7/2}}\Bigg(\chi_{+}\pi\bigg(-84 + \frac{147}{2}\nu - 3 \nu^{2}\bigg) + \chi_{-}\delta\pi\bigg(-84 + \frac{21}{2}\nu\bigg)\Bigg)\Bigg] \nn \\
&+ \frac{\bar{\varepsilon}^{7/2}\,\bar{\jmath}^{2}}{(\bar{\jmath}+1)^{3}}\Bigg[
\chi_{+}\bigg(-\frac{7}{32} + \frac{85}{64}\nu - \frac{73}{32}\nu^{2} + \frac{43}{64}\nu^{3}\bigg) + \chi_{-}\delta\bigg(-\frac{7}{32} + \frac{25}{32}\nu - \frac{7}{8}\nu^{2}\bigg) \nn \\
&\qquad\quad + \frac{1}{\bar{\jmath}^{1/2}}\Bigg(\chi_{+}\pi\bigg(-\frac{15}{2} + 21 \nu - \frac{27}{2}\nu^{2} + \frac{3}{2}\nu^{3}\bigg) + \chi_{-}\delta\pi\bigg(-\frac{15}{2} + \frac{39}{4}\nu - 3 \nu^{2}\bigg)\Bigg) \nn \\
&\qquad\quad + \frac{1}{\bar{\jmath}}\Bigg(\chi_{+}\bigg(-\frac{4275}{16} + \frac{17679}{32}\nu - \frac{427}{2}\nu^{2} + \frac{371}{32}\nu^{3}\bigg) + \chi_{-}\delta\bigg(-\frac{4275}{16} + \frac{1959}{8}\nu - \frac{491}{16}\nu^{2}\bigg)\Bigg) \nn \\
&\qquad\quad + \frac{1}{\bar{\jmath}^{3/2}}\Bigg(\chi_{+}\pi\bigg(-\frac{885}{2} + 780 \nu - \frac{495}{2}\nu^{2} + \frac{21}{2}\nu^{3}\bigg) + \chi_{-}\delta\pi\bigg(-\frac{885}{2} + \frac{1305}{4}\nu - 30 \nu^{2}\bigg)\Bigg) \nn \\
&\qquad\quad + \frac{1}{\bar{\jmath}^{2}}\Bigg(\chi_{+}\bigg(-\frac{103863}{32} + \frac{319969}{64}\nu - \frac{41543}{32}\nu^{2} + \frac{2419}{64}\nu^{3}\bigg) + \chi_{-}\delta\bigg(-\frac{103863}{32} + \frac{62723}{32}\nu - \frac{2009}{16}\nu^{2}\bigg)\Bigg) \nn \\
&\qquad\quad + \frac{1}{\bar{\jmath}^{5/2}}\Bigg(\chi_{+}\pi\bigg(-\frac{5535}{2} + \frac{32877}{8}\nu - \frac{2013}{2}\nu^{2} + \frac{105}{4}\nu^{3}\bigg) + \chi_{-}\delta\pi\bigg(-\frac{5535}{2} + \frac{12627}{8}\nu - \frac{363}{4}\nu^{2}\bigg)\Bigg) \nn \\
&\qquad\quad + \frac{1}{\bar{\jmath}^{3}}\Bigg(\chi_{+}\bigg(-8789 + \frac{48827}{4}\nu - 2649 \nu^{2} + \frac{103}{2}\nu^{3}\bigg) + \chi_{-}\delta\bigg(-8789 + \frac{17997}{4}\nu - \frac{401}{2}\nu^{2}\bigg)\Bigg) \nn \\
&\qquad\quad + \frac{1}{\bar{\jmath}^{7/2}}\Bigg(\chi_{+}\pi\bigg(-\frac{11445}{2} + \frac{62871}{8}\nu - \frac{3339}{2}\nu^{2} + \frac{123}{4}\nu^{3}\bigg) + \chi_{-}\delta\pi\bigg(-\frac{11445}{2} + \frac{23001}{8}\nu - \frac{489}{4}\nu^{2}\bigg)\Bigg) \nn \\
&\qquad\quad + \frac{1}{\bar{\jmath}^{4}}\Bigg(\chi_{+}\bigg(-8760 + 11544 \nu - 2254 \nu^{2} + 32 \nu^{3}\bigg) + \chi_{-}\delta\bigg(-8760 + 4104 \nu - 142 \nu^{2}\bigg)\Bigg) \nn \\
&\qquad\quad + \frac{1}{\bar{\jmath}^{9/2}}\Bigg(\chi_{+}\pi\bigg(-4875 + \frac{51231}{8}\nu - 1242 \nu^{2} + \frac{69}{4}\nu^{3}\bigg) + \chi_{-}\delta\pi\bigg(-4875 + \frac{18171}{8}\nu - \frac{309}{4}\nu^{2}\bigg)\Bigg) \nn \\
&\qquad\quad + \frac{1}{\bar{\jmath}^{5}}\Bigg(\chi_{+}\bigg(-2970 + \frac{15165}{4}\nu - 690 \nu^{2} + \frac{15}{2}\nu^{3}\bigg) + \chi_{-}\delta\bigg(-2970 + \frac{5265}{4}\nu - \frac{75}{2}\nu^{2}\bigg)\Bigg) \nn \\
&\qquad\quad + \frac{1}{\bar{\jmath}^{11/2}}\Bigg(\chi_{+}\pi\bigg(-1485 + \frac{15165}{8}\nu - 345 \nu^{2} + \frac{15}{4}\nu^{3}\bigg) + \chi_{-}\delta\pi\bigg(-1485 + \frac{5265}{8}\nu - \frac{75}{4}\nu^{2}\bigg)\Bigg)\Bigg]
\end{align}

The quadratic-in-spin piece reads
\begin{align}
\chi_{S^2} &= 2\arctan\!\left(\frac{1}{\sqrt{\bar{\jmath}}}\right)\Bigg\{\bar{\varepsilon}^{2}\Bigg[
\frac{1}{\bar{\jmath}^{2}}\Bigg(\frac{3}{2}\chi_{+}^{2} + 3\chi_{+}\chi_{-}\delta + \chi_{-}^{2}\bigg(\frac{3}{2} - 6 \nu\bigg)\Bigg)\Bigg] \nn \\
&\qquad\qquad + \bar{\varepsilon}^{3}\Bigg[
\frac{1}{\bar{\jmath}^{2}}\Bigg(\chi_{+}^{2}\bigg(\frac{33}{2} - 24 \nu + 6 \nu^{2}\bigg) + \chi_{+}\chi_{-}\delta\bigg(33 - 27 \nu\bigg) + \chi_{-}^{2}\bigg(\frac{33}{2} - 69 \nu + 9 \nu^{2}\bigg)\Bigg) \nn \\
&\qquad\qquad\qquad + \frac{1}{\bar{\jmath}^{3}}\Bigg(\chi_{+}^{2}\bigg(105 - \frac{225}{2}\nu + 30 \nu^{2}\bigg) + \chi_{+}\chi_{-}\delta\bigg(210 - 120 \nu\bigg) + \chi_{-}^{2}\bigg(105 - \frac{855}{2}\nu + 15 \nu^{2}\bigg)\Bigg)\Bigg] \nn \\
&\qquad\qquad + \bar{\varepsilon}^{4}\Bigg[
\frac{1}{\bar{\jmath}^{2}}\Bigg(\chi_{+}^{2}\bigg(\frac{351}{16} - \frac{291}{4}\nu + \frac{963}{16}\nu^{2} - 9 \nu^{3}\bigg) + \chi_{+}\chi_{-}\delta\bigg(\frac{351}{8} - 96 \nu + \frac{291}{8}\nu^{2}\bigg)\nn\\
&\qquad\qquad\qquad\qquad+ \chi_{-}^{2}\bigg(\frac{351}{16} - 111 \nu + \frac{1467}{16}\nu^{2} - 9 \nu^{3}\bigg)\Bigg) \nn \\
&\qquad\qquad\qquad + \frac{1}{\bar{\jmath}^{3}}\Bigg(\chi_{+}^{2}\bigg(\frac{3885}{4} - \frac{7815}{4}\nu + \frac{7905}{8}\nu^{2} - 90 \nu^{3}\bigg) + \chi_{+}\chi_{-}\delta\bigg(\frac{3885}{2} - 2385 \nu + \frac{1605}{4}\nu^{2}\bigg) \nn \\
&\qquad\qquad\qquad\qquad + \chi_{-}^{2}\bigg(\frac{3885}{4} - \frac{17265}{4}\nu + \frac{12885}{8}\nu^{2} - \frac{75}{2}\nu^{3}\bigg)\Bigg) \nn \\
&\qquad\qquad\qquad + \frac{1}{\bar{\jmath}^{4}}\Bigg(\chi_{+}^{2}\bigg(\frac{51975}{16} - 5145 \nu + \frac{33075}{16}\nu^{2} - 105 \nu^{3}\bigg) + \chi_{+}\chi_{-}\delta\bigg(\frac{51975}{8} - \frac{23835}{4}\nu + \frac{4515}{8}\nu^{2}\bigg) \nn \\
&\qquad\qquad\qquad\qquad + \chi_{-}^{2}\bigg(\frac{51975}{16} - \frac{27615}{2}\nu + \frac{46935}{16}\nu^{2} - \frac{105}{4}\nu^{3}\bigg)\Bigg)\Bigg]\Bigg\} \nn \\
&+ \frac{\bar{\varepsilon}^{2}}{\bar{\jmath}^{1/2}\,(\bar{\jmath}+1)}\Bigg[
2\chi_{+}^{2} + 4\chi_{+}\chi_{-}\delta + \chi_{-}^{2}\bigg(2 - 8 \nu\bigg) + \frac{1}{\bar{\jmath}^{1/2}}\Bigg(\frac{3}{2}\pi\chi_{+}^{2} + 3 \pi\chi_{+}\chi_{-}\delta + \chi_{-}^{2}\pi\bigg(\frac{3}{2} - 6 \nu\bigg)\Bigg) \nn \\
&\qquad\qquad\qquad + \frac{1}{\bar{\jmath}}\Bigg(3\chi_{+}^{2} + 6\chi_{+}\chi_{-}\delta + \chi_{-}^{2}\bigg(3 - 12 \nu\bigg)\Bigg) + \frac{1}{\bar{\jmath}^{3/2}}\Bigg(\frac{3}{2}\pi\chi_{+}^{2} + 3 \pi\chi_{+}\chi_{-}\delta + \chi_{-}^{2}\pi\bigg(\frac{3}{2} - 6 \nu\bigg)\Bigg)\Bigg] \nn \\
&+ \frac{\bar{\varepsilon}^{3}\,\bar{\jmath}^{1/2}}{(\bar{\jmath}+1)^{2}}\Bigg[
\chi_{+}^{2}\bigg(\frac{17}{4} - \frac{39}{4}\nu + 2 \nu^{2}\bigg) + \chi_{+}\chi_{-}\delta\bigg(\frac{17}{2} - \frac{23}{2}\nu\bigg) + \chi_{-}^{2}\bigg(\frac{17}{4} - \frac{75}{4}\nu + 7 \nu^{2}\bigg) \nn \\
&\qquad\qquad + \frac{1}{\bar{\jmath}^{1/2}}\Bigg(\chi_{+}^{2}\pi\bigg(\frac{33}{2} - 24 \nu + 6 \nu^{2}\bigg) + \chi_{+}\chi_{-}\delta\pi\bigg(33 - 27 \nu\bigg) + \chi_{-}^{2}\pi\bigg(\frac{33}{2} - 69 \nu + 9 \nu^{2}\bigg)\Bigg) \nn \\
&\qquad\qquad + \frac{1}{\bar{\jmath}}\Bigg(\chi_{+}^{2}\bigg(167 - 200 \nu + 52 \nu^{2}\bigg) + \chi_{+}\chi_{-}\delta\bigg(334 - 218 \nu\bigg) + \chi_{-}^{2}\bigg(167 - 686 \nu + 46 \nu^{2}\bigg)\Bigg) \nn \\
&\qquad\qquad + \frac{1}{\bar{\jmath}^{3/2}}\Bigg(\chi_{+}^{2}\pi\bigg(138 - \frac{321}{2}\nu + 42 \nu^{2}\bigg) + \chi_{+}\chi_{-}\delta\pi\bigg(276 - 174 \nu\bigg) + \chi_{-}^{2}\pi\bigg(138 - \frac{1131}{2}\nu + 33 \nu^{2}\bigg)\Bigg) \nn \\
&\qquad\qquad + \frac{1}{\bar{\jmath}^{2}}\Bigg(\chi_{+}^{2}\bigg(383 - 423 \nu + 112 \nu^{2}\bigg) + \chi_{+}\chi_{-}\delta\bigg(766 - 454 \nu\bigg) + \chi_{-}^{2}\bigg(383 - 1563 \nu + 68 \nu^{2}\bigg)\Bigg) \nn \\
&\qquad\qquad + \frac{1}{\bar{\jmath}^{5/2}}\Bigg(\chi_{+}^{2}\pi\bigg(\frac{453}{2} - 249 \nu + 66 \nu^{2}\bigg) + \chi_{+}\chi_{-}\delta\pi\bigg(453 - 267 \nu\bigg) + \chi_{-}^{2}\pi\bigg(\frac{453}{2} - 924 \nu + 39 \nu^{2}\bigg)\Bigg) \nn \\
&\qquad\qquad + \frac{1}{\bar{\jmath}^{3}}\Bigg(\chi_{+}^{2}\bigg(210 - 225 \nu + 60 \nu^{2}\bigg) + \chi_{+}\chi_{-}\delta\bigg(420 - 240 \nu\bigg) + \chi_{-}^{2}\bigg(210 - 855 \nu + 30 \nu^{2}\bigg)\Bigg) \nn \\
&\qquad\qquad + \frac{1}{\bar{\jmath}^{7/2}}\Bigg(\chi_{+}^{2}\pi\bigg(105 - \frac{225}{2}\nu + 30 \nu^{2}\bigg) + \chi_{+}\chi_{-}\delta\pi\bigg(210 - 120 \nu\bigg) + \chi_{-}^{2}\pi\bigg(105 - \frac{855}{2}\nu + 15 \nu^{2}\bigg)\Bigg)\Bigg] \nn \\
&+ \frac{\bar{\varepsilon}^{4}\,\bar{\jmath}^{3/2}}{(\bar{\jmath}+1)^{3}}\Bigg[
\chi_{+}^{2}\bigg(\frac{95}{64} - \frac{277}{32}\nu + \frac{799}{64}\nu^{2} - \frac{9}{4}\nu^{3}\bigg) + \chi_{+}\chi_{-}\delta\bigg(\frac{95}{32} - \frac{197}{16}\nu + \frac{319}{32}\nu^{2}\bigg) + \chi_{-}^{2}\bigg(\frac{95}{64} - \frac{307}{32}\nu + \frac{1015}{64}\nu^{2} - \frac{79}{16}\nu^{3}\bigg) \nn \\
&\qquad\qquad + \frac{1}{\bar{\jmath}^{1/2}}\Bigg(\chi_{+}^{2}\pi\bigg(\frac{351}{16} - \frac{291}{4}\nu + \frac{963}{16}\nu^{2} - 9 \nu^{3}\bigg) + \chi_{+}\chi_{-}\delta\pi\bigg(\frac{351}{8} - 96 \nu + \frac{291}{8}\nu^{2}\bigg) \nn \\
&\qquad\qquad\qquad\qquad + \chi_{-}^{2}\pi\bigg(\frac{351}{16} - 111 \nu + \frac{1467}{16}\nu^{2} - 9 \nu^{3}\bigg)\Bigg) \nn \\
&\qquad\qquad + \frac{1}{\bar{\jmath}}\Bigg(\chi_{+}^{2}\bigg(\frac{42027}{64} - \frac{50625}{32}\nu + \frac{61259}{64}\nu^{2} - \frac{445}{4}\nu^{3}\bigg) + \chi_{+}\chi_{-}\delta\bigg(\frac{42027}{32} - \frac{31905}{16}\nu + \frac{15019}{32}\nu^{2}\bigg) \nn \\
&\qquad\qquad\qquad\qquad + \chi_{-}^{2}\bigg(\frac{42027}{64} - \frac{97239}{32}\nu + \frac{100611}{64}\nu^{2} - \frac{1139}{16}\nu^{3}\bigg)\Bigg) \nn \\
&\qquad\qquad + \frac{1}{\bar{\jmath}^{3/2}}\Bigg(\chi_{+}^{2}\pi\bigg(\frac{16593}{16} - 2172 \nu + \frac{18699}{16}\nu^{2} - 117 \nu^{3}\bigg) + \chi_{+}\chi_{-}\delta\pi\bigg(\frac{16593}{8} - 2673 \nu + \frac{4083}{8}\nu^{2}\bigg) \nn \\
&\qquad\qquad\qquad\qquad + \chi_{-}^{2}\pi\bigg(\frac{16593}{16} - \frac{18597}{4}\nu + \frac{30171}{16}\nu^{2} - \frac{129}{2}\nu^{3}\bigg)\Bigg) \nn \\
&\qquad\qquad+ \frac{1}{\bar{\jmath}^{2}}\Bigg(\chi_{+}^{2}\bigg(\frac{14721}{2} - \frac{27377}{2}\nu + \frac{26235}{4}\nu^{2} - 540 \nu^{3}\bigg) + \chi_{+}\chi_{-}\delta\bigg(14721 - 16454 \nu + \frac{4951}{2}\nu^{2}\bigg) \nn \\
&\qquad\qquad\qquad\qquad + \chi_{-}^{2}\bigg(\frac{14721}{2} - \frac{64415}{2}\nu + \frac{41031}{4}\nu^{2} - 237 \nu^{3}\bigg)\Bigg) \nn \\
&\qquad\qquad + \frac{1}{\bar{\jmath}^{5/2}}\Bigg(\chi_{+}^{2}\pi\bigg(6228 - \frac{22449}{2}\nu + \frac{41697}{8}\nu^{2} - 402 \nu^{3}\bigg) + \chi_{+}\chi_{-}\delta\pi\bigg(12456 - \frac{53607}{4}\nu + \frac{7509}{4}\nu^{2}\bigg) \nn \\
&\qquad\qquad\qquad\qquad + \chi_{-}^{2}\pi\bigg(6228 - \frac{108357}{4}\nu + \frac{64323}{8}\nu^{2} - \frac{663}{4}\nu^{3}\bigg)\Bigg) \nn \\
&\qquad\qquad + \frac{1}{\bar{\jmath}^{3}}\Bigg(\chi_{+}^{2}\bigg(19517 - \frac{66407}{2}\nu + 14486 \nu^{2} - 960 \nu^{3}\bigg) + \chi_{+}\chi_{-}\delta\bigg(39034 - \frac{78261}{2}\nu + 4696 \nu^{2}\bigg) \nn \\
&\qquad\qquad\qquad\qquad + \chi_{-}^{2}\bigg(19517 - 83995 \nu + \frac{43361}{2}\nu^{2} - \frac{667}{2}\nu^{3}\bigg)\Bigg) \nn \\
&\qquad\quad + \frac{1}{\bar{\jmath}^{7/2}}\Bigg(\chi_{+}^{2}\pi\bigg(12681 - 21369 \nu + \frac{73809}{8}\nu^{2} - 594 \nu^{3}\bigg) + \chi_{+}\chi_{-}\delta\pi\bigg(25362 - \frac{100509}{4}\nu + \frac{11733}{4}\nu^{2}\bigg) \nn \\
&\qquad\qquad\qquad\qquad + \chi_{-}^{2}\pi\bigg(12681 - \frac{217929}{4}\nu + \frac{109791}{8}\nu^{2} - \frac{801}{4}\nu^{3}\bigg)\Bigg) \nn \\
&\qquad\qquad + \frac{1}{\bar{\jmath}^{4}}\Bigg(\chi_{+}^{2}\bigg(\frac{38535}{2} - \frac{62695}{2}\nu + \frac{52005}{4}\nu^{2} - 740 \nu^{3}\bigg) + \chi_{+}\chi_{-}\delta\bigg(38535 - 36550 \nu + \frac{7625}{2}\nu^{2}\bigg) \nn \\
&\qquad\qquad\qquad\qquad + \chi_{-}^{2}\bigg(\frac{38535}{2} - \frac{164545}{2}\nu + \frac{75465}{4}\nu^{2} - 215 \nu^{3}\bigg)\Bigg) \nn \\
&\qquad\qquad + \frac{1}{\bar{\jmath}^{9/2}}\Bigg(\chi_{+}^{2}\pi\bigg(\frac{171465}{16} - \frac{69555}{4}\nu + \frac{115035}{16}\nu^{2} - 405 \nu^{3}\bigg) + \chi_{+}\chi_{-}\delta\pi\bigg(\frac{171465}{8} - \frac{81045}{4}\nu + \frac{16755}{8}\nu^{2}\bigg) \nn \\
&\qquad\qquad\qquad\qquad + \chi_{-}^{2}\pi\bigg(\frac{171465}{16} - \frac{182955}{4}\nu + \frac{166575}{16}\nu^{2} - \frac{465}{4}\nu^{3}\bigg)\Bigg) \nn \\
&\qquad\qquad + \frac{1}{\bar{\jmath}^{5}}\Bigg(\chi_{+}^{2}\bigg(\frac{51975}{8} - 10290 \nu + \frac{33075}{8}\nu^{2} - 210 \nu^{3}\bigg) + \chi_{+}\chi_{-}\delta\bigg(\frac{51975}{4} - \frac{23835}{2}\nu + \frac{4515}{4}\nu^{2}\bigg) \nn \\
&\qquad\qquad\qquad\qquad + \chi_{-}^{2}\bigg(\frac{51975}{8} - 27615 \nu + \frac{46935}{8}\nu^{2} - \frac{105}{2}\nu^{3}\bigg)\Bigg) \nn \\
&\qquad\qquad + \frac{1}{\bar{\jmath}^{11/2}}\Bigg(\chi_{+}^{2}\pi\bigg(\frac{51975}{16} - 5145 \nu + \frac{33075}{16}\nu^{2} - 105 \nu^{3}\bigg) + \chi_{+}\chi_{-}\delta\pi\bigg(\frac{51975}{8} - \frac{23835}{4}\nu + \frac{4515}{8}\nu^{2}\bigg) \nn \\
&\qquad\qquad\qquad\qquad + \chi_{-}^{2}\pi\bigg(\frac{51975}{16} - \frac{27615}{2}\nu + \frac{46935}{16}\nu^{2} - \frac{105}{4}\nu^{3}\bigg)\Bigg)\Bigg]
\end{align}

The cubic-in-spin piece reads
\begin{align}
\chi_{S^3} &= 2\arctan\!\left(\frac{1}{\sqrt{\bar{\jmath}}}\right)\Bigg\{\bar{\varepsilon}^{7/2}\Bigg[
\frac{1}{\bar{\jmath}^{5/2}}\Bigg(\chi_{+}^{3}\bigg(-12 + 12 \nu\bigg) + \chi_{+}^{2}\chi_{-}\delta\bigg(-36 + 24 \nu\bigg) \nn \\
&\qquad\qquad\qquad\qquad\qquad\qquad\qquad + \chi_{+}\chi_{-}^{2}\bigg(-36 + 156 \nu - 48 \nu^{2}\bigg) + \chi_{-}^{3}\delta\bigg(-12 + 48 \nu\bigg)\Bigg) \nn \\
&\qquad\qquad\qquad\qquad\qquad + \frac{1}{\bar{\jmath}^{7/2}}\Bigg(\chi_{+}^{3}\bigg(-60 + 45 \nu\bigg) + \chi_{+}^{2}\chi_{-}\delta\bigg(-180 + 90 \nu\bigg)\nn \\
&\qquad\qquad\qquad\qquad\qquad\qquad\qquad + \chi_{+}\chi_{-}^{2}\bigg(-180 + 765 \nu - 180 \nu^{2}\bigg)  + \chi_{-}^{3}\delta\bigg(-60 + 240 \nu\bigg)\Bigg)\Bigg]\Bigg\} \nn \\
&+ \frac{\bar{\varepsilon}^{7/2}}{(\bar{\jmath}+1)^{2}}\Bigg[
\chi_{+}^{3}\bigg(-4 + 6 \nu\bigg) + \chi_{+}^{2}\chi_{-}\delta\bigg(-12 + 12 \nu\bigg) + \chi_{+}\chi_{-}^{2}\bigg(-12 + 54 \nu - 24 \nu^{2}\bigg) + \chi_{-}^{3}\delta\bigg(-4 + 16 \nu\bigg) \nn \\
&\qquad\quad + \frac{1}{\bar{\jmath}^{1/2}}\Bigg(\chi_{+}^{3}\pi\bigg(-12 + 12 \nu\bigg) + \chi_{+}^{2}\chi_{-}\delta\pi\bigg(-36 + 24 \nu\bigg) \nn \\
&\qquad\qquad\qquad + \chi_{+}\chi_{-}^{2}\pi\bigg(-36 + 156 \nu - 48 \nu^{2}\bigg)  + \chi_{-}^{3}\delta\pi\bigg(-12 + 48 \nu\bigg)\Bigg) \nn \\
&\qquad\quad + \frac{1}{\bar{\jmath}}\Bigg(\chi_{+}^{3}\bigg(-104 + 88 \nu\bigg) + \chi_{+}^{2}\chi_{-}\delta\bigg(-312 + 176 \nu\bigg) \nn \\
&\qquad\qquad\qquad  + \chi_{+}\chi_{-}^{2}\bigg(-312 + 1336 \nu - 352 \nu^{2}\bigg)   + \chi_{-}^{3}\delta\bigg(-104 + 416 \nu\bigg)\Bigg) \nn \\
&\qquad\quad + \frac{1}{\bar{\jmath}^{3/2}}\Bigg(\chi_{+}^{3}\pi\bigg(-84 + 69 \nu\bigg) + \chi_{+}^{2}\chi_{-}\delta\pi\bigg(-252 + 138 \nu\bigg) \nn \\
&\qquad\qquad\qquad  + \chi_{+}\chi_{-}^{2}\pi\bigg(-252 + 1077 \nu - 276 \nu^{2}\bigg) + \chi_{-}^{3}\delta\pi\bigg(-84 + 336 \nu\bigg)\Bigg) \nn \\
&\qquad\quad + \frac{1}{\bar{\jmath}^{2}}\Bigg(\chi_{+}^{3}\bigg(-224 + 174 \nu\bigg) + \chi_{+}^{2}\chi_{-}\delta\bigg(-672 + 348 \nu\bigg) \nn \\
&\qquad\qquad\qquad + \chi_{+}\chi_{-}^{2}\bigg(-672 + 2862 \nu - 696 \nu^{2}\bigg)  + \chi_{-}^{3}\delta\bigg(-224 + 896 \nu\bigg)\Bigg) \nn \\
&\qquad\quad + \frac{1}{\bar{\jmath}^{5/2}}\Bigg(\chi_{+}^{3}\pi\bigg(-132 + 102 \nu\bigg) + \chi_{+}^{2}\chi_{-}\delta\pi\bigg(-396 + 204 \nu\bigg) \nn \\
&\qquad\qquad\qquad + \chi_{+}\chi_{-}^{2}\pi\bigg(-396 + 1686 \nu - 408 \nu^{2}\bigg) + \chi_{-}^{3}\delta\pi\bigg(-132 + 528 \nu\bigg)\Bigg) \nn \\
&\qquad\quad + \frac{1}{\bar{\jmath}^{3}}\Bigg(\chi_{+}^{3}\bigg(-120 + 90 \nu\bigg) + \chi_{+}^{2}\chi_{-}\delta\bigg(-360 + 180 \nu\bigg)\nn \\
&\qquad\qquad\qquad   + \chi_{+}\chi_{-}^{2}\bigg(-360 + 1530 \nu - 360 \nu^{2}\bigg)  + \chi_{-}^{3}\delta\bigg(-120 + 480 \nu\bigg)\Bigg) \nn \\
&\qquad\quad + \frac{1}{\bar{\jmath}^{7/2}}\Bigg(\chi_{+}^{3}\pi\bigg(-60 + 45 \nu\bigg) + \chi_{+}^{2}\chi_{-}\delta\pi\bigg(-180 + 90 \nu\bigg) \nn \\
&\qquad\qquad\qquad + \chi_{+}\chi_{-}^{2}\pi\bigg(-180 + 765 \nu - 180 \nu^{2}\bigg)   + \chi_{-}^{3}\delta\pi\bigg(-60 + 240 \nu\bigg)\Bigg)\Bigg]
\end{align}

The quartic-in-spin piece reads

\begin{align}
\chi_{S^4} &= 2\arctan\!\left(\frac{1}{\sqrt{\bar{\jmath}}}\right)\Bigg\{\bar{\varepsilon}^{4}\Bigg[
\frac{1}{\bar{\jmath}^{3}}\Bigg(\frac{15}{4}\chi_{+}^{4} + 15\chi_{+}^{3}\chi_{-}\delta + \chi_{+}^{2}\chi_{-}^{2}\bigg(\frac{45}{2} - 90 \nu\bigg) + \chi_{+}\chi_{-}^{3}\delta\bigg(15 - 60 \nu\bigg) + \chi_{-}^{4}\bigg(\frac{15}{4} - 30 \nu + 60 \nu^{2}\bigg)\Bigg) \nn \\
&\qquad + \frac{1}{\bar{\jmath}^{4}}\Bigg(\frac{105}{8}\chi_{+}^{4} + \frac{105}{2}\chi_{+}^{3}\chi_{-}\delta + \chi_{+}^{2}\chi_{-}^{2}\bigg(\frac{315}{4} - 315 \nu\bigg) + \chi_{+}\chi_{-}^{3}\delta\bigg(\frac{105}{2} - 210 \nu\bigg) + \chi_{-}^{4}\bigg(\frac{105}{8} - 105 \nu + 210 \nu^{2}\bigg)\Bigg)\Bigg]\Bigg\} \nn \\
&+ \frac{\bar{\varepsilon}^{4}}{\bar{\jmath}^{1/2}\,(\bar{\jmath}+1)^{2}}\Bigg[
2\chi_{+}^{4} + 8\chi_{+}^{3}\chi_{-}\delta + \chi_{+}^{2}\chi_{-}^{2}\bigg(12 - 48 \nu\bigg) + \chi_{+}\chi_{-}^{3}\delta\bigg(8 - 32 \nu\bigg) + \chi_{-}^{4}\bigg(2 - 16 \nu + 32 \nu^{2}\bigg) \nn \\
&\qquad\quad + \frac{1}{\bar{\jmath}^{1/2}}\Bigg(\frac{15}{4}\pi\chi_{+}^{4} + 15 \pi\chi_{+}^{3}\chi_{-}\delta + \chi_{+}^{2}\chi_{-}^{2}\pi\bigg(\frac{45}{2} - 90 \nu\bigg) + \chi_{+}\chi_{-}^{3}\delta\pi\bigg(15 - 60 \nu\bigg) + \chi_{-}^{4}\pi\bigg(\frac{15}{4} - 30 \nu + 60 \nu^{2}\bigg)\Bigg) \nn \\
&\qquad\quad + \frac{1}{\bar{\jmath}}\Bigg(\frac{53}{2}\chi_{+}^{4} + 106\chi_{+}^{3}\chi_{-}\delta + \chi_{+}^{2}\chi_{-}^{2}\bigg(159 - 636 \nu\bigg) + \chi_{+}\chi_{-}^{3}\delta\bigg(106 - 424 \nu\bigg) + \chi_{-}^{4}\bigg(\frac{53}{2} - 212 \nu + 424 \nu^{2}\bigg)\Bigg) \nn \\
&\qquad\quad + \frac{1}{\bar{\jmath}^{3/2}}\Bigg(\frac{165}{8}\pi\chi_{+}^{4} + \frac{165}{2}\pi\chi_{+}^{3}\chi_{-}\delta + \chi_{+}^{2}\chi_{-}^{2}\pi\bigg(\frac{495}{4} - 495 \nu\bigg) \nn \\
&\qquad\qquad\qquad\quad + \chi_{+}\chi_{-}^{3}\delta\pi\bigg(\frac{165}{2} - 330 \nu\bigg)  + \chi_{-}^{4}\pi\bigg(\frac{165}{8} - 165 \nu + 330 \nu^{2}\bigg)\Bigg) \nn \\
&\qquad\quad + \frac{1}{\bar{\jmath}^{2}}\Bigg(\frac{205}{4}\chi_{+}^{4} + 205\chi_{+}^{3}\chi_{-}\delta + \chi_{+}^{2}\chi_{-}^{2}\bigg(\frac{615}{2} - 1230 \nu\bigg) + \chi_{+}\chi_{-}^{3}\delta\bigg(205 - 820 \nu\bigg) + \chi_{-}^{4}\bigg(\frac{205}{4} - 410 \nu + 820 \nu^{2}\bigg)\Bigg) \nn \\
&\qquad\quad + \frac{1}{\bar{\jmath}^{5/2}}\Bigg(30 \pi\chi_{+}^{4} + 120 \pi\chi_{+}^{3}\chi_{-}\delta + \chi_{+}^{2}\chi_{-}^{2}\pi\bigg(180 - 720 \nu\bigg)\nn \\
&\qquad\qquad\qquad\quad + \chi_{+}\chi_{-}^{3}\delta\pi\bigg(120 - 480 \nu\bigg) + \chi_{-}^{4}\pi\bigg(30 - 240 \nu + 480 \nu^{2}\bigg)\Bigg) \nn \\
&\qquad\quad + \frac{1}{\bar{\jmath}^{3}}\Bigg(\frac{105}{4}\chi_{+}^{4} + 105\chi_{+}^{3}\chi_{-}\delta + \chi_{+}^{2}\chi_{-}^{2}\bigg(\frac{315}{2} - 630 \nu\bigg) + \chi_{+}\chi_{-}^{3}\delta\bigg(105 - 420 \nu\bigg) + \chi_{-}^{4}\bigg(\frac{105}{4} - 210 \nu + 420 \nu^{2}\bigg)\Bigg) \nn \\
&\qquad\quad + \frac{1}{\bar{\jmath}^{7/2}}\Bigg(\frac{105}{8}\pi\chi_{+}^{4} + \frac{105}{2}\pi\chi_{+}^{3}\chi_{-}\delta + \chi_{+}^{2}\chi_{-}^{2}\pi\bigg(\frac{315}{4} - 315 \nu\bigg) \nn \\
&\qquad\qquad\qquad\quad + \chi_{+}\chi_{-}^{3}\delta\pi\bigg(\frac{105}{2} - 210 \nu\bigg)   + \chi_{-}^{4}\pi\bigg(\frac{105}{8} - 105 \nu + 210 \nu^{2}\bigg)\Bigg)\Bigg]
\end{align}

\bibliography{references.bib}

\end{document}